\documentclass[draft,tightenlines,nofootinbib,preprint,aps,eqsecnum,amsmath,amssymb]{revtex4}

\newcommand{\beq}{\begin{equation}}
\newcommand{\eeq}{\end{equation}}
\newcommand{\bea}{\begin{eqnarray}}
\newcommand{\eea}{\end{eqnarray}}
\newcommand{\cir}{{\buildrel \circ \over =}}
\newcommand{\sgn}{\epsilon}
\newcommand{\eo}{{}^4{\buildrel \circ \over E}}

\begin{document}

\title{The Einstein-Maxwell-Particle System in the York Canonical Basis
of ADM Tetrad Gravity: III) The Post-Minkowskian  N-Body Problem,
its Post-Newtonian Limit in Non-Harmonic 3-Orthogonal Gauges and
Dark Matter as an Inertial Effect.}

\medskip

\author{David Alba}

\affiliation{Dipartimento di Fisica\\
Universita' di Firenze\\Polo Scientifico, via Sansone 1\\
 50019 Sesto Fiorentino, Italy\\
 E-mail ALBA@FI.INFN.IT}

\author{Luca Lusanna}

\affiliation{ Sezione INFN di Firenze\\ Polo Scientifico\\ Via Sansone 1\\
50019 Sesto Fiorentino (FI), Italy\\ Phone: 0039-055-4572334\\
FAX: 0039-055-4572364\\ E-mail: lusanna@fi.infn.it}

\today

\begin{abstract}

We conclude the study of the Post-Minkowskian linearization of ADM
tetrad gravity in the York canonical basis for asymptotically
Minkowskian space-times in the family of non-harmonic 3-orthogonal
gauges parametrized by the York time ${}^3K(\tau, \vec \sigma)$ (the
inertial gauge variable, not existing in Newton gravity, describing
the general relativistic remnant of the freedom in clock
synchronization in the definition of the instantaneous 3-spaces). As
matter we consider only N scalar point particles with a Grassmann
regularization of the self-energies and with a ultraviolet cutoff
making possible the PM linearization and the evaluation of the PM
solution for the gravitational field.

We study in detail all the properties of these PM space-times
emphasizing their dependence on the gauge variable ${}^3{\cal
K}_{(1)} = {1\over {\triangle}}\, {}^3K_{(1)}$ (the non-local York
time): Riemann and Weyl tensors, 3-spaces, time-like and null
geodesics, red-shift and luminosity distance. Then we study the
Post-Newtonian (PN) expansion of the PM equations of motion of the
particles. We find that in the two-body case at the 0.5PN order
there is a damping (or anti-damping) term depending only on
${}^3{\cal K}_{(1)}$. This open the possibility to explain dark
matter in Einstein theory as a relativistic inertial effect: the
determination of ${}^3{\cal K}_{(1)}$ from the masses and  rotation
curves of galaxies would give information on how to find a PM
extension of the existing PN Celestial frame (ICRS) used as
observational convention in the 4-dimensional description of stars
and galaxies. Dark matter would describe the difference between the
inertial and gravitational masses seen in the non-Euclidean
3-spaces, without a violation of their equality in the 4-dimensional
space-time as required by the equivalence principle.

\end{abstract}

\maketitle

\vfill\eject

\section{Introduction}

In Refs.\cite{1,2}, quoted as papers I and II respectively, we
studied Hamiltonian ADM tetrad gravity in asymptotically Minkowskian
space-times in the York canonical basis defined in Ref.\cite{3} and
its Hamiltonian Post-Minkoskian (HPM) linearization in a family of
non-harmonic 3-orthogonal gauges. Since in this formulation the
instantaneous 3-spaces are well defined, we have control on the
general relativistic remnant of the gauge freedom in clock
synchronization, whose relevance for gravitational physics will be
investigated in this paper, where the matter consists only of N
scalar point particles (without the transverse electro-magnetic
field present in papers I and II), in the Post-Minkowskian (PM)
approximation.

\medskip

The definition of 3-spaces, a pre-requisite for the formulation of
the Cauchy problem for the field equations, is done by using radar
4-coordinates $\sigma^A = (\sigma^{\tau} = \tau ; \sigma^r)$, $A =
\tau ,r$, adapted to the admissible 3+1 splitting of the space-time
and centered on an arbitrary time-like observer $x^{\mu}(\tau)$
(origin of the 3-coordinates $\sigma^r$): they define a non-inertial
frame centered on the observer, so that they are {\it observer and
frame- dependent}. The time variable $\tau$ is an arbitrary
monotonically increasing function of the proper time given by the
atomic clock carried by the observer. The instantaneous 3-spaces
identified by this convention for clock synchronization are denoted
$\Sigma_{\tau}$. The transformation $\sigma^A \mapsto x^{\mu} =
z^{\mu}(\tau, \sigma^r)$ to world 4-coordinates defines the
embedding $z^{\mu}(\tau, \vec \sigma)$ of the Riemannian
instantaneous 3-spaces $\Sigma_{\tau}$ into the space-time. By
choosing  world 4-coordinates centered on the time-like observer,
whose world-line is the time axis, we have $x^{\mu}(\tau) =
(x^o(\tau); 0)$: the condition $x^o(\tau) = const.$ is equivalent to
$\tau = const.$ and identifies the instantaneous 3-space
$\Sigma_{\tau}$. If the time-like observer coincides with an
asymptotic inertial observer $x^{\mu}(\tau) = x^{\mu}_o +
\epsilon^{\mu}_{\tau}\, \tau$ with $\epsilon^{\mu}_{\tau} = (1; 0)$,
$\epsilon^{\mu}_r = (0; \delta^i_r)$, $x^{\mu}_o = (x^o_o; 0)$, then
the natural embedding describing the given 3+1 splitting of
space-time is $z^{\mu}(\tau, \sigma^r) = x^{\mu}_o +
\epsilon^{\mu}_A\, \sigma^A$ and the world 4-metric is
${}^4g_{\mu\nu} = \epsilon^A_{\mu}\, \epsilon^B_{\nu}\, {}^4g_{AB}$
($\epsilon^A_{\mu}$ are flat asymptotic cotetrads,
$\epsilon^A_{\mu}\, \epsilon^{\mu}_B = \delta^A_B$,
$\epsilon^A_{\mu}\, \epsilon^{\nu}_A = \delta^{\mu}_{\nu}$).

\medskip

From now on we shall denote the curvilinear 3-coordinates $\sigma^r$
with the notation $\vec \sigma$ for the sake of simplicity. Usually
the convention of sum over repeated indices is used, except when
there are too many summations.

\bigskip

The 4-metric ${}^4g_{AB}$ has signature $\sgn\, (+---)$ with $\sgn =
\pm$ (the particle physics, $\sgn = +$, and general relativity,
$\sgn = -$, conventions). Flat indices $(\alpha )$, $\alpha = o, a$,
are raised and lowered by the flat Minkowski metric
${}^4\eta_{(\alpha )(\beta )} = \sgn\, (+---)$. We define
${}^4\eta_{(a)(b)} = - \sgn\, \delta_{(a)(b)}$ with a
positive-definite Euclidean 3-metric. On each instantaneous 3-space
$\Sigma_{\tau}$ we have that the 4-metric has a
direction-independent limit to the flat Minkowski 4-metric (the
asymptotic background) at spatial infinity ${}^4g_{AB}(\tau, \vec
\sigma) \rightarrow {}^4\eta_{AB(asym)} = \sgn\, (+---)$.

\bigskip

After a review of the York canonical basis and of the HPM
linearization in Subsections A and B respectively, we will outline
the new results of this paper in Subsection C.

\subsection{The York Canonical Basis}

In the York canonical basis of ADM tetrad gravity of paper I

\beq
 \begin{minipage}[t]{4 cm}
\begin{tabular}{|ll|ll|l|l|l|} \hline
$\varphi_{(a)}$ & $\alpha_{(a)}$ & $n$ & ${\bar n}_{(a)}$ &
$\theta^r$ & $\tilde \phi$ & $R_{\bar a}$\\ \hline
$\pi_{\varphi_{(a)}} \approx0$ &
 $\pi^{(\alpha)}_{(a)} \approx 0$ & $\pi_n \approx 0$ & $\pi_{{\bar n}_{(a)}} \approx 0$
& $\pi^{(\theta )}_r$ & $\pi_{\tilde \phi}$ & $\Pi_{\bar a}$ \\
\hline
\end{tabular}
\end{minipage}
 \label{1.1}
 \eeq

\noindent the family of non-harmonic 3-orthogonal gauges is the
family of Schwinger time gauges where we have

\bea
 &&\alpha_{(a)}(\tau ,\vec \sigma ) \approx 0,\qquad
 \varphi_{(a)}(\tau ,\vec \sigma ) \approx 0,\nonumber \\
 &&\theta^i(\tau ,\vec \sigma) \approx 0,\qquad
 \pi_{\tilde \phi}(\tau ,\vec \sigma) = {{c^3}\over {12\pi\, G}}\,
{}^3K(\tau ,\vec \sigma) \approx {{c^3}\over {12\pi\, G}}\, F(\tau,
\vec \sigma),
 \label{1.2}
 \eea

\noindent where $F(\tau, \vec \sigma)$ is an arbitrary numerical
function parametrizing the residual gauge freedom in clock
synchronization, namely in the choice of the non-dynamical aspect of
the instantaneous 3-spaces $\Sigma_{\tau}$.

\bigskip

In the York canonical basis we have (${}^4E^A_{(\alpha)}$ are
arbitrary tetrads; $\eo^A_{(\alpha)}$ and ${}^4{\buildrel \circ
\over {\bar E}}^A_{(\alpha)}$, $\eo^A_{(a)} =
R_{(a)(b)}(\alpha_{(e)})\, {}^4{\buildrel \circ \over {\bar
E}}^A_{(b)}$, are tetrads adapted to the 3-spaces; ${}^3e^r_{(a)}$,
${}^3e_{(a)r}$ and ${}^3{\bar e}^r_{(a)}$, ${}^3{\bar e}_{(a)r}$,
with ${}^3e^r_{(a)} = R_{(a)(b)}(\alpha_{(e)})\, {}^3{\bar
e}^r_{(b)}$, ${}^3e_{(a)r} = R_{(a)(b)}(\alpha_{(e)})\, {}^3{\bar
e}_{(b)r}$, are triads and cotriads on the 3-spaces $\Sigma_{\tau}$;
for the shift function we have $n_{(a)} = R_{(a)(b)}(\alpha_{(e)})\,
{\bar n}_{(a)}$)

\begin{eqnarray*}
  {}^4E^A_{(\alpha )} &=& {}^4{\buildrel \circ \over {\bar E}}^A_{(o)}\,
 L^{(o)}{}_{(\alpha )}(\varphi_{(c)}) + {}^4{\buildrel \circ \over
 {\bar E}}^A_{(b)}\, R^T_{(b)(a)}(\alpha_{(c)})\,
 L^{(a)}{}_{(\alpha )}(\varphi_{(c)}) \approx {}^4{\buildrel \circ \over {\bar E}}^A_{(o)},
 \nonumber \\
 &&{}\nonumber \\
  {}^4{\buildrel \circ \over {\bar E}}^A_{(o)}
 &=& \eo^A_{(o)} = {1\over {1 + n}}\, (1; - {\bar n}_{(a)}\,
 {}^3{\bar e}^r_{(a)}) = l^A,\qquad {}^4{\buildrel \circ \over
 {\bar E}}^A_{(a)} = (0; {}^3{\bar e}^r_{(a)}) \approx \eo^A_{(a)},
  \nonumber \\
 &&{}\nonumber  \\
 {}^4{\buildrel \circ \over {\bar E}}^{(o)}_A
 &=& \eo^{(o)}_A = (1 + n)\, (1; \vec 0) = \sgn\, l_A,\qquad
 {}^4{\buildrel \circ \over {\bar E}}^{(a)}_A
 = ({\bar n}_{(a)}; {}^3{\bar e}_{(a)r}) \approx \eo^{(a)}_A,\nonumber \\
 &&{}\nonumber \\
  {}^4g_{AB} &=& {}^4E^{(\alpha )}_A\, {}^4\eta_{(\alpha
 )(\beta )}\, {}^4E^{(\beta )}_B = \eo^{(\alpha )}_A\, {}^4\eta_{(\alpha
 )(\beta )}\, \eo^{(\beta )}_B = {}^4{\buildrel \circ \over {\bar E}}_A^{(\alpha)}\,
 {}^4\eta_{(\alpha)(\beta)}\, {}^4{\buildrel \circ \over {\bar E}}_B^{(\beta)},
 \end{eqnarray*}

\bea
 {}^4g_{\tau\tau} &=& \sgn\, \Big[(1 + n)^2 - \sum_a\,
 {\bar n}_{(a)}^2\Big],\nonumber \\
 {}^4g_{\tau r} &=& - \sgn\, \sum_a\, {\bar n}_{(a)}\, {}^3{\bar e}_{(a)r} =
 - \sgn\, {\tilde \phi}^{1/3}\,  Q_r,\nonumber \\
 {}^4g_{rs} &=& - \sgn\, {}^3g_{rs} = - \sgn\, \sum_a\, {}^3{\bar e}_{(a)r}\,
 {}^3{\bar e}_{(a)s} = - \sgn\, \phi^4\,
 {}^3{\hat g}_{rs}\, =\, - \sgn\, {\tilde \phi}^{2/3}\,  Q^2_r\,
 \delta_{rs},\nonumber \\
 &&{}\nonumber \\
 &&Q_a\, =\, e^{ \Gamma^{(1)}_a}\, =\, e^{\sum_{\bar a}^{1,2}\, \gamma_{\bar aa}\, R_{\bar a}},
 \qquad
 \tilde \phi = \phi^6 = \sqrt{\gamma} =
 \sqrt{det\, {}^3g} = {}^3\bar e,\nonumber \\
  &&{}\nonumber \\
 {}^3{\bar e}_{(a)r} &=& {\tilde \phi}^{1/3}\, Q_a\,
 \delta_{ra},\qquad
 {}^3{\bar e}^r_{(a)} = {\tilde \phi}^{- 1/3}\, Q^{-1}_a\,
 \delta_{ra}.
 \label{1.3}
 \eea

The set of numerical parameters $\gamma_{\bar aa}$ satisfies
\cite{3} $\sum_u\, \gamma_{\bar au} = 0$, $\sum_u\, \gamma_{\bar a
u}\, \gamma_{\bar b u} = \delta_{\bar a\bar b}$, $\sum_{\bar a}\,
\gamma_{\bar au}\, \gamma_{\bar av} = \delta_{uv} - {1\over 3}$. A
different York canonical basis is associated to each solution of
these equations. Let us remember \cite{4} that to avoid coordinate
singularities we must always have $N(\tau, \vec \sigma) = 1 +
n(\tau, \vec \sigma) > 0$ (3-spaces at different times do not
intersect each other), $\sgn\, {}^4g_{\tau\tau}(\tau, \vec \sigma) >
0$ (no rotating disk pathology) and ${}^3g_{rs}(\tau, \vec \sigma)$
with three distinct positive eigenvalues (M$\o$ller conditions).

\subsection{The HPM Linearization}

The standard decomposition used for the weak field approximation in
the harmonic gauges  is

\bea
 {}^4g_{\mu\nu} &=& {}^4\eta_{\mu\nu} + h_{\mu\nu},\qquad
  |h_{\mu\nu}|, |\partial_{\alpha}\, h_{\mu\nu}|, |\partial_{\alpha}\,
 \partial_{\beta}\, h_{\mu\nu}| << 1,\nonumber \\
 &&{}
 \label{1.4}
 \eea

\noindent where ${}^4\eta_{\mu\nu}$ is the flat metric in an
inertial frame of the background Minkowski space-time. This is
equivalent to take a 3+1 splitting of our space-time with an
inertial foliation, having Euclidean instantaneous 3-spaces, against
the equivalence principle and against the fact (explicitly shown in
paper I) that each solution of Einstein's equations has an
associated dynamically selected preferred 3+1 splitting. The final
$x^o = constant$ 3-spaces have a non-zero extrinsic curvature but
are not well defined in this formalism and depend on the initial
data at $\tau \rightarrow - \infty$ needed to solve the wave
equations in the linearized harmonic gauges for the lapse and shift
functions (see paper II).

\bigskip

Instead the HPM-linearization of paper II  of the Hamilton-Dirac
equations in the (non-harmonic) 3-orthogonal Schwinger time gauges
(\ref{1.2}) uses as background the asymptotic Minkowski 4-metric
existing in our asymptotically Minkowskian space-times. By using
radar 4-coordinates adapted to an admissible 3+1 splitting of
space-time, we put

\bea
 {}^4g_{AB}(\tau, \sigma^r) &=& {}^4g_{(1)AB}(\tau, \sigma^r) +
 O(\zeta^2)\, \rightarrow {}^4\eta_{AB(asym)}\,\, at\,  spatial\, infinity,\nonumber \\
 &&{}\nonumber \\
 &&{}^4g_{(1)AB}(\tau, \sigma^r) = {}^4\eta_{AB (asym)} + {}^4h_{(1)AB}(\tau,
 \sigma^r),\nonumber \\
 &&{}^4h_{(1)AB}(\tau, \sigma^r) = O(\zeta) \rightarrow 0\,\, at\,
   spatial\, infinity,
 \label{1.5}
 \eea

\noindent where $\zeta << 1$ is a small dimensionless  parameter,
the small perturbation ${}^4h_{(1)AB}$ has no intrinsic meaning in
the bulk and ${}^3g_{(1)rs}(\tau, \sigma^r) = - \sgn\,
{}^4g_{(1)rs}(\tau, \sigma^r)$ is the positive-definite 3-metric on
the instantaneous (non-Euclidean) 3-space $\Sigma_{\tau}$.

The asymptotic 4-metric allows to define both a flat d'Alambertian
$\Box = \partial_{\tau}^2 - \triangle$ and a flat Laplacian
$\triangle = \sum_r\, \partial_r^2$ on $\Sigma_{\tau}$ ($\partial_A
= {{\partial}\over {\partial\, \sigma^A}}$). We will also need the
flat distribution  $c(\vec \sigma, {\vec \sigma}^{'}) = {1\over
{\Delta}}\,\, \delta^3(\vec \sigma, {\vec \sigma}^{'}) = -
\frac{1}{4\pi}\,\frac{1}{|\vec \sigma - {\vec \sigma}^{'}|}$ with
$|\vec \sigma - {\vec \sigma}^{'}| = \sqrt{\sum_u\, (\sigma^u -
\sigma^{'\, u})^2}$, where $\delta^3(\vec \sigma, {\vec
\sigma}^{'})$ is the Dirac delta function on the 3-manifold
$\Sigma_{\tau}$.

Therefore we will solve the constraints and the Hamilton-Dirac
equations in a fictitious Euclidean inertial 3-space identified by
the asymptotic Minkowski metric and the solution will describe the
gravitational field in our well defined Riemannian 3-spaces modulo
corrections of order $O(\zeta^2)$. The instantaneous 3-spaces will
deviated from flat Euclidean 3-spaces by curvature effects of order
$O(\zeta)$, in accord with the equivalence principle.

\bigskip

We assume that the dimensionless configurational tidal variables
$R_{\bar a}$ in the York canonical basis satisfy the following
conditions

\bea
 &&| R_{\bar a}(\tau ,\vec \sigma ) = R_{(1)\bar a}(\tau,
 \vec \sigma) |  = O(\zeta) << 1,\nonumber \\
 &&{}\nonumber \\
 &&|\partial_u\, R_{\bar a}(\tau ,\vec \sigma )| \sim {1\over L}
 O(\zeta),\qquad |\partial_u\, \partial_v\, R_{\bar a}(\tau ,\vec
 \sigma )| \sim {1\over {L^2}} O(\zeta),\nonumber \\
 && |\partial_{\tau}\, R_{\bar a}| = {1\over L}\, O(\zeta),\qquad
 |\partial^2_{\tau}\, R_{\bar a}| = {1\over {L^2}}\, O(\zeta),\qquad
 |\partial_{\tau}\, \partial_u\, R_{\bar a}| = {1\over {L^2}}\, O(\zeta),
 \nonumber \\
 &&{}\nonumber \\
 &&\Rightarrow\,\, Q_a(\tau, \vec \sigma) = e^{\sum_{\bar a}\,
 \gamma_{\bar aa}\, R_{\bar a}(\tau, \vec \sigma)} = 1 +
 \Gamma^{(1)}_a(\tau, \vec \sigma) + O(\zeta^2),\nonumber \\
 &&\qquad \Gamma_a^{(1)} = \sum_{\bar a}\, \gamma_{\bar aa}\,
 R_{\bar a},\qquad \sum_a\, \Gamma^{(1)}_a = 0,\qquad R_{\bar a} =
 \sum_a\, \gamma_{\bar aa}\, \Gamma_a^{(1)},
 \label{1.6}
 \eea

\noindent where $L$ is a {\it big enough characteristic length
interpretable as the reduced wavelength $\lambda / 2\pi$ of the
resulting gravitational waves (GW)}. Therefore the tidal variables
$R_{\bar a}$ are slowly varying over the length $L$ and times $L/c$.
This also implies that the Riemann tensor ${}^4R_{ABCD}$, the Ricci
tensor ${}^4R_{AB}$ and the scalar 4-curvature ${}^4R$ behave as
${1\over {L^2}}\, O(\zeta)$. Also the intrinsic 3-curvature scalar
of the instantaneous 3-spaces $\Sigma_{\tau}$ is of order ${1\over
{L^2}}\, O(\zeta)$. To simplify the notation we use $R_{\bar a}$ for
$R_{(1)\bar a}$ in the rest of the paper. As shown in paper II, this
condition defines a weak field approximation.

\bigskip

Eq.(\ref{1.5}) can be implemented if we add the following
assumptions

\bea
 \tilde \phi &=& \phi^6 = \sqrt{det\, {}^3g_{rs}} = 1 +
 6\, \phi_{(1)} + O(\zeta^2),\nonumber \\
 N &=& 1 + n = 1 + n_{(1)} +
 O(\zeta^2),\qquad {\bar n}_{(a)} = {\bar n}_{(1)(a)} + O(\zeta^2),
 \nonumber \\
 &&{}\nonumber \\
 &&\Downarrow \nonumber \\
 &&{}\nonumber \\
  {}^4g_{(1)\tau\tau} &=& \sgn +   {}^4h_{(1)\tau\tau} = \sgn\, (1 + 2\, n_{(1)}) =
  \sgn + O(\zeta),\nonumber \\
  {}^4g_{(1)\tau r} &=&   {}^4h_{(1)\tau r} = - \sgn\, {\bar n}_{(1)(r)} = O(\zeta),\nonumber \\
  {}^4g_{(1)rs} &=&- \sgn\, \delta_{rs} +  {}^4h_{(1)rs} = -  \sgn\, [1 + 2\,
  (\Gamma_r^{(1)} + 2\, \phi_{(1)})]\, \delta_{rs} = - \sgn\, \delta_{rs} + O(\zeta),
 \label{1.7}
 \eea

\noindent while the triads and cotriads become ${}^3{\bar
e}^r_{(1)(a)} = \delta^r_a\, (1 - \Gamma^{(1)}_r - 2\, \phi_{(1)}) +
O(\zeta^2)$ and ${}^3{\bar e}_{(1)(a)r} = \delta_{ra}\, (1 +
\Gamma^{(1)}_r + 2\, \phi_{(1)}) + O(\zeta^2)$, respectively.

\bigskip

As shown in paper II, with these assumptions we have \footnote{Let
us remark that everywhere $\Pi_{(1) \bar a}$ appears in the
combination ${G\over {c^3}}\, \Pi_{(1) \bar a} = {1\over L}\,
O(\zeta)$, which behaves like $\partial_{\tau}\, R_{\bar a}$, i.e.
it varies slowly over $L$.}

 \bea
 {{8\pi\, G}\over {c^3}}\, \Pi_{\bar a}(\tau, \vec \sigma)\, &=&
  {{8\pi\, G}\over {c^3}}\, \Pi_{(1) \bar a}(\tau, \vec \sigma) =
 {1\over L}\, O(\zeta) \cir
 \Big[\partial_{\tau}\, R_{\bar a} - \sum_a\, \gamma_{\bar aa}\,
 \partial_a\, {\bar n}_{(1)(a)}\Big](\tau, \vec \sigma) + {1\over L}\,
 O(\zeta^2),\nonumber \\
 &&{}\nonumber \\
 &&\sigma_{(a)(a)} = \sigma_{(1)(a)(a)} = - {{8\pi\, G}\over
 {c^3}}\, \sum_{\bar a}\, \gamma_{\bar aa}\, \Pi_{(1) \bar a} +
 {1\over L}\, O(\zeta^2).
 \label{1.8}
 \eea

\noindent where $\sigma_{(a)(a)}$ are the diagonal elements of the
shear $\sigma_{(a)(b)}$ of the congruence of Eulerian observers,
whose 4-velocity is the unit normal to the 3-spaces $\Sigma_{\tau}$
as 3-sub-manifolds of space-time (see Subsection IID of paper I).
For the non-diagonal elements of the shear, for the momenta
$\pi_i^{(\theta)}$ and for the extrinsic curvature the assumptions
of paper II are

\begin{eqnarray*}
 &&\sigma_{(a)(b)}{|}_{a\not= b} = \sigma_{(1)(a)(b)}{|}_{a\not= b}
 = {1\over L}\, O(\zeta),\nonumber \\
 &&\Rightarrow\quad {{8\pi\, G}\over {c^3}}\, \pi_i^{(\theta)} =
 {1\over L}\, O(\zeta^2) = \sum_{a\not= b}\, (\Gamma^{(1)}_a -
 \Gamma^{(1)}_b)\, \epsilon_{iab}\, \sigma_{(1)(a)(b)} + {1\over
 L}\, O(\zeta^3),
  \end{eqnarray*}

 \bea
 &&{}^3K = {{12\pi\, G}\over {c^3}}\, \pi_{\tilde \phi} =
 {}^3K_{(1)} = {{12\pi\, G}\over {c^3}}\, \pi_{(1) \tilde \phi} =
 {1\over L}\, O(\zeta),\nonumber \\
 &&{}\nonumber \\
 &&\Downarrow\nonumber \\
 &&{}\nonumber \\
 {}^3K_{rs} &=& {}^3K_{(1)rs} = {1\over L}\, O(\zeta) =\nonumber \\
 &=& (1 - \delta_{rs})\,
 \sigma_{(1)(r)(s)} + \delta_{rs}\, \Big[{1\over 3}\, {}^3K_{(1)} -
 \partial_{\tau}\, \Gamma_r^{(1)} + \sum_a\, (\delta_{ra} - {1\over 3})\,
 \partial_a\, {\bar n}_{(1)(a)}\Big] + {1\over L}\, O(\zeta^2).\nonumber \\
 &&{}
 \label{1.9}
 \eea

\bigskip

Let us now consider our matter, i.e. positive-energy scalar
particles described by the 3-coordinates $\eta_i^r(\tau)$, $i =
1,..,N$, such that their world-lines are $x^{\mu}_i(\tau) =
z^{\mu}(\tau, {\vec \eta}_i(\tau))$. $\kappa_{ir}(\tau)$ are the
canonically conjugate 3-momenta. We have $\eta^r_i(\tau) = O(1)$ and
${\dot \eta}^r_i(\tau) = {{d\, \eta^r_i(\tau)}\over {d\tau}}\,
{\buildrel {def}\over =}\, {{v^r_i(t)}\over c} = O(1)$ since $\tau =
c\, t$ (in the non-relativistic limit we have $ {\dot {\vec \eta}}_i
= {\vec v}_i/c = O(1)\, \rightarrow_{c \rightarrow \infty}\, 0$).

\bigskip

As shown in paper II, to get a consistent approximation we must
introduce a {\it ultraviolet cutoff} $M$ on the masses and momenta
of the particles so that the mass density and the mass current
density (see the next Section for the energy-momentum of the
particles) satisfy the following requirements

\bea
 {\cal M}(\tau, \vec \sigma) &=& {\cal
 M}^{(UV)}_{(1)}(\tau, \vec \sigma) + {\cal R}_{(2)}(\tau, \vec \sigma),\nonumber \\
 &&{}\nonumber \\
 && m_i = M\, O(\zeta),\qquad
 \int d^3\sigma\, {\cal M}_{(1)}^{(UV)}(\tau, \vec
 \sigma) = Mc\, O(\zeta),\qquad \int d^3\sigma\, {\cal R}_{(2)}(\tau, \vec \sigma)
 = Mc\, O(\zeta^2),\nonumber \\
 &&{}\nonumber \\
 &&{}\nonumber \\
 {\cal M}_r(\tau, \vec \sigma) &=&
 {\cal M}_{(1)r}(\tau, \vec \sigma),\qquad
 \int d^3\sigma\, {\cal M}^{(UV)}_{(1)r}(\tau, \vec \sigma) =
 Mc\, O(\zeta).
 \label{1.10}
 \eea

\medskip

Here $M$ is a finite mass defining the ultraviolet cutoff: $M\, c^2$
gives an estimate of the weak ADM energy of the 3-universe contained
in the instantaneous 3-spaces $\Sigma_{\tau}$. The associated length
scale is the gravitational radius $R_M = 2M\, {G\over {c^2}} \approx
10^{-29}\, M$.
\medskip

The description of particles in our approximation will be reliable
only if their masses and momenta are less of $Mc\, O(\zeta)$ and at
distances $r$ from the particles satisfying $r = |\vec \sigma -
{\vec \eta}_i(\tau)| >> R_M$ (that is at each instant we must
enclose each particle in a sphere of radius $R_M$ and our
approximation is not reliable inside these spheres).

\bigskip

Therefore for the particles the validity of the weak field
approximation requires

\beq
 {\vec \eta}_i(\tau) = O(1),\qquad {{{\vec \kappa}_i(\tau)}\over {m_i c}} =
 O(1),\qquad {{{\vec \kappa}_i(\tau)}\over {Mc}} =
 O(\zeta),\qquad {{m_i}\over M} \leq  O(\zeta).
 \label{1.11}
 \eeq

\medskip
Our results are equivalent to a re-summation of the post-Newtonian
expansions valid for small rest masses  still having relativistic
velocities (${{{\check {\vec \kappa}}^2_i}\over {m_i^2\, c^2}} =
O(1)$, ${{{\vec v}_i}\over c}= O(1)$).\bigskip

Since, as said in Subsection IIE of paper I, we have that the matter
energy-momentum tensor satisfies $\nabla_A\, T^{AB}(\tau, \vec
\sigma) \equiv 0$ due to the Bianchi identities and since
${}^4g_{AB} = {}^4\eta_{AB(asym)} + O(\zeta)$, we must have
$\partial_A\, T_{(1)}^{AB}(\tau, \vec \sigma) \equiv 0 +
\partial_A\, {\cal R}^{AB}_{(2)}$. At the lowest order this implies

 \bea
  &&\partial_{\tau}\, {\cal M}^{(UV)}_{(1)} + \partial_r\,
 {\cal M}_{(1)r}^{(UV)} = 0 + \partial_A\, {\cal R}^{A\tau}_{(2)},\nonumber \\
 &&\partial_{\tau}\, {\cal M}^{(UV)}_{(1)r} + \partial_s\, T^{rs}_{(1)} =
 0 + \partial_A\, {\cal R}^{Ar}_{(2)},
 \label{1.12}
 \eea

\noindent as in inertial frames in Minkowski space-time. The
equation $\partial_A\, T_{(1)}^{AB}(\tau, \vec \sigma)\, \equiv\, 0
+ \partial_A\, {\cal R}^{AB}_{(2)}$ implies $\partial_A\,
\Big(T_{(1)}^{AB}(\tau, \vec \sigma)\, \sigma^C - T^{AC}_{(1)}(\tau,
\vec \sigma)\, \sigma^B\Big)\, \equiv\, 0 + \partial_A\, {\cal
R}^{ABC}_{(2)}$ (angular momentum conservation).

\bigskip

\bigskip

In conclusion, since the weak field linearized solution can be
trusted only at distances $d >> R_M$ from the particles, the GW's
described by our linearization must have a wavelength satisfying
$\lambda \approx L > d >> R_M$ (with the weak field approximation we
have $\lambda << {}^4{\cal R}$ without the slow motion assumption).
\medskip

If all the particles are contained in a compact set of radius $l_c$
(the source), the frequency $\nu = {c\over {\lambda}}$ of the
emitted GW's will be of the order  of the typical frequency
$\omega_s$ of the motion inside the source, where the typical
velocities are of the order $v \approx \omega_s\, l_c$. As a
consequence we get $\nu = {c\over {\lambda}} \approx \omega_s
\approx v/l_c$ or $\lambda \approx {c\over v}\, l_c >> R_M$, so that
we get ${v\over c} \approx {{l_c}\over {\lambda}} << {{l_c}\over
{R_M}}$ and $l_c >> R_M$ if ${v\over c} = O(1)$.
\medskip

If the velocities of the particles become non-relativistic, i.e. in
the slow motion regime with $v << c$ (for binary systems with total
mass m and held together by weak gravitational forces we have also
${v\over c} \approx \sqrt{{{R_m}\over {l_c}}} << 1$), we have
$\lambda >> l_c$ and we can have $l_c \approx R_M$.
\medskip

As shown in paper II, this HPM linearization allows to get a
consistent description of GW's in non-harmonic 3-orthogonal gauges
reproducing their known properties in harmonic gauges.

\subsection{Outline of the Paper}

In this paper we look in detail at the properties of the PM
space-times identified by our HPM solution and we will study the
equations of motion of the particles. It will be shown how all the
relevant gravitational quantities depend upon the York time, which
is the general relativistic remnant of the special relativistic
gauge freedom in clock synchronization. It will turn out that they
(with the only exception of the ADM Lorentz generators) depend upon
the gradients of the spatially non-local function ${}^3{\cal
K}_{(1)}(\tau, \vec \sigma) = {1\over {\triangle}}\,
{}^3K_{(1)}(\tau, \vec \sigma)$ (the non-local York time) of the
lowest order component ${}^3K_{(1)}(\tau, \vec \sigma)$ of the York
time. This study will be done in our family of 3-orthogonal gauges,
where the Riemannian instantaneous 3-spaces $\Sigma_{\tau}$ have a
diagonal 3-metric but still depend on the arbitrary numerical
function $F(\tau, \vec \sigma)$ determining the inertial gauge
variable ${}^3K_{(1)}(\tau, \vec \sigma)$.

\medskip

We will determine the explicit dependence  of the proper time of
time-like observers, of the time-like and null geodesics, of the
redshift of light and of the luminosity distance upon the York time
in these PM space-times.

\medskip

Then we will study  the consequences of the HPM linearization on the
equations of motion for the particles and we will make their
Post-Newtonian (PN) expansion at all ${n\over 2}PN$ orders. In
particular we will show that at the astrophysical level there is a
0.5PN contribution to dark matter coming from the relativistic
inertial effect connected to the choice of the function ${}^3{\cal
K}_{(1)}$. In the non-Euclidean 3-space there is an effective
inertial mass different from the gravitational mass, even if the
equality on inertial and gravitational masses holds in the
4-dimensional space-time in accord with the equivalence principle.

\bigskip

In Section II we rewrite the solutions of the constraints and of the
equations for the lapse and shift functions (all determined by
elliptic equations inside the 3-spaces $\Sigma_{\tau}$) in
3-orthogonal gauges, given in paper II, restricting them to the case
in which the matter consists only of point particles. An equal time
development of the retarded solution for the GW's is also given.
Then, after the expression of the ADM Poincare' generators we give
the Christoffel symbols and the Riemann and Weyl tensors of the PM
space-times. Finally the expression of the 4-spin and 3-spin
connections and the expression of the Ashtekar's variables in the
York canonical basis are given.

\medskip

In Section III we show properties of the PM space-times and of their
Riemannian 3-spaces in 3-orthogonal gauges: the proper time of
time-like observers; the 3-distance, the 3-curvature and the
extrinsic curvature of 3-spaces; the PM time-like 4-geodesics. Also
the comparison of the HPM solution for the 4-metric in 3-orthogonal
gauges with the astronomical conventions for the 4-metric of the
Solar System in a suitable harmonic gauge is given.

\medskip

Section IV is devoted to the PM null  geodesics, the red-shift, the
geodesic deviation equation and the luminosity distance of PM
space-times.
\medskip

In Section V we give the PM equations of motion for the particles
and discuss their qualitative properties and differences from the
case of charged particles plus the electro-magnetic field in
Minkowski space-time. After a discussion of the problem of the
center of mass and of the relative variables  in the PM space-times
(using the two-body problem as an example), we study the PN
expansion of the equations of motion and  we compare a 1PN binary
system in 3-orthogonal gauges with the standard one in harmonic
gauges. In Appendix A there is  the 1PN expression of the ADM
Poincare' generators with terms till order $O(\zeta^2)$ included.

\medskip

In Section VI we show that the non-local York time allows to
describe the dark matter present in the masses of galaxies and their
rotation curves as a relativistic inertial effect absent in Newton
gravity.

\medskip

In the Conclusions we comment on the gauge problem in GR, discussed
in the Conclusions of paper II, and on how our explanation of dark
matter may help in finding the PM extension of the Celestial
reference frame (ICRS) \cite{5} outside the Solar System. Then we
show which lines of development are opened by our formulation,
especially in cosmology and in particular for the possibility to
reformulate the back-reaction approach for the elimination of dark
energy, based on averages of scalar quantities in 3-volumes of
3-space, in the York canonical basis.

\vfill\eject

\section{The PM Solution for the Gravitational Field}

In this Section we review the results of paper II when the matter
consists only of point particles. At this order the HPM linearized
solution in the family of 3-orthogonal gauges  depends on the York
time ${}^3K$ only through the following function ${}^3{\cal K} =
{1\over {\triangle}}\, {}^3K$. Only the linearized ADM Lorentz
generators have also a dependence on ${}^3K$. We give also the equal
time development of the retarded solution for GW's. The metric,
Christoffel symbols, spin connection, Riemann and Weyl tensors of PM
space-times are given. Finally Ashtekar's variables are expressed in
the York canonical basis and their PM limit is found.

\subsection{The Energy-Momentum of the Particles}

From Eqs.(3.9), (3.10) and (3.12) of paper II we get the following
expression for the energy-momentum of the particles (in the
following equations the notation ${M\over {L^3}}\, O(\zeta^2)$ means
$ \sum_{i=1}^N\, \delta^3(\vec \sigma, {\vec \eta}_i(\tau ))\, M\,
O(\zeta^2)$)

\begin{eqnarray*}
 {\cal M}^{(UV)}_{(1)}(\tau, \vec \sigma) &=& T^{\tau\tau}_{(1)}(\tau, \vec \sigma) =
 \sum_{i=1}^N\, \delta^3(\vec \sigma
 ,{\vec \eta}_i(\tau ))\, \eta_i\, \sqrt{m^2_i\, c^2 +
 {\vec \kappa}^2_i(\tau )} =\nonumber \\
 &=&\sum_{i=1}^N\, \delta^3(\vec \sigma
 ,{\vec \eta}_i(\tau ))\, \eta_i\, {{m_ic}\over {\sqrt{1
 - {\dot {\vec \eta}}_i^2(\tau)}}} + {M\over {L^3}}\, O(\zeta^2),\nonumber \\
 && M_{(1)} c = q^{|\tau\tau} = \sum_{i=1}^N\, \eta_i\, \sqrt{m_i^2c^2 +
 {\vec \kappa}_i^2(\tau)} = \sum_{i=1}^N\, \eta_i\,
 {{m_ic}\over {\sqrt{1 - {\dot {\vec \eta}}_i^2(\tau)}}}
 + M\, O(\zeta^2),
 \end{eqnarray*}

 \begin{eqnarray*}
 {\cal M}^{(UV)}_{(2)}(\tau, \vec \sigma) &=&-
 \sum_{i=1}^N\, \delta^3(\vec \sigma
 ,{\vec \eta}_i(\tau ))\, \eta_i\, \Big(
 {{ 2\, \phi_{(1)}\, {\vec \kappa}^2_i(\tau )
  +  \sum_a\, \Gamma_a^{(1)}\, \kappa^2_{ia}(\tau )}\over
  {\sqrt{m^2_i\, c^2 +  {\vec \kappa}^2_i(\tau )}}}\Big)(\tau
  ,\vec \sigma ) =\nonumber \\
  &=&- \sum_{i=1}^N\, \delta^3(\vec \sigma ,{\vec \eta}_i(\tau ))\, \eta_i\,
 m_i\, c\, \Big( {{ 2\, \phi_{(1)}\, {\dot {\vec \eta}}^2_i(\tau )
  +  \sum_a\, \Gamma_a^{(1)}\, ({\dot \eta}^a_i(\tau))^2(\tau )}\over
  {\sqrt{1 -  {\dot {\vec \eta}}^2_i(\tau )}}}\Big)(\tau
  ,\vec \sigma ),
  \end{eqnarray*}

\begin{eqnarray*}
   {\cal M}^{(UV)}_{(1)r}(\tau ,\vec \sigma )&=& T^{\tau r}_{(1)}(\tau, \vec \sigma) =
   \sum_{i=1}^N\, \delta^3(\vec \sigma ,{\vec \eta}_i(\tau ))\,
  \eta_i\, \kappa_{ir}(\tau ) = \sum_{i=1}^N\, \delta^3(\vec \sigma,
  {\vec \eta}_i(\tau ))\, \eta_i\, {{m_ic\, {\dot \eta}^r_i(\tau)}
  \over {\sqrt{1 - {\dot {\vec \eta}}_i^2(\tau)}}} + M\, O(\zeta^2),
  \nonumber \\
 &&q^{r|\tau s} = \sum_{i=1}^N\, \eta_i\, \eta^r_i(\tau)\,
 \kappa_{is}(\tau) = \sum_{i=1}^N\, \eta_i\, {{m_ic\, \eta^r_i(\tau)\,
 {\dot \eta}^s_i(\tau)}\over {\sqrt{1 - {\dot {\vec \eta}}_i^2(\tau)}}}
 + M\, O(\zeta^2),
 \end{eqnarray*}

\begin{eqnarray*}
  T_{(1)}^{rs}(\tau, \vec \sigma)&=& \sum_{i=1}^N\, \delta^3(\vec \sigma,
  {\vec \eta}_i(\tau))\, \eta_i\, {{\kappa_{ir}(\tau)\,
  \kappa_{is}(\tau)}\over {\sqrt{m^2_i\, c^2 +  {\vec \kappa}^2_i(\tau
  )}}} =\nonumber \\
  &=&\sum_{i=1}^N\, \delta^3(\vec \sigma,
  {\vec \eta}_i(\tau))\, \eta_i\, {{m_ic\, {\dot \eta}^r_i(\tau)\,
  {\dot \eta}^s_i(\tau)}\over {\sqrt{1 - {\dot {\vec \eta}}_i^2(\tau)}}}
 + {M\over {L^3}}\, O(\zeta^2),\nonumber \\
  &&q^{|rs} = \sum_i\, \eta_i\, {{\kappa_{ir}(\tau)\,
  \kappa_{is}(\tau)}\over {\sqrt{m^2_i\, c^2 +  {\vec \kappa}^2_i(\tau
  )}}} = \sum_{i=1}^N\, \eta_i\, {{m_ic\, {\dot \eta}^r_i(\tau)\,
  {\dot \eta}^s_i(\tau)}\over {\sqrt{1 - {\dot {\vec \eta}}_i^2(\tau)}}}
 + M\, O(\zeta^2),
 \end{eqnarray*}

  \bea
  &&{1\over {\triangle}}\, {\cal M}_{(1)}^{(UV)}(\tau, \vec \sigma)
  = - \sum_{i=1}^N\, \eta_i\, {{\sqrt{m_i^2c^2 + {\vec \kappa}^2_i(\tau)}}\over
  {4\pi\, |\vec \sigma - {\vec \eta}_i(\tau)|}},\qquad
   {1\over {\triangle}}\, \sum_a\, T_{(1)}^{aa}(\tau, \vec \sigma)
  = - \sum_{i=1}^N\, \eta_i\,  {{{{{\vec \kappa}^2_i(\tau)}\over
  {\sqrt{m_i^2c^2 + {\vec \kappa}^2_i(\tau)}}}}\over {4\pi\,
  |\vec \sigma - {\vec \eta}_i(\tau)|}},\nonumber \\
  &&{1\over {\triangle}}\, {\cal M}_{(1)r}^{(UV)}(\tau, \vec \sigma)
  = - \sum_{i=1}^N\, \eta_i\, {{\kappa_{ir}(\tau)}\over {4\pi\,
  |\vec \sigma - {\vec \eta}_i(\tau)|}},\nonumber \\
  &&{{\partial_a}\over {\triangle}}\, \sum_c\, \partial_c\, {\cal
  M}_{(1)c}^{(UV)}(\tau, \vec \sigma) = - \sum_{i=1}^N\, \eta_i\,
  \sum_c\, \kappa_{ic}(\tau)\, \int {{d^3\sigma_1}\over
 {(4\pi)^2\, |\vec \sigma - {\vec \sigma}_1|\, |{\vec \sigma}_1
 - {\vec \eta}_i(\tau)|^3}}\, \Big(\delta_{ac} -\nonumber \\
 &-& 3\, {{(\sigma_1^a
 - \eta_i^a(\tau))\, (\sigma_1^c - \eta_i^c(\tau))}\over
 {|{\vec \sigma}_1 - {\vec \eta}_i(\tau)|^2}}\Big), \nonumber \\
 &&{}
 \label{2.1}
 \eea

\noindent where we used  ${\vec \kappa}_i = {{m_ic\, {\dot {\vec
\eta}}_i}\over {\sqrt{1 - {\dot {\vec \eta}}_i^2}}} + Mc\, O(\zeta)$
and $\sqrt{m_i^2c^2 + {\vec \kappa}_i^2} = {{m_ic}\over {\sqrt{1 -
{\dot {\vec \eta}}_i^2}}} + Mc\, O(\zeta)$. We have also given the
second order of the mass function. The quantities $q^{|\tau\tau}$,
$q^{r|\tau s}$ and $q^{|rs}$ are the mass monopole, the momentum
dipole and the stress tensor monopole respectively (see Appendix B
of paper II).

\subsection{The Solution of the Super-Hamiltonian and
Super-Momentum Constraints and the Lapse and Shift Functions for the
Family of 3-Orthogonal Gauges}

From Eqs.(4.6), (4.7), (4.16) and (4.17) of paper II we get the
following expressions for the solutions: a) ${\tilde
\phi}_{(1)}(\tau, \vec \sigma)$ of the super-Hamiltonian constraint;
b) $N(\tau, \vec \sigma) = 1 + n_{(1)}(\tau, \vec \sigma)$ and
${\bar n}_{(1)(a)}(\tau, \vec \sigma)$ of the equations for the
lapse and shift functions in the family of 3-orthogonal gauges; c)
$\sigma_{(1)(a)(b)}{|}_{a \not= b}(\tau, \vec \sigma)$ (the
off-diagonal terms of the shear of the congruence of Eulerian
observers) of the super-momentum constraints (see Eq.(\ref{1.9}) for
$\pi_i^{(\theta)}$):

\bea
 {\tilde \phi}(\tau, \vec \sigma) &=& 1 + 6\, \phi_{(1)}(\tau, \vec \sigma)
 + O(\zeta^2),\nonumber \\
 &&{}\nonumber \\
 \phi_{(1)}(\tau, \vec \sigma) \, &\cir& \Big[- {{2\pi\, G}\over
 {c^3}}\, {1\over {\triangle}}\, {\cal M}^{(UV)}_{(1)} +
 {1\over 4}\, \sum_c\, {{\partial_c^2}\over {\triangle}}\,
 \Gamma_c^{(1)}\Big](\tau, \vec \sigma) \cir\nonumber \\
 &&{}\nonumber \\
 &\cir& {G\over {2c^3}}\, \sum_i\, \eta_i\, {{\sqrt{m_i^2\, c^2 +
 {\vec \kappa}^2_i(\tau)}}\over
 { |\vec \sigma - {\vec \eta}_i(\tau)|}} -
  {1\over {16\pi}}\, \int d^3\sigma_1\, {{\sum_a\, \partial_{1a}^2\,
 \Gamma_a^{(1)}(\tau, {\vec \sigma}_1)}\over { |\vec \sigma -
 {\vec \sigma}_1|  }} =\nonumber \\
 &=& {G\over {2c^2}}\, \sum_i\, \eta_i\, {{ {{m_i}\over {\sqrt{1 -
 {\dot {\vec \eta}}^2_i(\tau)}}}  }\over
 { |\vec \sigma - {\vec \eta}_i(\tau)|}} -
  {1\over {16\pi}}\, \int d^3\sigma_1\, {{\sum_a\, \partial_{1a}^2\,
 \Gamma_a^{(1)}(\tau, {\vec \sigma}_1)}\over { |\vec \sigma -
 {\vec \sigma}_1|  }},\nonumber \\
 &&{}
 \label{2.2}
 \eea

\bigskip

\bea
 n_{(1)}(\tau, \vec \sigma)\, &\cir& \Big[{{4\pi\, G}\over {c^3}}\,
 {1\over {\triangle}}\, \Big({\cal M}^{(UV)}_{(1)} + \sum_a\,
 T_{(1)}^{aa}\Big) -  \partial_{\tau}\,
 {}^3{\cal K}_{(1)}\Big](\tau, \vec \sigma)\, \cir\nonumber \\
 &&{}\nonumber \\
 &\cir& - {G\over {c^3}}\, \sum_i\, \eta_i\, {{\sqrt{m_i^2\, c^2 +
 {\vec \kappa}^2_i(\tau)}}\over { |\vec \sigma -
 {\vec \eta}_i(\tau)|}}\, \Big(1 +
 {{{\vec \kappa}^2_i}\over {m_i^2\, c^2 + {\vec \kappa}_i^2}}\Big) -
  \partial_{\tau}\, {}^3{\cal K}_{(1)}(\tau, \vec \sigma) =
  \nonumber \\
 &=& - {G\over {c^2}}\, \sum_i\, \eta_i\, {{ {{m_i}\over {\sqrt{1 -
 {\dot {\vec \eta}}^2_i(\tau)}}} }\over
 { |\vec \sigma - {\vec \eta}_i(\tau)|}}\, (1 + {\dot {\vec \eta}}^2_i(\tau))
 - \partial_{\tau}\, {}^3{\cal K}_{(1)}(\tau, \vec \sigma),\nonumber \\
 &&{}
 \label{2.3}
 \eea

\medskip

\begin{eqnarray*}
 {\bar n}_{(1)(a)}(\tau, \vec \sigma) \, &\cir& \Big[\partial_a
 \, {}^3{\cal K}_{(1)} + {{4\pi\, G}\over {c^3}}\, {1\over
 {\triangle}}\, \Big(4\, {\cal M}_{(1) a}^{(UV)} -
 {{\partial_a}\over {\triangle}}\,\, \sum_c\,
 \partial_c\, {\cal M}_{(1) c}^{(UV)}\Big) +\nonumber \\
 &+& {1\over 2}\, {{\partial_a}\over {\triangle}}\, \partial_{\tau}\, \Big(4\,
 \Gamma_a^{(1)} - \sum_c\, {{\partial_c^2}\over {\triangle}}\,
 \Gamma_c^{(1)}\Big) \Big](\tau, \vec \sigma) \cir\nonumber \\
 &&{}\nonumber \\
 &\cir& \partial_a\, {}^3{\cal K}_{(1)}(\tau, \vec \sigma) -
 {{G}\over {c^3}}\, \sum_i\, {{\eta_i}\over {|\vec \sigma - {\vec \eta}_i(\tau)|}}\,
 \Big( \frac{7}{2}\kappa_{ia}(\tau) +\nonumber \\
 &-&\frac{1}{2} {{(\sigma^a - \eta^a_i(\tau))\, {\vec \kappa}_i(\tau) \cdot (\vec \sigma
 - {\vec \eta}_i(\tau))}\over {|\vec \sigma - {\vec \eta}_i(\tau)|^2}} \Big)
 -\nonumber \\
 &-& \int {{d^3\sigma_1}\over {4\pi\, |\vec \sigma - {\vec \sigma}_1|}}\,
 \partial_{1a}\, \partial_{\tau}\, \Big[ 2\,
 \Gamma_a^{(1)}(\tau, {\vec \sigma}_1) -
  \int d^3\sigma_2\, {{\sum_c\,  \partial_{2c}^2\,
 \Gamma_c^{(1)}(\tau, {\vec \sigma}_2)}\over {8\pi\, |{\vec \sigma}_1 -
 {\vec \sigma}_2|}}\Big] =\end{eqnarray*}

\bea
 &=& \partial_a \, {}^3{\cal K}_{(1)}(\tau, \vec \sigma) -
 {{ G}\over {c^2}}\, \sum_i\, {{\eta_i}\over {|\vec \sigma - {\vec \eta}_i(\tau)|}}\,
 {{m_i\, }\over {\sqrt{1 - {\dot {\vec
 \eta}}^2_i(\tau)}}}\, \Big(\frac{7}{2}{\dot \eta}^a_i(\tau) +\nonumber \\
 &-&\frac{1}{2} {{(\sigma^a - \eta^a_i(\tau))\, {\dot {\vec \eta}}_i(\tau)
 \cdot (\vec \sigma - {\vec \eta}_i(\tau))}\over {|\vec \sigma -
 {\vec \eta}_i(\tau)|^2}} \Big) -\nonumber \\
 &-& \int {{d^3\sigma_1}\over {4\pi\, |\vec \sigma - {\vec \sigma}_1|}}\,
 \partial_{1a}\, \partial_{\tau}\, \Big[ 2\,
 \Gamma_a^{(1)}(\tau, {\vec \sigma}_1) -
  \int d^3\sigma_2\, {{\sum_c\,  \partial_{2c}^2\,
 \Gamma_c^{(1)}(\tau, {\vec \sigma}_2)}\over {8\pi\, |{\vec \sigma}_1 -
 {\vec \sigma}_2|}}\Big].\nonumber \\
 &&{}
 \label{2.4}
 \eea
\bigskip

\begin{eqnarray*}
 \sigma_{(1)(a)(b)}{|}_{a \not= b}(\tau, \vec \sigma) &\cir& {1\over 2}\,
 \Big(\partial_a\, {\bar n}_{(1)(b)} + \partial_b\, {\bar
 n}_{(1)(a)}\Big){|}_{a \not= b}(\tau, \vec \sigma)\, \cir\nonumber \\
 &&{}\nonumber \\
 &\cir& \Big[ \partial_a\, \partial_b\, {}^3{\cal K}_{(1)} + {{8\pi\,
 G}\over {c^3}}\, \Big[{1\over {\triangle}}\, \Big(\partial_a\,
 {\cal M}^{(UV)}_{(1) b} + \partial_b\, {\cal M}^{(UV)}_{(1) a}\Big)
  - {1\over 2}\, {{\partial_a\, \partial_b}\over {\triangle}}\, \sum_c\,
  {{\partial_c}\over {\triangle}}\, {\cal M}^{(UV)}_{(1) c}\Big] +\nonumber \\
 &+& \partial_{\tau}\, {{\partial_a\, \partial_b}\over {\triangle}}\,
 \Big(\Gamma_a^{(1)} + \Gamma_b^{(1)} - {1\over 2}\, \sum_c\,
 {{\partial_c^2}\over {\triangle}}\, \Gamma_c^{(1)}\Big) \Big](\tau, \vec \sigma)
 \cir \end{eqnarray*}

\bea
 &\cir&- {1\over 2}\, \sum_d\, (\delta_{ad}\, \partial_b + \delta_{bd}\,
 \partial_a)\quad
 \Big( {{G}\over {c^3}}\, \sum_i\, {{\eta_i}\over {|\vec \sigma - {\vec \eta}_i(\tau)|}}\,
 \Big(\frac{7}{2}\kappa_{id}(\tau) +\nonumber \\
 &-&\frac{1}{2} {{(\sigma^d - \eta^d_i(\tau))\, {\vec \kappa}_i(\tau)
 \cdot (\vec \sigma - {\vec \eta}_i(\tau))}\over {|\vec \sigma - {\vec \eta}_i(\tau)|^2}}
 \Big) +\nonumber \\
 &+& \int {{d^3\sigma_1}\over {4\pi\, |\vec \sigma - {\vec \sigma}_1|}}\,
 \partial_{1d}\, \Big[ 2\, \partial_{\tau}\,
 \Gamma_d^{(1)}(\tau, {\vec \sigma}_1) +
  \int d^3\sigma_2\, {{\sum_c\, \partial_{\tau}\, \partial_{2c}^2\,
 \Gamma_c^{(1)}(\tau, {\vec \sigma}_2)}\over {8\pi\, |{\vec \sigma}_1 -
 {\vec \sigma}_2|}}\Big]\,\, \Big) + \partial_a\, \partial_b\,
 {}^3{\cal K}_{(1)}(\tau, \vec \sigma).\nonumber \\
 &&{}
 \label{2.5}
 \eea
\medskip

The action-at-a-distance part of the solution is explicitly shown.
Only the PM volume element ${\tilde \phi}_{(1)} = 1 + 6\,
\phi_{(1)}$ is independent from the York time. Eq.(\ref{2.4})
describes gravito-magnetism in these PM space-times \footnote{ The
gravito-magnetic potential ${\vec A}_G$ has the components $A_{G(r)}
\sim c^2\, {\bar n}_{(1)(r)}$. The gravito-magnetic field $B_{G(r)}
= c\, \Omega_{G(r)} =(\vec \partial \times {\vec A}_G)_r$ is
proportional to the second term in the Christoffel symbol
${}^4\Gamma^u_{(1)\tau r}$ given in Eq.(\ref{2.15}). Instead the
gravito-electric potential is $\Phi_G = - {{c^2}\over 4}\, n_{(1)} =
- {{8\pi\, G}\over c}\, {1\over {\triangle}}\, ({\cal
M}_{(1)}^{(UV)} + \sum_a\, T_{(1)}^{aa}) + {{c^2}\over 4}\,
\partial_{\tau}\, {}^3{\cal K}_{(1)}$). Both $A_{G(r)}$ and $\Phi_G$
depend on the non-local York time.}: it has an inertial gauge part
$\partial_a\, {}^3{\cal K}_{(1)}$.

\subsection{The HPM Gravitational Waves}

By using Eqs.(7.1) and (7.2) of paper II, the retarded solution for
the tidal variables with the matter restricted to point particles is
(see Eq. (3.12) of paper II for $T^{rs}$; the TT projector ${\cal
P}_{rsuv}$ is defined in Eqs.(6.7) of paper II; the spatial operator
${\tilde M}^{-1}_{ab}$ is defined in Eqs. (6.24) and (6.25) of paper
II) \footnote{One could study the radiative fields
$\Gamma^{(1)}_a(\tau, \vec \sigma)$ at null infinity ($|\vec \sigma|
\rightarrow \infty$ with the retarded time $\tau - |\vec \sigma|$
fixed) to see whether terms in $ln\, |\vec \sigma|$ appear like in
the standard approach to GW's in harmonic gauges (see Section 5.3.4
of Ref.\cite{6}), but this will done elsewhere. }

\bea
 \Gamma_a^{(1)}(\tau, \vec \sigma) &=& \sum_{\bar a}\, \gamma_{\bar aa}\,
 R_{\bar a}(\tau, \vec \sigma) \cir \nonumber \\
 &\cir& - {2\, G\over {c^2}}\, \sum_{b}\, {\tilde
 M}^{-1}_{ab}(\vec \sigma)\, \sum_i\, \eta_i\, m_i\, \sum_{uv}\,
 {\cal P}_{bbuv}(\vec \sigma)\nonumber \\
 && \int d^3\sigma_1\, {{{\dot \eta}^u_i(\tau -
 |\vec \sigma - {\vec \sigma}_1|)\, {\dot \eta}^v_i(\tau - |\vec \sigma -
 {\vec \sigma}_1|)}\over {\sqrt{1 - {\dot \eta}^2_i(\tau - |\vec \sigma -
 {\vec \sigma}_1|)}}}\, {{\delta^3({\vec \sigma}_1 - {\vec \eta}_i(\tau -
 |\vec \sigma - {\vec \sigma}_1|))}\over {|\vec \sigma - {\vec \sigma}_1|}}
 + O(\zeta^2).\nonumber \\
 &&{}
 \label{2.6}
 \eea
\medskip

By making a equal time development of the retarded kernel like in
Ref.\cite{7} for the extraction of the Darwin potential from the
Lienard-Wiechert solution (see Eqs. (5.1)-(5.21) of Ref.\cite{7}
with $\sum_s\, P_{\perp}^{rs}(\vec \sigma)\, {\dot \eta}^s_i(\tau)\,
\mapsto \sum_{uv}\, {\cal P}_{bbuv}(\vec \sigma)\, {{{\dot
\eta}^u_i(\tau)\, {\dot \eta}_i^v(\tau)}\over {\sqrt{1 - {\dot
\eta}_i^2(\tau)}}}$) and by using the fact that ${\ddot
\eta}_i^r(\tau) = O(\zeta)$ (see also Section V) we get the
following expression of the HPM GW's from point masses

\bea
 \Gamma^{(1)}_a(\tau, \vec \sigma) &\cir& - {2\, G\over {c^2}}\, \sum_{b}\, {\tilde
 M}^{-1}_{ab}(\vec \sigma)\, \sum_i\, \eta_i\, m_i\, \sum_{uv}\,
 {\cal P}_{bbuv}(\vec \sigma)\, {{{\dot
 \eta}^u_i(\tau)\, {\dot \eta}_i^v(\tau)}\over {\sqrt{1 - {\dot
 \eta}_i^2(\tau)}}}\nonumber \\
 && \Big[|\vec \sigma - {\vec \eta}_i(\tau)|^{- 1}
  + \sum_{m=1}^{\infty}\, {{1}\over {(2m)!}}\,
 \Big({\dot {\vec \eta}}_i(\tau) \cdot {{\partial}\over {\partial\,
 \vec \sigma}}\Big)^{2m}\, |\vec \sigma - {\vec \eta}_i(\tau)|^{2m -
 1}\Big] + O(\zeta^2).
 \label{2.7}
 \eea

As shown in paper II the multipolar expansion of the TT 3-metric and
of the tidal variables reproduces the quadrupolar emission formula
($q^{uv|\tau\tau}$ is the mass quadrupole; the TT projector
$\Lambda_{abcd}$ is defined in Eqs.(7.17) of paper II)

\bea
 {}^4h^{TT}_{(1)rs}(\tau, \vec \sigma) &\cir&
  - \sgn\, {{2\, G}\over {c^3}}\, \sum_{uv}\, \Lambda_{rsuv}(n)\,
 {{\partial^2_{\tau}\, q^{uv|\tau\tau}((\tau - |\vec \sigma|))}\over
 {|\vec \sigma|}} + (higher\, multipoles) + O(1/r^2),\nonumber \\
 &&{}\nonumber \\
 R_{\bar a}(\tau, \vec \sigma) &\cir&
  - {G\over {c^3}}\, \sum_{ab}\, \gamma_{\bar aa}\, {\tilde
 M}^{-1}_{ab}(\vec \sigma)\, {{\sum_{uv}\, {\cal P}_{bbuv}\,
 \partial^2_{\tau}\, q^{uv | \tau\tau}(\tau - |\vec \sigma|)}\over
 {|\vec \sigma|}} + (higher\, multipoles) +O(1/r^2),\nonumber \\
 &&{}\nonumber \\
 q^{uv | \tau\tau}(\tau - |\vec \sigma|) &=& \int d^3\sigma_1\,
 \sigma_1^u\, \sigma_1^v\, {\cal M}^{(UV)}_{(1)}(\tau - |\vec \sigma|,
 {\vec \sigma}_1) =\nonumber \\
 &=& \sum_{i=1}^N\, \eta_i\, \eta_i^u(\tau - |\vec \sigma|)\, \eta_i^v(\tau
 - |\vec \sigma|)\,  \sqrt{m_i^2c^2 + {\vec \kappa}_i^2(\tau - |\vec
 \sigma|)} =\nonumber \\
 &=& \sum_{i=1}^N\, \eta_i\, {{m_ic\,\eta_i^u(\tau - |\vec \sigma|)\,
 \eta_i^v(\tau - |\vec \sigma|)}\over {\sqrt{1 - {\dot {\vec \eta}}_i^2(\tau
 - |\vec \sigma|)}}}.
 \label{2.8}
 \eea

\medskip

Eq.(4.18) of paper II gives the following expression for the tidal
momenta of Eqs.(\ref{1.8}) (namely for diagonal elements
$\sigma_{(1)(a)(a)}$ of the shear, see Eq.(\ref{1.8}))

\bea
 {{8\pi\, G}\over {c^3}}\, \Pi_{\bar a}(\tau, \vec \sigma) &\cir&
 \partial_{\tau}\, R_{\bar a}(\tau, \vec \sigma) - \sum_a\,
 \gamma_{\bar aa}\, \Big[\partial_{\tau}\, {{\partial_a^2}\over
 {2\, \triangle}}\, (4\, \Gamma_a^{(1)} - {1\over {\triangle}}\,
 \sum_c\, \partial_c^2\, \Gamma_c^{(1)}) +\nonumber \\
 &+& {{4\pi\, G}\over {c^3}}\, {1\over {\triangle}}\, (4\, \partial_a\,
 {\cal M}^{(UV)}_{(1)a} - {{\partial_a^2}\over {\triangle}}\, \sum_c\,
 \partial_c\, {\cal M}^{(UV)}_{(1)c}) + \partial_a^2\,
 {}^3{\cal K}_{(1)}\Big] =\nonumber \\
 &&{}\nonumber \\
 &=& \Big(\sum_{\bar b}\, M_{\bar a\bar b}\, \partial_{\tau} \, R_{\bar
 b} - \sum_a\, \gamma_{\bar aa}\, \Big[{{4\pi\, G}\over {c^3}}\,
 {1\over {\triangle}}\, (4\, \partial_a\,
 {\cal M}^{(UV)}_{(1)a} - {{\partial_a^2}\over {\triangle}}\, \sum_c\,
 \partial_c\, {\cal M}^{(UV)}_{(1)c}) +\nonumber \\
 &+&  \partial_a^2\, {}^3{\cal K}_{(1)} \Big]\Big)(\tau, \vec \sigma),\nonumber \\
 &&{}\nonumber \\
 && M_{\bar a\bar b} = \delta_{\bar a\bar b} - \sum_a\, \gamma_{\bar
 aa}\, {{\partial_a^2}\over {\triangle}}\, \Big(2\, \gamma_{\bar ba} -
 {1\over 2}\, \sum_b\, \gamma_{\bar bb}\, {{\partial_b^2}\over
 {\triangle}}\Big),
 \label{2.9}
 \eea

While the tidal variables $R_{\bar a}$ do not depend on the York
time, the tidal momenta $\Pi_{\bar a}$ depend upon it.

\subsection{The Weak ADM Poincare' Generators}

The HPM linearization of the ADM Poincare' generators is given in
Eqs. (4-21)-(4.24) of paper II. The final expression of the
generators is ($\sqrt{m^2_i\, c^2 + {\vec \kappa}_i^2(\tau)} = m_i\,
c /\sqrt{1 - {\dot {\vec \eta}}_i^2(\tau)}$; the spatial operator
$M_{\bar a\bar b}$ is given in Eq.(\ref{2.6}); $L$ is the GW
wavelength of Section III of paper II)

\bea
 {1\over c}\, {\hat E}_{ADM}\, &=&\,  M_{(1)}\, c + {1\over c}\, {\hat E}_{ADM (2)}  +
  Mc\, O(\zeta^3) =\nonumber \\
  &=&\sum_i\, \eta_i\, \sqrt{m_i^2\, c^2 + {\vec \kappa}_i^2(\tau)}
  -\nonumber \\
  &-& \sum_i\, \eta_i\, {{ {\vec \kappa}_i^2(\tau)\, \Big[{G\over {c^3}}\,
  \sum_{j \not= i}\, \eta_j\, {{\sqrt{m_j^2\, c^2 + {\vec \kappa}_j^2(\tau)}}\over
  {|{\vec \eta}_i(\tau) - {\vec \eta}_j(\tau)|}} + \sum_c\, {{\partial_c^2}\over
  {2\, \triangle}}\, \Gamma_c^{(1)}(\tau, {\vec \eta}_i(\tau))\Big]
  +  \sum_c\, \kappa^2_{ic}(\tau)\, \Gamma_c^{(1)}(\tau, {\vec \eta}_i(\tau))}\over
  {\sqrt{m_i^2\, c^2 + {\vec \kappa}_i^2(\tau)}}} -\nonumber \\
  &-& {G\over {c^3}}\, \sum_{i > j}\, \eta_i\, \eta_j\, {{\sqrt{m_i^2\, c^2 + {\vec \kappa}_i^2(\tau)}\,
  \sqrt{m_j^2\, c^2 + {\vec \kappa}_j^2(\tau)}}\over {|{\vec \eta}_i(\tau) -
  {\vec \eta}_j(\tau)|}} + {{ G}\over {c^3}}\, \sum_{i > j}\,
  \eta_i\, \eta_j\, \Big({{4\, {\vec \kappa}_i(\tau) \cdot {\vec \kappa}_j(\tau)}\over
  {|{\vec \eta}_i(\tau) - {\vec \eta}_j(\tau)|}} -\nonumber \\
  &-&{1\over {4\, \pi}}\, \sum_{rs}\, \kappa_{ir}(\tau)\,
  \kappa_{js}(\tau)\, \int d^3\sigma\, {{(\sigma^r - \eta_i^r(\tau))\, (\sigma^s -
  \eta_j^s(\tau))}\over {|\vec \sigma - {\vec \eta}_i(\tau)|^3\, |\vec \sigma -
  {\vec \eta}_j(\tau)|^3}} \Big) +\nonumber \\
  &+&{{c^3}\over {16\pi G}}\, \sum_{\bar a\bar b}\, \int d^3\sigma\,
  \Big[\partial_{\tau}\, R_{\bar a}\, M_{\bar a\bar b}\,
 \partial_{\tau}\, R_{\bar b} + \sum_a\, \partial_a\,
 R_{\bar a}\, M_{\bar a\bar b}\, \partial_a\, R_{\bar b}\Big] \,
 \Big)(\tau, \vec \sigma) -\nonumber \\
 &-& \sum_i\, \eta_i\, {\vec \kappa}_i(\tau) \cdot \vec \partial\,
 {}^3{\cal K}_{(1)}(\tau, {\vec \eta}_i(\tau)) + Mc\, O(\zeta^3),
 \label{2.10}
 \eea

 \bea
  {\hat P}^{r}_{ADM}\, &=&  p^r_{(1)} + p^r_{(2)} + Mc\, O(\zeta^3)
 =\nonumber \\
 &=&\sum_i\, \eta_i\, \kappa_{ir}(\tau) - {{c^3}\over {8\pi\, G}}\,
 \int d^3\sigma\, \sum_{\bar a\bar b}\, \Big(\partial_r\, R_{\bar a}\, M_{\bar a\bar b}\,
 \partial_{\tau}\, R_{\bar b}\Big)(\tau, \vec \sigma) +\nonumber \\
 &+& \sum_i\, \eta_i\, \sum_a\, \kappa_{ia}(\tau)\, {{\partial_r\, \partial_a}\over {\triangle}}\,
 \Big(\sum_c\, {{\partial_c^2}\over
 {2\, \triangle}}\, \Gamma_c^{(1)} - 2\, \Gamma_a^{(1)}\Big)(\tau, {\vec \eta}_i(\tau))
 -\nonumber \\
 &-& \sum_i\, \eta_i\, \sqrt{m_i^2\, c^2 + {\vec
 \kappa}_i^2(\tau)}\,
 \partial_r\, {}^3{\cal K}_{(1)}(\tau, {\vec \eta}_i(\tau))
 + Mc\, O(\zeta^3) \approx 0,
 \label{2.11}
 \eea

 \bea
 {\hat J}^{rs}_{ADM} &=& j^{rs}_{(1)} + j^{rs}_{(2)} + Mc\, O(\zeta^3)
 =\nonumber \\
 &=&\sum_i\, \eta_i\, \Big(\eta_i^r(\tau)\, \kappa_{is}(\tau) -
 \eta^s_i(\tau)\, \kappa_{ir}(\tau)\Big) -\nonumber \\
  &-&2\, \sum_i\, \eta_i\, \sum_u\, \kappa_{iu}(\tau)\, \Big(\eta_i^r(\tau)\, {{\partial}\over
 {\partial\, \eta_i^s}} - \eta_i^s(\tau)\, {{\partial}\over {\partial\, \eta_i^r}}\Big)
 \nonumber \\
 &&{{\partial_u}\over {\triangle}}\, \Big(\Gamma_u^{(1)}(\tau, {\vec \eta}_i(\tau)) -
 {1\over 4}\, \sum_c\, {{\partial_c^2}\over {\triangle}}\,
 \Gamma_c^{(1)} \Big) +\nonumber \\
 &+&2\, \sum_i\, \eta_i\, \Big[\kappa_{ir}(\tau)\, {{\partial_s}\over {\triangle}}\,
 \Big(\Gamma_s^{(1)} - {1\over 4}\, \sum_c\, {{\partial_c^2}\over {\triangle}}\,
 \Gamma_c^{(1)}\Big) -\nonumber \\
 &-&  \kappa_{is}(\tau)\, {{\partial_r}\over {\triangle}}\, \Big(\Gamma_r^{(1)}
 - {1\over 4}\, \sum_c\, {{\partial_c^2}\over {\triangle}}\,
 \Gamma_c^{(1)}  \Big)\Big](\tau, {\vec \eta}_i(\tau)) -\nonumber \\
 &-& {{c^3}\over {8\pi\, G}}\, \int d^3\sigma\,
 \Big[\sum_{\bar a\bar b}\, (\sigma^r\, \partial_s
 - \sigma^s\, \partial_r)\, R_{\bar a}\, M_{\bar a\bar b}\,
 \partial_{\tau}\, R_{\bar b} + 2\, {}^3K_{(1)}\, \partial_r\,
 \partial_s\, (\Gamma^{(1)}_s - \Gamma_r^{(1)}) +\nonumber \\
 &+& 2\, (\partial_{\tau}\, \Gamma_r^{(1)} + \partial_{\tau}\, \Gamma_s^{(1)} -
 {1\over 2}\, \sum_c\, {{\partial_c^2}\over {\triangle}}\,
 \partial_{\tau}\, \Gamma_c^{(1)})\,
 {{\partial_r\, \partial_s}\over {\triangle}}\, (\Gamma_s^{(1)} - \Gamma_r^{(1)})
 \Big](\tau, \vec \sigma) +\nonumber \\
 &+& \sum_i\, \eta_i\, \sqrt{m_i^2\, c^2 + {\vec
 \kappa}_i^2(\tau)}\, \Big(\eta_i^r(\tau)\, {{\partial}\over {\partial \eta_i^s}}
 - \eta_i^s(\tau)\, {{\partial}\over {\partial\, \eta_i^r}}\Big)\,
 {}^3{\cal K}_{(1)}(\tau, {\vec \eta}_i(\tau)) + Mc\, L\, O(\zeta^3),
 \nonumber \\
 &&{}
 \label{2.12}
 \eea

 \begin{eqnarray*}
 {\hat J}^{\tau r}_{ADM}\, &=& - {\hat J}^{r\tau}_{ADM} =
 j^{\tau r}_{(1)} + j^{\tau r}_{(2)}   + Mc\, L\,  O(\zeta^3)
 =\nonumber \\
 &=& -\sum_i\,\eta_i\,\eta^r_i(\tau)\,\sqrt{m_i^2c^2+\vec{\kappa}_i^2(\tau)}-\\
 &-&\frac{G}{c^3}\sum_{i\neq i}\,\eta_i\eta_j
 \frac{\vec{\kappa}_i^2(\tau)\,\sqrt{m_j^2c^2+\vec{\kappa}_j^2(\tau)}}
 {\sqrt{m_i^2c^2+\vec{\kappa}_i^2(\tau)}}
 \frac{\eta^r_i(\tau)}{\mid\vec{\eta}_i(\tau) - \vec{\eta}_j(\tau)\mid}+
 \nonumber \\
 &+&\sum_i\,\eta_i\,\eta_i^r(\tau)\, \sum_a\,\frac{\kappa_{ia}^2(\tau)\, \Big(
 \Gamma_a^{(1)} + \frac{1}{2}\,\sum_c\,\frac{\partial_c^2}{\Delta}\,
 \Gamma_c^{(1)} \Big)(\tau, {\vec \eta}_i(\tau))}{\sqrt{m_i^2c^2 +
 \vec{\kappa}_i^2(\tau)}} -\\
 &-&\int d^3\sigma\,\sigma^r\,\Big[ \frac{c^3}{16\pi
 G}\sum_{a,b}\,(\partial_a\, \Gamma_b^{(1)}(\tau, \vec \sigma))^2 -\nonumber \\
 &-&\frac{c^3}{8\pi G}\sum_a\,\partial_a\, \Gamma_a^{(1)}(\tau, \vec \sigma)\,
 \partial_a\Big( \Gamma_a^{(1)} - \frac{1}{2}\, \sum_c\,\frac{\partial_c^2}{\Delta}\,
 \Gamma_c^{(1)}\Big)(\tau, \vec \sigma) -\\
 &-&\frac{c^3}{32\pi
 G}\sum_a\, \partial_a\, \Big(\sum_c\,\frac{\partial_c^2}{\Delta}\,
 \Gamma_c^{(1)}(\tau, \vec \sigma)\Big)
 \partial_a\, \Big(\sum_d\,\frac{\partial_d^2}{\Delta}\,
 \Gamma_d^{(1)}(\tau, \vec \sigma)\Big)+\\
 &+&\frac{1}{2}\sum_a\,
 \sum_i\,\eta_i\,\sqrt{m_i^2c^2+\vec{\kappa}_i^2(\tau)}\frac{\sigma^a -
 \eta_i^a(\tau)}{4\pi\mid\vec{\sigma} - \vec{\eta}_i(\tau)\mid^3}
 \,\partial_a\, \Big(\Gamma_a^{(1)}-\frac{1}{2}\,\sum_c\,
 \frac{\partial_c^2}{\Delta}\, \Gamma_c^{(1)}(\tau, \vec \sigma)\Big)+\\
 &+&\frac{2\, G}{\pi\, c^3} \sum_{i\neq j}\,\eta_i\eta_j\,
 \sqrt{m_i^2c^2+\vec{\kappa}_i^2(\tau)}\sqrt{m_j^2c^2+\vec{\kappa}_j^2(\tau)}
 \frac{(\vec{\sigma} - \vec{\eta}_i(\tau))\cdot(\vec{\sigma} - \vec{\eta}_j(\tau))}{\mid\vec{\sigma}
 - \vec{\eta}_i(\tau)\mid^3\mid\vec{\sigma} - \vec{\eta}_j(\tau)\mid^3} +\\
 &+&\frac{c^3}{16\pi
 G}\sum_{a,b}\,\Big(\widetilde{M}_{ab}(\vec \sigma)\,\partial_\tau\,
 \Gamma^{(1)}_b(\tau, \vec \sigma)\Big)^2 +\nonumber \\
 &+&\frac{c^3}{16\pi G}\sum_{a\neq b}\,\Big[
 \frac{\partial_a\partial_b\partial_\tau}{\Delta}\Big(
 \Gamma^{(1)}_a + \Gamma^{(1)}_b - \frac{1}{2}\,
 \sum_c\,\frac{\partial_c^2}{\Delta}\, \Gamma_c^{(1)}
 \Big)(\tau, \vec \sigma)\Big]^2-\\
 &-&{1\over {2\pi}}\, \sum_{a,b}\,\Big(\widetilde{M}_{ab}(\vec \sigma)\,\partial_\tau\,
 \Gamma^{(1)}_b(\tau, \vec \sigma)\Big)\,
 \sum_i\,\eta_i\,\frac{\kappa_{ia}(\tau)\,(\sigma^a - \eta_i^a(\tau))}
 {\mid\vec{\sigma} - \vec{\eta}_i(\tau)\mid^3}+\\
 &+&{1\over {2\pi}}\, \sum_{a\neq b}\,
 \frac{\partial_a\partial_b\partial_\tau}{\Delta}\Big(
 \Gamma^{(1)}_a+\Gamma^{(1)}_b-\frac{1}{2}\,\sum_c\,\frac{\partial_c^2\Gamma_c^{(1)}}{\Delta}
 \Big)(\tau, \vec \sigma)\,
 \sum_i\,\eta_i\,\frac{\kappa_{ia}(\tau)\, (\sigma^a - \eta_i^a(\tau))}
 {\mid\vec{\sigma} - \vec{\eta}_i(\tau)\mid^3}-\\
 \end{eqnarray*}

\begin{eqnarray*}
 &-&\frac{c^3}{8\pi
 G}\sum_{a,b}\,\Big(\widetilde{M}_{ab}(\vec \sigma)\,\partial_\tau\,
 \Gamma^{(1)}_b(\tau, \vec \sigma)\Big)\,
 \frac{\partial_a^2}{\Delta}\Big( {}^3K_{(1)}(\tau, \vec \sigma) - \frac{
 G}{c^3}\, \sum_i\,\eta_i\,\frac{\vec{\kappa}_i(\tau)\cdot(\vec{\sigma} -
 \vec{\eta}_i(\tau))}{\mid\vec{\sigma} - \vec{\eta}_i(\tau)\mid^3}\Big)
 +\\
 &+&\frac{c^3}{8\pi G}\sum_{a\neq b}\,
 \frac{\partial_a\partial_b\partial_\tau}{\Delta}\Big(
 \Gamma^{(1)}_a+\Gamma^{(1)}_b-\frac{1}{2}\,\sum_c\,\frac{\partial_c^2}{\Delta}
 \, \Gamma_c^{(1)}\Big)(\tau, \vec \sigma)\nonumber \\
 && \frac{\partial_a\partial_b}{\Delta}\, \Big(
 {}^3K_{(1)}(\tau, \vec \sigma)) -\frac{ G}{c^3}\, \sum_i\,\eta_i\,
 \frac{\vec{\kappa}_i(\tau)\cdot(\vec{\sigma} -
 \vec{\eta}_i(\tau))}{\mid\vec{\sigma} - \vec{\eta}_i(\tau)\mid^3}\Big)+\\
 &+&\frac{c^3}{16\pi
 G}\sum_{a,b}\,\Big[\frac{\partial_a\partial_b}{\Delta}\Big(
 {}^3K_{(1)}(\tau, \vec \sigma) - \frac{G}{c^3}\, \sum_i\,\eta_i\,
 \frac{\vec{\kappa}_i(\tau)\cdot(\vec{\sigma} -
 \vec{\eta}_i(\tau))}{\mid\vec{\sigma} - \vec{\eta}_i(\tau)\mid^3}\Big)\Big]^2+\\
 &+&{1\over {2\pi}}\, \sum_{a,b}\,
 \sum_i\,\eta_i\,\frac{\kappa_{ib}(\tau)\, (\sigma^a - \eta_i^a(\tau))}
 {\mid\vec{\sigma} - \vec{\eta}_i(\tau)\mid^3}\,
 \frac{\partial_a\partial_b}{\Delta}\, \Big(
 {}^3K_{(1)}(\tau, \vec \sigma) - \frac{G}{c^3}\, \sum_j\,\eta_j\,
 \frac{\vec{\kappa}_j(\tau)\cdot(\vec{\sigma} -
 \vec{\eta}_j(\tau))}{\mid\vec{\sigma} - \vec{\eta}_j(\tau)\mid^3}\Big)-\\
 &-&\frac{c^3}{48\pi G}\Big( {}^3K_{(1)}(\tau, \vec \sigma) +
 \frac{3 G}{c^3}\, \sum_i\,\eta_i\,\frac{\vec{\kappa}_i(\tau)\cdot(\vec{\sigma} -
 \vec{\eta}_i(\tau))}{\mid\vec{\sigma} - \vec{\eta}_i(\tau)\mid^3}\Big)^2
 -\frac{c^3}{24\pi G}\Big({}^3K_{(1)}(\tau, \vec \sigma)\Big)^2+\\
 &+&\frac{G}{2\pi\, c^3}\, \sum_{i\ne j} \eta_i\eta_j\,\frac{
 \vec{\kappa}_i(\tau)\cdot\vec{\kappa}_j(\tau)\,
 (\vec{\sigma} - \vec{\eta}_j(\tau))\cdot(\vec{\sigma} - \vec{\eta}_i(\tau)) +
 \vec{\kappa}_i(\tau)\cdot(\vec{\sigma} - \vec{\eta}_j(\tau))\,
 \vec{\kappa}_j(\tau)\cdot(\vec{\sigma} - \vec{\eta}_i(\tau))}
 {\mid\vec{\sigma} - \vec{\eta}_i(\tau)\mid^3\mid\vec{\sigma} - \vec{\eta}_j(\tau)\mid^3}
 \,\,\Big] +
 \end{eqnarray*}

\bea
 &+&\int d^3\sigma\,\Big[ \frac{3}{8\pi}\,
 \sum_i\,\eta_i\frac{\sqrt{m_i^2c^2+\vec{\kappa}_i^2(\tau)}}
 {\mid\vec{\sigma} - \vec{\eta}_i(\tau)\mid}
 \,\partial_r\, \Gamma^{(1)}_r + \frac{3c^3}{16\pi
 G}\partial_r\, \Gamma^{(1)}_r\, \Big(\sum_c\,
 \frac{\partial_c^2}{\Delta}\, \Gamma_c^{(1)}
 \Big)\Big](\tau, \vec \sigma)+\nonumber \\
 &-&\int d^3\sigma\,\partial_r\,\Big[ \frac{c^3}{16\pi
 G}\Big[2\Big(\Gamma^{(1)}_r(\tau, \vec \sigma)\Big)^2
 - \sum_s\,\Big(\Gamma^{(1)}_s(\tau, \vec \sigma)\Big)^2 -
 \frac{1}{2}\Big(\sum_c\,\frac{\partial_c^2}{\Delta}\,
 \Gamma_c^{(1)}(\tau, \vec \sigma)\Big)^2\Big]-\nonumber \\
 &-&\frac{G}{8\pi\, c^3}\, \sum_{i\neq
 j}\,\eta_i\eta_j\,\frac{\sqrt{m_i^2c^2+\vec{\kappa}_i^2(\tau)}\sqrt{m_j^2c^2+
 \vec{\kappa}_j^2(\tau)}}{\mid\vec{\sigma} - \vec{\eta}_i(\tau)\mid\mid\vec{\sigma} -
 \vec{\eta}_j(\tau)\mid}\,\,\Big] + Mc\, L\, O(\zeta^3) \approx 0. \nonumber \\
 &&{}
 \label{2.13}
 \eea

At the lowest order they reduce to the special relativistic internal
Poincare' generators in the rest-frame instant form of Ref.\cite{4}.
They are $p^o_{(1)} = M_{(1)} c = \sum_i\, \eta_i\, \sqrt{m_i^2 c^2
+ {\vec \kappa}_i^2(\tau)}$, $p^r_{(1)} = \sum_i\, \eta_i\,
\kappa_{ir}(\tau) \approx 0$, $j^{rs}_{(1)} = \sum_i\, \eta_i\,
\Big(\eta^r_i(\tau)\, \kappa_{is}(\tau) - \eta^s_i(\tau)\,
\kappa_{ir}(\tau)\Big)$, $j_{(1)}^{\tau r} = \sum_i\, \eta_i\,
\eta^r_i(\tau)\, \sqrt{m_i^2 c^2 + {\vec \kappa}_i^2(\tau)} \approx
0$. The conditions $j_{(1)}^{\tau r} \approx 0$ and $p^r_{(1)}
\approx 0$ are the rest-frame conditions eliminating the 3-center of
mass and its conjugate 3-momentum inside the 3-spaces of the rest
frame. As shown in Ref.\cite{4} in special relativity (and also in
PM canonical gravity) there is a decoupled external (canonical but
not covariant) 4-center of mass to be used as collective variable.
\medskip

Eqs.(\ref{2.12}) and (\ref{2.13}) show that the Lorentz generators
depend both on the local and non-local York times at the second
order.\medskip

For the effective Hamiltonian for 3-orthogonal gauges, not used in
this paper, see Eq.(4.26) of paper II.

\subsection{The 4-Metric, the Triads and Cotriads, the
$\Sigma_{\tau}$-Adapted Tetrads and Cotetrads}

Eqs. (\ref{1.5}) and (\ref{1.7}) imply the following expression for
triads, cotriads, tetrads, cotetrads and the 4-metric ($l^A_{(o)} =
(1; 0)$; $\sgn\, l_{(o)A} = (1; 0)$)

\bea
 {}^3{\bar e}^r_{(1)(a)} &=& \delta^r_a\, (1 - \Gamma^{(1)}_r - 2\,
 \phi_{(1)}),\qquad
 {}^3{\bar e}_{(1)(a)r} = \delta_{ar}\, (1 + \Gamma^{(1)}_r +
 2\, \phi_{(1)}),\nonumber \\
 &&{}\nonumber \\
 {}^4{\buildrel \circ \over {\bar E}}^A_{(1)(o)} &=&
 l^A_{(o)} + l^A_{(1)} = \Big(1 - n_{(1)}; - \delta^r_a\,
 {\bar n}_{(1)(a)}\Big),\qquad {}^4{\buildrel \circ \over {\bar E}}^A_{(1)(a)} =
 \Big(0; {}^3{\bar e}^r_{(1)(a)}\Big),\nonumber \\
 {}^4{\buildrel \circ \over {\bar E}}^{(o)}_{(1)A} &=&
 \sgn\, (l_{(o)A} + l_{(1)A}) = (1 + n_{(1)})\,
 \Big(1; 0\Big),\qquad {}^4{\buildrel \circ \over {\bar E}}^{(a)}_{(1)A} =
 \Big({\bar n}_{(1)(a)}; {}^3{\bar e}_{(a)r}\Big),\nonumber \\
 &&{}\nonumber \\
 \sgn\, {}^4g_{(1)\tau\tau} &=& 1 + 2\, n_{(1)} = 1 +
  {{8\pi\, G}\over {c^3}}\, {1\over {\triangle}}\,
 \Big({\cal M}_{(1)}^{(UV)} + \sum_a\, T_{(1)}^{aa}\Big) -
 2\, \partial_{\tau}\, {}^3{\cal K}_{(1)},\nonumber \\
 \sgn\, {}^4g_{(1)\tau r} &=& - {\bar n}_{(1)(r)} =
 - \partial_r\, {}^3{\cal K}_{(1)} -
 {{4\pi\, G}\over {c^3}}\, {1\over
 {\triangle}}\, \Big(4\, {\cal M}_r^{(UV)} -
 {{\partial_r}\over {\triangle}}\,\, \sum_c\,
 \partial_c\, {\cal M}_c^{(UV)}\Big) -\nonumber \\
 &-& {1\over 2}\, {{\partial_r}\over {\triangle}}\, \partial_{\tau}\, \Big(4\,
 \Gamma_r^{(1)} - \sum_c\, {{\partial_c^2}\over {\triangle}}\,
 \Gamma_c^{(1)}\Big),\nonumber \\
 - \sgn\, {}^4g_{(1)rs} &=& {}^3g_{(1)rs} = \delta_{rs}\, [1 +
 2\, (\Gamma^{(1)}_r + 2\, \phi_{(1)})] =\nonumber \\
 &=& \delta_{rs}\, \Big[1 - {{8\pi\, G}\over {c^3}}\, {1\over
 {\triangle}}\, {\cal M}_{(1)}^{(UV)} + 2\, \Gamma_r^{(1)} +
 \sum_c\, {{\partial_c^2}\over {\triangle}}\, \Gamma_c^{(1)}\Big].
 \nonumber \\
 &&{}
 \label{2.14}
 \eea

\noindent The tetrads ${}^4{\buildrel \circ \over {\bar
E}}^A_{(1)(\alpha)}$ are adapted to the 3-spaces: ${}^4{\buildrel
\circ \over {\bar E}}^A_{(1)(o)} = l^A_{(o)} + l^A_{(1)}$ is the
normal to $\Sigma_{\tau}$. While the triads and the 3-metric in
$\Sigma_{\tau}$ are independent from the non-local York time, the
4-metric components ${}^4g_{(1)\tau\tau}$, ${}^4g_{(1)\tau r}$ and
the tetrads depend upon it.

\subsection{The PM 4-Christoffel Symbols and the PM 4-Riemann
and 4-Weyl Tensors}

By using the PM linearized 4-metric  given in Eq.(\ref{2.14}) we can
evaluate the Christoffel symbols and the Riemann and Weyl tensors of
these PM space-times and study the properties of the Riemannian
instantaneous 3-spaces. While the terms containing ${\cal
M}_{(1)}^{(UV)}$, ${\cal M}_{(1) r}^{(UV)}$, $T_{(1)}^{rs}$,
correspond to action-at-a-distance contributions, the terms
containing $\Gamma_a^{(1)} = \sum_{\bar a}\, \gamma_{\bar ar}\,
R_{\bar a}$ denote retarded GW contributions. The non-fixed gauge
part is given by the terms depending upon ${}^3{\cal K}_{(1)} =
{1\over {\triangle}}\, {}^3K_{(1)}$.

\subsubsection{The  PM Christoffel Symbols}

The  Christoffel symbols and their linearization have the following
expressions in our gauges \footnote{For the evaluation of
${}^4\Gamma^u_{\tau\tau}$ we need the kinematical Hamilton equations
for the cotriads given in Ref.\cite{8}. If we are in a Schwinger
time-gauge with $\alpha_{(a)}(\tau, \vec \sigma) \not= 0$, we have
to add to ${}^4\Gamma^u_{\tau\tau}$ the term $- {1\over 2}\,
\sum_{abc}\, {}^3e^u_{(a)}\, \epsilon_{(a)(b)(c)}\, {\hat
\mu}_{(b)}\, n_{(c)}$, where ${\hat \mu}_{(a)}(\tau, \vec \sigma)
\cir \partial_{\tau}\, \alpha_{(a)}(\tau, \vec \sigma)$ is the Dirac
multiplier in front of the rotation primary constraint.}
(${}^3\Gamma^u_{rs}$ is the Christoffel symbol built with the
3-metric ${}^3g_{rs}$ of the 3-space)

\begin{eqnarray*}
 {}^4\Gamma^A_{BC} &=& {1\over 2}\, {}^4g^{AE}\, (\partial_B\, {}^4g_{CE} +
 \partial_C\, {}^4g_{BE} - \partial_E\, {}^4g_{BC}) =\nonumber \\
 &=&{}^4\Gamma^A_{(1)BC} + O(\zeta^2) = {1\over 2}\, {}^4\eta^{AE}\, (\partial_B\,
 {}^4g_{(1)CE} + \partial_C\, {}^4g_{(1)BE} - \partial_E\,
 {}^4g_{(1)BC}) + O(\zeta^2),\end{eqnarray*}

 \begin{eqnarray*}
 {}^4\Gamma^{\tau}_{\tau\tau} &=&{1\over {1 + n}}\, \Big(\partial_{\tau}\, n +
 {\bar n}_{(a)}\, {}^3{\bar e}^r_{(a)}\, \partial_r\, n -
 {\bar n}_{(a)}\, {\bar n}_{(b)}\,   {}^3{\bar e}^s_{(a)}\,
 {}^3K_{rs}\, {}^3{\bar e}^r_{(b)} \Big) =\nonumber \\
 &=&{}^4\Gamma^{\tau}_{(1)\tau\tau} + O(\zeta^2) = \partial_{\tau}\, n_{(1)} =
  {{4\pi\, G}\over {c^3}}\, {1\over {\triangle}}\, \partial_{\tau}\,
 \Big({\cal M}_{(1)}^{(UV)} + \sum_a\, T_{(1)}^{aa}\Big) -
  \partial_{\tau}^2\, {}^3{\cal K}_{(1)} + O(\zeta^2),\nonumber \\
 {}^4\Gamma^{\tau}_{\tau r} &=& {1\over {1 + n}}\, \Big(\partial_r\, n
 - {}^3K_{rs}\, {}^3e^s_{(a)}\, {\bar n}_{(a)}\Big)
  =\nonumber \\
 &=& {}^4\Gamma^{\tau}_{(1)\tau r} + O(\zeta^2) = \partial_r\, n_{(1)} +
 {{4\pi\, G}\over {c^3}}\, {{\partial_r}\over {\triangle}}\,
 \Big({\cal M}_{(1)}^{(UV)} + \sum_a\, T_{(1)}^{aa}\Big) -
 \partial_r\, \partial_{\tau}\, {}^3{\cal K}_{(1)} + O(\zeta^2), \nonumber \\
 {}^4\Gamma^{\tau}_{rs} &=& - {1\over {1 + n}}\, {}^3K_{rs}
 =\nonumber \\
 &=& {}^4\Gamma^{\tau}_{(1)rs} + O(\zeta^2) = - {1\over 2}\, (\partial_r\,
 {\bar n}_{(1)(s)} + \partial_s\, {\bar n}_{(1)(r)}) + \delta_{rs}\,
 \partial_{\tau}\, (\Gamma_r^{(1)} + 2\, \phi_{(1)}) + O(\zeta^2) =\nonumber \\
 &=& -{{4\pi\, G}\over {c^3}}\, {1\over {\triangle}}\, \Big(2\,
 (\partial_r\, {\cal M}_{(1)s}^{(UV)} + \partial_s\, {\cal M}_{(1)r}^{(UV)})
 - {{\partial_r\, \partial_s}\over {\triangle}}\, \sum_c \partial_c\,
 {\cal M}_{(1)c}^{(UV)} + \delta_{rs}\, \partial_{\tau}\, {\cal M}_{(1)}^{(UV)}\Big)
 +\nonumber \\
 &+& \delta_{rs}\, \partial_{\tau}\, \Gamma_r^{(1)} - {{\partial_r\,
 \partial_s}\over {\triangle}}\, \partial_{\tau}\, (\Gamma_r^{(1)} + \Gamma_s^{(1)})
 + {1\over 2}\, (\delta_{rs} + {{\partial_r\, \partial_s}\over {\triangle}})\,
 \partial_{\tau}\, \sum_c\, {{\partial_c^2}\over {\triangle}}\, \Gamma_c^{(1)}
 -\nonumber \\
 &-& \partial_r\, \partial_s\, {}^3{\cal K}_{(1)} + O(\zeta^2),
 \end{eqnarray*}

\bea
 {}^4\Gamma^u_{\tau\tau} &\cir& \sum_a\, {}^3{\bar e}^u_{(a)}\,
 \Big[\partial_{\tau}\, {\bar n}_{(a)} - {{\partial_{\tau}\, n}\over {1 +
 n}}\, {\bar n}_{(a)} + \sum_{bcrs}\, {\bar n}_{(b)}\, {\bar n}_{(c)}\, {}^3{\bar
 e}^r_{(a)}\, {}^3{\bar e}^s_{(b)}\, (\partial_s\, {}^3{\bar e}_{(c)r} -
 \partial_r\, {}^3{\bar e}_{(c)s}) +\nonumber \\
 &+& \sum_{bcrs}\, {}^3e^r_{(b)}\, {}^3e^s_{(c)}\, n_{(b)}\, n_{(c)}\, {{n_{(a)}}\over {1 + n}}\,
 {}^3K_{rs} + (1 + n)\, \sum_{rb}\, {}^3e^r_{(b)}\, \Big(\delta_{(a)(b)} - {{n_{(a)}\,
 n_{(b)}}\over {(1 + n)^2}}\Big)\, \partial_r\, n -\nonumber \\
 &-& {1\over 2}\, \sum_{rb}\, n_{(b)}\, \Big({}^3e^r_{(a)}\, \partial_r\, n_{(b)} + {}^3e^r_{(b)}\,
 \partial_r\, n_{(a)}\Big) \Big] =\nonumber \\
 &=& {}^4\Gamma^u_{(1)\tau\tau} + O(\zeta^2) = \partial_{\tau}\, {\bar n}_{(1)(u)} +
 \partial_u\, n_{(1)} + O(\zeta^2) =\nonumber \\
 &=& {{4\pi\, G}\over {c^3}}\, \Big[{{\partial_{\tau}}\over {\triangle}}\,
 (4\, {\cal M}_{(1)u}^{(UV)} - {{\partial_u}\over {\triangle}}\,
 \sum_c\, \partial_c\, {\cal M}_{(1)c}^{(UV)}) + {{\partial_u}\over
 {\triangle}}\, ({\cal M}_{(1)}^{(UV)} + \sum_a\, T_{(1)}^{aa})\Big] +\nonumber \\
 &+& {1\over 2}\, {{\partial_u}\over {\triangle}}\,
 \partial_{\tau}^2\, (4\, \Gamma_u^{(1)} - \sum_c\,
 {{\partial_c^2}\over {\triangle}}\, \Gamma_c^{(1)}) + O(\zeta^2),\nonumber \\
 {}^4\Gamma^u_{\tau r} &=& {}^3{\bar e}^u_{(a)}\, \Big[-
 {{\partial_r\, n}\over {1 + n}}\, {\bar n}_{(a)} - (1 + n)\, (\delta_{(a)(b)} -
 {{{\bar n}_{(a)}\, {\bar n}_{(b)}}\over {(1 + n)^2}})\,
 {}^3K_{rs}\, {}^3{\bar e}^s_{(b)} + {}^3e^s_{(a)}\, {}^3e_{(b)r}\,
 \partial_s\, n_{(b)} -\nonumber \\
 &-&{1\over 2}\, \Big({}^3{\bar e}^v_{(a)}\, (\partial_r\, {}^3{\bar e}_{(b)v} -
 \partial_v\, {}^3{\bar e}_{(b)r}) - {}^3{\bar e}^v_{(b)}\, (\partial_r\, {}^3{\bar e}_{(a)v}
 - \partial_v\, {}^3{\bar e}_{(a)r}) +\nonumber \\
 &+&{}^3{\bar e}^v_{(a)}\, {}^3{\bar e}_{(c)r}\, {}^3{\bar
 e}^s_{(b)}\, (\partial_v\, {}^3{\bar e}_{(c)s} - \partial_s\, {}^3{\bar e}_{(c)v})
 \Big)\, {\bar n}_{(b)}\Big] =\nonumber \\
 &=&{}^4\Gamma^u_{(1)\tau r} + O(\zeta^2) = \delta_{ur}\, \partial_{\tau}\,
 (\Gamma_r^{(1)} + 2\, \phi_{(1)}) + {1\over 2}\, (\partial_r\,
 {\bar n}_{(1)(u)} - \partial_u\, {\bar n}_{(1)(r)}) + O(\zeta^2) =\nonumber \\
 &=& {{8\pi\, G}\over {c^3}}\, {1\over {\triangle}}\, \Big(\partial_r\,
 {\cal M}_{(1)(u)}^{(UV)} - \partial_u\, {\cal M}_{(1)(r)}^{(UV)} -
 {1\over 2}\, \delta_{ur}\, \partial_{\tau}\, {\cal M}_{(1)}^{(UV)}\Big)
 +\nonumber \\
 &+&\delta_{ur}\, \partial_{\tau}\, \Big(\Gamma_r^{(1)} + {1\over 2}\,
 \sum_c\, {{\partial_c^2}\over {\triangle}}\Big) - {{\partial_r\, \partial_u}\over
 {\triangle}}\, \partial_{\tau}\, (\Gamma_r^{(1)} - \Gamma_u^{(1)}) + O(\zeta^2),\nonumber \\
 {}^4\Gamma^u_{rs} &=& {}^3\Gamma^u_{rs} + {{{\bar n}_{(a)}}\over {1 + n}}\, {}^3{\bar
 e}^u_{(a)}\, {}^3K_{rs} =\nonumber \\
 &=& {}^4\Gamma^u_{(1)rs} + O(\zeta^2) = {}^3\Gamma^u_{(1)rs} + O(\zeta^2) =\nonumber \\
 &=& \delta_{ur}\, \partial_s\, (\Gamma^{(1)}_u + 2\, \phi_{(1)})
 + \delta_{us}\, \partial_r\, (\Gamma^{(1)}_u + 2\, \phi_{(1)})
 - \delta_{rs}\, \partial_u\, (\Gamma^{(1)}_r + 2\, \phi_{(1)}) +
 O(\zeta^2) =\nonumber \\
 &=& - {{4\pi\, G}\over {c^3}}\, {{\delta_{ur}\, \partial_s + \delta_{us}\,
 \partial_r - \delta_{rs}\, \partial_u}\over {\triangle}}\, {\cal
 M}_{(1)}^{(UV)} +\nonumber \\
 &+& (\delta_{ur}\, \partial_s + \delta_{us}\, \partial_r)\, \Gamma_u^{(1)}
 - \delta_{rs}\, \partial_u\, \Gamma_r^{(1)} +
  {{\delta_{ur}\, \partial_s + \delta_{us}\,
 \partial_r - \delta_{rs}\, \partial_u}\over {2\, \triangle}}\,
 \sum_c\, \partial_c^2\, \Gamma_c^{(1)} + O(\zeta^2).\nonumber \\
 &&{}
 \label{2.15}
 \eea

 \medskip

\subsubsection{The PM Riemann and Ricci Tensors}

The 4-Riemann tensor and its linearization have the following
expressions

\begin{eqnarray*}
 {}^4R_{ABCD} &=& {}^4g_{AE}\, {}^4R^E{}_{BCD} =\nonumber \\
 &=&- {1\over 2}\, \Big(\partial_A\, \partial_c\, {}^4g_{BD} + \partial_B\, \partial_D\, {}^4g_{AC}
 - \partial_A\, \partial_D\, {}^4g_{BC} - \partial_B\, \partial_C\, {}^4g_{AD}\Big) +\nonumber \\
 &+& {}^4g_{EF}\, \Big({}^4\Gamma^E_{AD}\, {}^4\Gamma^F_{BC} - {}^4\Gamma^E_{AC}\,
  {}^4\Gamma^F_{BD}\Big) =\nonumber \\
 &=&{}^4R_{(1)ABCD} + O(\zeta^2) =  {}^4\eta_{AE}\, {}^4R^E_{(1)BCD} + O(\zeta^2) =\nonumber \\
 &=& - {1\over 2}\, (\partial_A\, \partial_C\, {}^4g_{(1)BD} +
 \partial_B\, \partial_D\, {}^4g_{(1)AC} -
 \partial_A\, \partial_D\, {}^4g_{(1)BC} -
 \partial_B\, \partial_C\, {}^4g_{(1)AD}) + O(\zeta^2),
 \end{eqnarray*}

 \begin{eqnarray*}
 {}^4R_{(1)rsuv} &=& - \sgn\, {}^3R_{(1)rsuv} =\nonumber \\
 &=& - \sgn\, \Big[\delta_{rv}\, \partial_s\, \partial_u\, (\Gamma^{(1)}_r + 2\, \phi_{(1)}) -
 \delta_{ru}\, \partial_s\, \partial_v\, (\Gamma^{(1)}_r + 2\,
 \phi_{(1)})  +\nonumber \\
 &+& \delta_{su}\, \partial_r\, \partial_v\, (\Gamma^{(1)}_s + 2\, \phi_{(1)})
 - \delta_{sv}\, \partial_r\, \partial_u\, (\Gamma^{(1)}_s + 2\,
 \phi_{(1)})\Big] =\nonumber \\
 &=& - \sgn\, \Big[- {{4\pi\, G}\over {c^3}}\, {{(\delta_{rv}\,
 \partial_u - \delta_{ru}\, \partial_v)\, \partial_s - (\delta_{sv}\,
 \partial_u - \delta_{su}\, \partial_v)\, \partial_r}\over {\triangle}}\,
 {\cal M}_{(1)}^{(UV)} +\nonumber \\
 &+& (\delta_{rv}\, \partial_u - \delta_{ru}\, \partial_v)\, \partial_s\, \Gamma_r^{(1)}
 - (\delta_{sv}\, \partial_u - \delta_{su}\, \partial_v)\, \partial_r\,
 \Gamma_s^{(1)} +\nonumber \\
 &+&{{(\delta_{rv}\,
 \partial_u - \delta_{ru}\, \partial_v)\, \partial_s - (\delta_{sv}\,
 \partial_u - \delta_{su}\, \partial_v)\, \partial_r}\over {2\, \triangle}}\,
 \sum_c\, \partial_c^2\, \Gamma_c^{(1)} \Big],
 \end{eqnarray*}

\begin{eqnarray*}
 {}^4R_{(1)\tau ruv} &=& \sgn\, \Big[(\delta_{rv}\, \partial_u -
 \delta_{ru}\, \partial_v)\, \partial_{\tau}\, (\Gamma_r^{(1)}
 + 2\, \phi_{(1)}) + {1\over 2}\, \partial_r\, (\partial_v\, {\bar n}_{(1)(u)} -
 \partial_u\, {\bar n}_{(1)(v)})\Big] =\nonumber \\
 &=& \sgn\, \Big[{{4\pi\, G}\over {c^3}}\, {1\over {\triangle}}\,
 \Big((\delta_{ru}\, \partial_v - \delta_{rv}\, \partial_u)\,
 \partial_{\tau}\, {\cal M}_{(1)}^{(UV)} - 2\, \partial_r\, (\partial_u\,
 {\cal M}_{(1)v}^{(UV)} - \partial_v\, {\cal M}_{(1)u}^{(UV)})\Big) +\nonumber \\
 &+& \partial_{\tau}\, \Big((\delta_{rv}\, \partial_u - \delta_{ru}\,
 \partial_v)\, (\Gamma_r^{(1)} + {1\over 2}\, \sum_c\, {{\partial_c^2}\over
 {\triangle}}\, \Gamma_c^{(1)}) + {{\partial_r\, \partial_u\, \partial_v}\over
 {\triangle}}\, (\Gamma_u^{(1)} - \Gamma_v^{(1)})\Big)\Big],
 \end{eqnarray*}

\bea
 {}^4R_{(1)\tau r\tau s} &=& - {{\sgn}\over 2}\, \Big(2\,
 \partial_r\, \partial_s\, n_{(1)} - 2\, \delta_{rs}\, \partial^2_{\tau}\,
 (\Gamma^{(1)}_r + 2\, \phi_{(1)}) + \partial_{\tau}\, (\partial_r\,
 {\bar n}_{(1)(s)} + \partial_s\, {\bar n}_{(1)(r)})\Big) =\nonumber \\
 &=& \sgn\, \Big[- {{4\pi\, G}\over {c^3}}\, {1\over {\triangle}}\,
 \Big(\partial_r\, \partial_s\, ({\cal M}_{(1)}^{(UV)} + \sum_a\, T_{(1)}^{aa})
 + \delta_{rs}\, \partial_{\tau}^2\, {\cal M}_{(1)}^{(UV)}
 +\nonumber \\
 &+& \partial_{\tau}\, \Big[2\, (\partial_r\, {\cal M}_{(1)s}^{(UV)}
 + \partial_s\, {\cal M}_{(1)r}^{(UV)}) - {{\partial_r\, \partial_s}\over
 {\triangle}} \sum_c\, \partial_c\, {\cal M}_{(1)c}^{(UV)}\Big] \Big)
 +\nonumber \\
 &+& \partial_{\tau}^2\, \Big(\delta_{rs}\, (\Gamma_r^{(1)} + {1\over 2}\,
 \sum_c\, {{\partial_c^2}\over {\triangle}}\, \Gamma_c^{(1)})
 - {{\partial_r\, \partial_s}\over {\triangle}}\, (\Gamma_r^{(1)} +
 \Gamma_s^{(1)} - {1\over 2}\, \sum_c\, {{\partial_c^2}\over {\triangle}}\,
 \Gamma_c^{(1)})\Big) \Big] =\nonumber \\
 &=&\sgn\, \Big[- {{4\pi\, G}\over {c^3}}\, {1\over {\triangle}}\,
 \Big(\partial_r\, \partial_s\, ({\cal M}_{(1)}^{(UV)} + \sum_a\, T_{(1)}^{aa})
 + \delta_{rs}\, \partial_{\tau}^2\, {\cal M}_{(1)}^{(UV)}
 +\nonumber \\
 &+& \partial_{\tau}\, \Big[2\, (\partial_r\, {\cal M}_{(1)s}^{(UV)}
 + \partial_s\, {\cal M}_{(1)r}^{(UV)}) - {{\partial_r\, \partial_s}\over
 {\triangle}} \sum_c\, \partial_c\, {\cal M}_{(1)c}^{(UV)}\Big] \Big)
 -\nonumber \\
 &-& {1\over 2}\, \partial^2_{\tau}\, {}^4h_{(1)rs}^{TT}.
 \label{2.16}
 \eea

\noindent The final expression of ${}^4R_{(1)\tau r\tau s}$ has been
obtained by using Eq.(6.12) of paper II and has been used in
Eq.(7.39) of paper II. Let us remark that the Riemann tensor does
not depend upon the York time ${}^3K$.

\medskip
For the 4-Ricci tensor and the  4-curvature scalar we have ($\Box =
\partial_{\tau}^2 - \triangle$)

\begin{eqnarray*}
 {}^4R_{(1)AB} &=& {}^4\eta^{EF}\, {}^4R_{(1)EAFB} = \sgn\,
 \Big({}^4R_{(1)\tau A\tau B} - \sum_r\, {}^4R_{(1)rArB}\Big),\nonumber \\
 {}^4R_{(1)} &=& {}^4\eta^{AB}\, {}^4R_{(1)AB} = \sgn\,
 \Big({}^4R_{(1)\tau\tau} - \sum_r\, {}^4R_{(1)rr}\Big),
 \end{eqnarray*}

\begin{eqnarray*}
 {}^4R_{(1)\tau\tau} &=& - 6\, \partial^2_{\tau}\, \phi_{(1)} +
 \triangle\, n_{(1)} + \sum_r\, \partial_{\tau}\, \partial_r\, {\bar
 n}_{(1)(r)} =\nonumber \\
 &=& {{4\pi\, G}\over {c^3}}\, \Big((1 + 3\, {{\partial_{\tau}^2}\over
 {\triangle}})\, {\cal M}_{(1)}^{(UV)} + \sum_a\, T_{(1)}^{aa} +
 3\, {{\partial_{\tau}}\over {\triangle}}\, \sum_c\, \partial_c\,
 {\cal M}_{(1)c}^{(UV)}\Big),\nonumber \\
 {}^4R_{(1)\tau r} &=& \partial_{\tau}\, \partial_r\, (\Gamma_r^{(1)}
 - 4\, \phi_{(1)}) + {1\over 2}\, \sum_s\, \partial_s\,
 (\partial_r\, {\bar n}_{(1)(s)} - \partial_s\, {\bar n}_{(1)(r)})
 =\nonumber \\
 &=& {{8\pi\, G}\over {c^3}}\, \Big({{\partial_{\tau}\, \partial_r}
 \over {\triangle}}\, {\cal M}_{(1)}^{(UV)} +  \sum_s\,
 {{\partial_s}\over {\triangle}}\, (\partial_r\, {\cal M}_{(1)s}^{(UV)}
 - \partial_s\, {\cal M}_{(1)r}^{(UV)})\Big), \nonumber \\
 {}^4R_{(1)rs} &=& \partial_r\, \partial_s\, (- n_{(1)} + \Gamma^{(1)}_r +
 \Gamma^{(1)}_s - 2\, \phi_{(1)}) + \delta_{rs}\, (\partial^2_{\tau} -
 \triangle)\, (\Gamma^{(1)}_r + 2\, \phi_{(1)}) -\nonumber \\
 &-& {1\over 2}\, \partial_{\tau}\, (\partial_r\, {\bar n}_{(1)(s)} +
 \partial_s\, {\bar n}_{(1)(r)}) =\nonumber \\
 &=& - {1\over 2}\, \Box\, {}^4h_{(1)rs}^{TT} + {{4\pi\, G}\over {c^3}}\,
 \Big(- \delta_{rs}\, {{\Box}\over {\triangle}}\, {\cal M}_{(1)}^{(UV)} -
 {{\partial_r\, \partial_s}\over {\triangle}}\, \sum_a\, T_{(1)}^{aa}
 -\nonumber \\
 &-& 2\, {{\partial_{\tau}}\over {\triangle}}\, (\partial_r\,
 {\cal M}_{(1)s}^{(UV)} + \partial_s\, {\cal M}_{(1)r}^{(UV)}) +
 {{\partial_r\, \partial_s\, \partial_{\tau}}\over {\triangle^2}}\, \sum_c\,
 \partial_c\, {\cal M}_{(1)c}^{(UV)}\Big),
 \end{eqnarray*}

 \bea
 {}^4R_{(1)} &=&  2\,  \Big(- \sum_r\, \partial_r^2\, \Gamma^{(1)}_r
 + \triangle\, n_{(1)} + \sum_r\, \partial_{\tau}\, \partial_r\, {\bar n}_{(1)(r)}
 + 8\, \triangle\, \phi_{(1)} - 12\, \partial_{\tau}^2\, \phi_{(1)}\Big)
 =\nonumber \\
 &=& - {{8\pi\, G}\over {c^3}}\, \Big((1 - 3\, {{\partial^2_{\tau}}\over
 {\triangle}})\, {\cal M}_{(1)}^{(UV)} - \sum_a\, T_{(1)}^{aa} - 3\,
 {{\partial_{\tau}}\over {\triangle}}\,
 \sum_c\, \partial_c\, {\cal M}_{(1)c}^{(UV)}\Big). \nonumber \\
 &&{}
 \label{2.17}
 \eea

 \medskip

By using Eqs.(\ref{1.12}), it can be checked that Einstein equations
${}^4R_{AB} - {1\over 2}\, {}^4g_{AB}\, {}^4R \cir {{8\pi\, G}\over
{c^3}}\, T_{AB}$ are verified, namely we have ${}^4R_{(1)AB} -
{1\over 2}\, {}^4\eta_{AB}\, {}^4R_{(1)} \cir {{8\pi\, G}\over
{c^3}}\, T_{(1)AB} + O(\zeta^2)$.

\subsubsection{The PM Weyl Tensor}

For the Weyl tensor and its electric and magnetic components with
respect to the Eulerian observers, whose unit 4-velocity $l^A$ is
the normal to the 3-space $\Sigma_{\tau}$ with the zeroth order
expression $l^A_{(o)} = (l^{\tau}_{(o)}1; l^r_{(o)} = 0)$ [see
Eqs.(\ref{2.14})], we have

\begin{eqnarray*}
  {}^4C_{ABCD} &=& {}^4R_{ABCD} - {1\over 2}\, ({}^4g_{AC}\, {}^4R_{BD} +
{}^4g_{BD}\, {}^4R_{AC} - {}^4g_{AD}\, {}^4R_{BC} - {}^4g_{BC}\, {}^4R_{AD}) +\nonumber \\
 &+&{1\over 6}\, ({}^4g_{AC}\, {}^4g_{BD} - {}^4g_{AD}\, {}^4g_{BC})\, {}^4R
 = {}^4C_{(1)ABCD} + O(\zeta^2),\nonumber \\
 &&{}\nonumber\\
 &&{}^4C_{ABCD} = {}^4C_{CDAB} = - {}^4C_{BACD} = - {}^4C_{ABDC},\nonumber  \\
 &&{}^4C_{ABCD} + {}^4C_{ADBC} + {}^4C_{ACDB} = 0,
 \end{eqnarray*}

 \begin{eqnarray*}
  {}^4C_{(1)\tau r\tau s} &=& {}^4R_{(1)\tau r\tau s} - {{\sgn}\over 2}\,
 \Big({}^4R_{(1)rs} - \delta_{rs}\, {}^4R_{(1)\tau\tau}\Big) - {1\over
 6}\, \delta_{rs}\, {}^4R_{(1)} =\nonumber \\
 &=& - {1\over 4}\, (\Box - \triangle)\, {}^4h_{(1)rs}^{TT} +
 {{4\pi\, G}\over {c^3}}\, \Big[({1\over 3}\, \delta_{rs} - {{\partial_r\,
 \partial_s}\over {\triangle}})\, ({\cal M}_{(1)}^{(UV)} + {1\over 2}\,
 \sum_a\, T_{(1)}^{aa}) +\nonumber \\
 &+&{1\over 2}\, (\delta_{rs} + {{\partial_r\, \partial_s}\over {\triangle}})\,
  \sum_c\, {{\partial_c\, \partial_{\tau}}\over {\triangle}}\, {\cal M}_{(1)c}^{(UV)}
 - {{\partial_{\tau}}\over {\triangle}}\, (\partial_r\, {\cal M}_{(1)s}^{(UV)} +
 \partial_s\, {\cal M}_{(1)r}^{(UV)}) \Big],\nonumber \\
 &&{}\nonumber \\
  {}^4C_{(1)\tau ruv} &=& {}^4R_{(1)\tau ruv} + {{\sgn}\over 2}\,
 \Big(\delta_{rv}\, {}^4R_{(1)\tau u} - \delta_{ru}\,
 {}^4R_{(1)\tau v}\Big) =\nonumber \\
 &=&\partial_{\tau}\, \Big[(\delta_{rv}\, \partial_u - \delta_{ru}\, \partial_v)\,
 (\Gamma_r^{(1)} + {1\over 2}\, \sum_c\, {{\partial_c^2}\over
 {\triangle}}\, \Gamma_c^{(1)}) + {{\partial_r\, \partial_u\, \partial_v}\over
 {\triangle}}\, (\Gamma_u^{(1)} - \Gamma_v^{(1)})\Big] +\nonumber \\
 &+& {{4\pi\, G}\over {c^3}}\, {1\over {\triangle}}\,
 \Big[\sum_c\, [\delta_{ru}\,  (\delta_{vc} - \partial_v\, \partial_c) -
 \delta_{rv}\, (\delta_{uc} - \partial_u\, \partial_c)]\, {\cal
 M}_{(1)c}^{(UV)} - 2\, \partial_r\, (\partial_u\, {\cal M}_{(1)v}^{(UV)} -
 \partial_v\, {\cal M}_{(1)u}^{(UV)}) \Big],
 \end{eqnarray*}

\begin{eqnarray*}
  {}^4C_{(1)rsuv} &=& {}^4R_{(1)rsuv} + {{\sgn}\over 2}\,
 \Big(\delta_{ru}\, {}^4R_{(1)sv} + \delta_{sv}\, {}^4R_{(1)ru} -\nonumber \\
 &-& \delta_{rv}\, {}^4R_{(1)su} - \delta_{su}\, {}^4R_{(1)rv}\Big) + {1\over
 6}\, \Big(\delta_{ru}\, \delta_{sv} - \delta_{rv}\, \delta_{su}\Big)\,
 {}^4R_{(1)} =\nonumber \\
 &=& - {1\over 4}\, \Box\, (\delta_{ru}\, {}^4h_{(1)rv}^{TT} + \delta_{sv}\, {}^4h_{(1)ru}^{TT} -
 \delta_{rv}\, {}^4h_{(1)su}^{TT} - \delta_{su}\, {}^4h_{(1)rv}^{TT}) -\nonumber \\
 &-&(\delta_{rv}\, \partial_u\, \partial_s + \delta_{su}\, \partial_v\, \partial_r
 - \delta_{ru}\, \partial_v\, \partial_s - \delta_{sv}\, \partial_u\, \partial_r)\,
 (\Gamma_s^{(1)} + {1\over 2}\, \sum_c\, {{\partial_c^2}\over {\triangle}}\,
 \Gamma_c^{(1)}) +\nonumber \\
 &+& {{4\pi\, G}\over {c^3}}\, \Big[({2\over 3}\, (\delta_{ru}\, \delta_{sv} - \delta_{rv}\,
 \delta_{su}) + \delta_{rv}\, {{\partial_u\, \partial_s}\over {\triangle}} + \delta_{su}\,
 {{\partial_v\, \partial_r}\over {\triangle}} - \delta_{ru}\, {{\partial_v\, \partial_s}
 \over {\triangle}} - \delta_{sv}\, {{\partial_u\, \partial_r}\over {\triangle}})\, {\cal M}_{(1)}^{(UV)}
 +\nonumber \\
 &+& ({1\over 3}\, (\delta_{ru}\, \delta_{sv} - \delta_{rv}\,
 \delta_{su}) + 2\,\delta_{rv}\, {{\partial_u\, \partial_s}\over {\triangle}} + 2\, \delta_{su}\,
 {{\partial_v\, \partial_r}\over {\triangle}} - 2\, \delta_{ru}\, {{\partial_v\, \partial_s}
 \over {\triangle}} - 2\, \delta_{sv}\, {{\partial_u\, \partial_r}\over {\triangle}})\,
   \sum_a\, T_{(1)}^{aa} +\nonumber \\
 &+& (\delta_{ru}\, \delta_{sv} - \delta_{rv}\,
 \delta_{su} - 2\, \delta_{rv}\, {{\partial_u\, \partial_s}\over {\triangle}} - 2\, \delta_{su}\,
 {{\partial_v\, \partial_r}\over {\triangle}} + 2\, \delta_{ru}\, {{\partial_v\, \partial_s}
 \over {\triangle}} + 2\, \delta_{sv}\, {{\partial_u\, \partial_r}\over {\triangle}})\,
 \sum_c\, {{\partial_{\tau}\, \partial_c}\over {\triangle}}\, {\cal M}_{(1)c}^{(UV)}
 +\nonumber \\
 &+& {{\partial_{\tau}}\over {\triangle}}\, \Big(\delta_{ru}\, (\partial_s\, {\cal M}_{(1)v}^{(UV)}
 + \partial_v\, {\cal M}_{(1)s}^{(UV)}) + \delta_{sv}\, (\partial_r\, {\cal M}_{(1)u}^{(UV)}
 + \partial_u\, {\cal M}_{(1)r}^{(UV)}) -\nonumber \\
 &-& \delta_{rv}\, (\partial_s\, {\cal M}_{(1)u}^{(UV)} + \partial_u\, {\cal M}_{(1)s}^{(UV)}) -
 \delta_{su}\, (\partial_r\, {\cal M}_{(1)v}^{(UV)} + \partial_v\, {\cal M}_{(1)r}^{(UV)})
 \Big)\Big],
 \end{eqnarray*}

\bea
 E^A_{(1)B} &=&{}^4\eta^{AC}\, {}^4C_{(1)CEBF}\, l_{(o)}^E\,
 l_{(o)}^F = {}^4\eta^{AC}\, {}^4C_{(1)C\tau B\tau} = - \sgn\,
 \sum_{rs}\, \delta^{Ar}\, \delta_{Bs}\, {}^4C_{(1)\tau r\tau s},\nonumber \\
 H_{(1)AB} &=& {1\over 2}\, \epsilon_{AMCD}\, l_{(o)}^M\, {}^4\eta^{DE}\,
 {}^4\eta^{CF}\, {}^4C_{(1)EFBG}\, l_{(o)}^G = {1\over 2}\, \sum_{rsuv}\, \delta_{As}\,
 \delta_{Br}\, \epsilon_{suv}\, C_{(1)\tau nrs} =\nonumber \\
 &=&\sum_{rsuv}\, \delta_{As}\, \delta_{Br}\, \Big(\partial_{\tau}\,
 \Big[\epsilon_{rsu}\, \partial_u\, (\Gamma_r^{(1)} + {1\over 2}\, \sum_c\,
 {{\partial_c^2}\over {\triangle}}\, \Gamma_c^{(1)}) + \epsilon_{suv}\,
 {{\partial_r\, \partial_u\, \partial_v}\over {\triangle}}\, \Gamma_u^{(1)}\Big]
 -\nonumber \\
 &-&{{4\pi\, G}\over {c^3}}\, {1\over {\triangle}}\,
 \Big[\epsilon_{rsu}\, \sum_c\, (\delta_{uc} - \partial_u\, \partial_c)\,
 {\cal M}_{(1)c}^{(UV)} + \epsilon_{suv}\, \partial_r\, (\partial_u\, {\cal M}_{(1)v}^{(UV)}
 - \partial_v\, {\cal M}_{(1)u}^{(UV)}) \Big]\Big).\nonumber \\
 &&{}
 \label{2.18}
 \eea

\noindent Their Newtonian limit, in particular the vanishing of
$H_{(1)AB}$, is consistent with Ref.\cite{9}.

\subsection{The 4-Spin and 3-Spin Connections}

The 4-spin connection ${}^4\omega_A{}^{(\alpha)}_{(\beta)}$
associated with the general tetrads ${}^4E^A_{(\alpha)}$ is
connected with the $\Sigma_{\tau}$-adapted one ${}^4{\buildrel \circ
\over \omega}_A{}^{(\alpha)}_{(\beta)}$ by means of the Lorentz
boosts with parameters $\varphi_{(a)}$ \cite{8}. The expressions of
these 4-spin connections are (${\hat \mu}_{(a)}(\tau, \vec \sigma)\,
\cir\, \partial_{\tau}\, \alpha_{(a)}(\tau, \vec \sigma)$ are the
Dirac multipliers in front of the primary first class rotation
constraints; they vanish in the gauges $\alpha_{(a)}(\tau, \vec
\sigma) \approx 0$; ${}^3K_{rs}$ is given in Eqs.(2.3) of paper II)

\begin{eqnarray*}
 {}^4\omega_A{}^{(\alpha )}{}_{(\beta )} &=& {}^4E^{(\alpha )}_B\,
 \Big(\partial_A\, {}^4E^B_{(\beta )} + {}^4\Gamma^B_{AC}\,
 {}^4E^C_{(\beta )}\Big) =\nonumber \\
 &=& \Big[L(\varphi_{(a)})\, {}^4{\buildrel o\over \omega}_A\,
 L^{-1}(\varphi_{(a)}) + \partial_A\, L(\varphi_{(a)})\,
 L^{-1}(\varphi_{(a)})\Big]^{(\alpha )}{}_{(\beta )},\nonumber \\
 &&{}\nonumber \\
 {}^4{\buildrel o\over \omega}_A{}^{(\alpha )}{}_{(\beta )} &=&
 \eo^{(\alpha )}_B\, \Big(\partial_A\, \eo^B_{(\beta )} +
 \Gamma^B_{AC}\, \eo^C_{(\beta )}\Big),
 \end{eqnarray*}

\bea
 {}^4{\buildrel o\over \omega}_{\tau}{}^{(o )}{}_{(a )} &=&
 \sgn\, {}^4{\buildrel o\over \omega}_{\tau}{}_{(o )(a )}
 = {}^3e^r_{(a)}\, \partial_r\, n - {}^3e^r_{(a)}\, {}^3K_{rs}\,
 {}^3e^s_{(b)}\, n_{(b)},\nonumber \\
 {}^4{\buildrel o\over \omega}_{\tau}{}^{(a )}{}_{(b )} &=&
 - \sgn\, {}^4{\buildrel o\over \omega}_{\tau}{}_{(a )(b )} =
 - \epsilon_{(a)(b)(c)}\, {\hat  \mu}_{(c)} +\nonumber \\
 &+& \sum_u\, \Big({}^3e^u_{(b)}\, \partial_u\, n_{(a)} -
 {}^3e^u_{(a)}\, \partial_u\, n_{(b)} + \sum_{sc}\, n_{(c)}\,
 {}^3e^s_{(c)}\, {}^3e^u_{(b)}\, \partial_s\, {}^3e_{(a)u}
 \Big) -\nonumber \\
 &-& {1\over 2}\, \sum_c\, n_{(c)}\, \sum_{su}\, \Big[{}^3e^s_{(c)}\,
 \Big({}^3e^u_{(b)}\, \partial_u\, {}^3e_{(a)s} -
 {}^3e^u_{(a)}\, \partial_u\, {}^3e_{(b)s}\Big) - {}^3e^u_{(a)}\,
 {}^3e^s_{(b)}\, \Big(\partial_u\, {}^3e_{(c)s} -
 \partial_s\, {}^3e_{(c)u}\Big) \Big],\nonumber \\
 {}^4{\buildrel o\over \omega}_r{}^{(o )}{}_{(a )} &=&
 \sgn\, {}^4{\buildrel o\over \omega}_r{}_{(o )(a )}
 = - \sgn\, {}^4{\buildrel o\over \omega}_r{}_{(a )(o)}
 = - {}^3K_{rs}\, {}^3e^s_{(a)},\nonumber \\
 {}^4{\buildrel o\over \omega}_r{}^{(a )}{}_{(b )} &=&
 - \sgn\, {}^4{\buildrel o\over \omega}_r{}_{(a )(b )}
 = {}^3\omega_{r(a)(b)}. \nonumber \\
 &&{}
 \label{2.19}
 \eea

\bigskip

The 3-spin connection \footnote{ It is defined by the vanishing of
the generalized covariant derivative acting on both types of indices
of the triad: $\partial_r\, {}^3e^u_{(a)} + {}^3\Gamma^u_{rs}\,
{}^3e^s_{(a)} + {}^3\omega_{r(a)(b)}\, {}^3e^u_{(b)} =
{}^3\nabla_r\,\, {}^3e^u_{(a)} + {}^3\omega_{r(a)(b)}\,
{}^3e^u_{(b)} = 0$, so that the 3-Christoffel symbols (see also the
last of Eqs.(\ref{2.15})) have the expression ${}^3\Gamma^u_{rs} =
{1\over 2}\, \Big[{}^3{\bar e}^u_{(a)}\,
 \Big({}^3{\bar e}_{(b)r}\, {}^3{\bar \omega}_{s(a)(b)} +
 {}^3{\bar e}_{(b)s}\, {}^3{\bar \omega}_{r(a)(b)}\Big)
 - \Big({}^3{\bar e}_{(a)r}\, \partial_s\, {}^3{\bar e}^u_{(a)} +
 {}^3{\bar e}_{(a)s}\, \partial_r\, {}^3{\bar
 e}^u_{(a)}\Big)\Big]$.} ${}^3\omega_{r(a)(b)} = {}^4{\buildrel \circ \over
\omega}_{r(a)(b)}$ with ${}^3\omega_{r(a)(b)}$ depending  on triads
and cotriads and also on the angles $\alpha_{(a)}$  is connected to
the 3-spin connection ${}^3{\bar \omega}_{r(a)(b)}$ in Schwinger
time gauges by local SO(3) rotations $R(\alpha_{(a)})$ ($R^T =
R^{-1}$) and has the expression in gauges near the 3-orthogonal ones
having $\theta^i(\tau, \vec \sigma) = \theta^i_{(1)}(\tau, \vec
\sigma) = O(\zeta) \not= 0$ (we use $V_{ra}(\theta^i) = \delta_{ra}
- \sum_i\, \epsilon_{rai}\, \theta^i_{(1)} + O(\zeta^2)$, see before
Eqs.(2,9) of paper I)

\begin{eqnarray*}
 {}^3\omega_{r(a)(b)} &=& \epsilon_{(a)(b)(c)}\, {}^3\omega_{r(c)}
 = \Big[R(\alpha_{(e)})\, {}^3{\bar \omega}_r\, R^T(\alpha_{(e)}) +
 R(\alpha_{(e)})\, \partial_r\, R^T(\alpha_{(e)})\Big]_{(a)(b)},
 \end{eqnarray*}

\bea
 {}^3{\bar \omega}_{r(a)} &=& {1\over 2}\, \sum_{bc}\,
 \epsilon_{(a)(b)(c)}\, {}^3{\bar \omega}_{r(b)(c)} =\nonumber \\
 &=& {1\over 2}\, \sum_{bcu}\, \epsilon_{(a)(b)(c)}\, {}^3{\bar e}^u_{(b)}\,
 \Big[\partial_r\, {}^3{\bar e}_{(c)u} - \partial_u\, {}^3{\bar e}_{(c)r}  +
 \sum_{dv}\, {}^3{\bar e}^v_{(c)}\, {}^3{\bar e}_{(d)r}\, \partial_v\,
 {}^3{\bar e}_{(d)u}\Big] =\nonumber \\
 &&{}\nonumber \\
 &=&{}^3{\bar \omega}_{(1)r(a)} + O(\zeta^2) =\nonumber \\
 &=& {1\over 2}\, \sum_{bc}\, \epsilon_{(a)(b)(c)}\, \Big[(\delta_{rb}\,
 \partial_c - \delta_{rc}\, \partial_b)\, (\Gamma_r^{(1)} + 2\, \phi_{(1)})
 - \sum_i\, \epsilon_{(b)(c)(i)}\, \theta^i_{(1)}\Big] + O(\zeta^2),\nonumber \\
 &&{}\nonumber \\
 {}^3{\bar \omega}_{(1)r(a)(b)} &=& \epsilon_{(a)(b)(c)}\, {}^3{\bar
 \omega}_{(1)r(c)}.
 \label{2.20}
 \eea

\medskip

 Once we know the PM 3-spin connection ${}^3{\bar \omega}_{(1)r(a)(b)}$ in the 3-orthogonal
Schwinger time gauges, we can go to  non-Schwinger gauges near the
3-orthogonal ones (with $\alpha_{(a)}(\tau, \vec \sigma) \not= 0$,
$\varphi_{(a)}(\tau, \vec \sigma) \not= 0$, $\theta^i(\tau, \vec
\sigma) = \theta^i_{(1)}(\tau, \vec \sigma) = O(\zeta) \not= 0$) and
find (we use Eqs. (\ref{1.9}), (\ref{2.5}) and (\ref{2.14}) and
$n_{(a)} = \sum_b\, R_{(a)(b)}(\alpha_{(c)})\, {\bar n}_{(b)}$,
${}^3e^r_{(a)} = \sum_b\, R_{(a)(b)}(\alpha_{(c)})\, {}^3{\bar
e}^r_{(b)}$)

\bea
 {}^4{\buildrel o\over \omega}_r{}^{(a )}{}_{(b )} &=&
  {}^3\omega_{r(a)(b)} = \epsilon_{(a)(b)(c)}\, {}^3\omega_{r(c)}
 = \Big[R(\alpha_{(e)})\, \partial_r\, R^T(\alpha_{(e)})
 + R(\alpha_{(e)})\, {}^3{\bar \omega}_r\, R^T(\alpha_{(e)})
 \Big]_{(a)(b)} =\nonumber \\
 &=&\Big[R(\alpha_{(e)})\, \partial_r\, R^T(\alpha_{(e)})
 + R(\alpha_{(e)})\, {}^3{\bar \omega}_{(1)r}\, R^T(\alpha_{(e)})
 \Big]_{(a)(b)} + O(\zeta^2),\nonumber \\
 &&{}\nonumber \\
 {}^4{\buildrel o\over \omega}_r{}^{(o)}{}_{(a)} &=& \sum_b\,
 R_{(a)(b)}(\alpha_{(e)})\, {}^3K_{(1)rs} + O(\zeta^2),\nonumber \\
 &&{}\nonumber \\
 {}^4{\buildrel o\over \omega}_{\tau}{}^{(a )}{}_{(b )} &=&
 - \sum_c\, \epsilon_{(a)(b)(c)}\, \partial_{\tau}\, \alpha_{(c)} +
 \sum_{cd}\, R_{(a)(d)}(\alpha_{(e)})\, R_{(b)(c)}(\alpha_{(e)})\,
 (\partial_c\, {\bar n}_{(1)(d)} - \partial_d\, {\bar n}_{(1)(c)})
 +\nonumber \\
 &+& \sum_{cd}\, \Big(R_{(b)(c)}(\alpha_{(e)})\, \partial_c\,
 R_{(a)(d)} - R_{(a)(c)}\, \partial_c\, R_{(b)(d)}\Big)\,
 {\bar n}_{(1)(d)} + O(\zeta^2),\nonumber \\
 &&{}\nonumber \\
 {}^4{\buildrel o\over \omega}_{\tau}{}^{(o)}{}_{(a)} &=&
 \sum_b\, R_{(a)(b)}(\alpha_{(e)})\, \partial_b\, n_{(1)} +
 O(\zeta^2).
 \label{2.21}
 \eea

\noindent From Eqs. (\ref{3.5}) we have ${}^3K_{(1)rs} =
\sigma_{(1)(r)(s)}{|}_{r \not= s} + \delta_{rs}\, \Big({1\over 3}\,
{}^3K - \partial_{\tau}\, \Gamma_r^{(1)} + \partial_r\, {\bar
n}_{(1)(r)} - \sum_a\, \partial_a\, {\bar n}_{(1)(a)}\Big)$ with
$\sigma_{(1)(r)(s)}{|}_{r \not= s}$ given in Eq. (\ref{2.5}) after
the solution of the constraints.
\bigskip

For the relation between the 4-field strength and the 4-curvature
tensors and between their 3-dimensional analogues we have

\bea
 {}^4\Omega_{AB}{}^{(\alpha )}{}_{(\beta )}&=&
 {}^4E^{(\gamma )}_A\, {}^4E^{(\delta )}_B\, {}^4\Omega ^{(\alpha
 )}{}_{(\beta )(\gamma )(\delta )}={}^4R^{C}{}_{DAB}\,
 {}^4E^{(\alpha )}_C\, {}^4E^D_{(\beta )}=\nonumber \\
 &=&\partial_A {}^4\omega_B{}^{(\alpha )}{}_{ (\beta )}-\partial_B\,
 {}^4\omega_A{}^{(\alpha )}{} _{ (\beta )} + {}^4\omega_A{}^{(\alpha
 )}{}_{ (\gamma )}\, {}^4\omega_B{}^{(\gamma )}{} _{ (\beta )} -
 {}^4\omega_B{}^{(\alpha )}{}_{ (\gamma )}\, {}^4\omega_A
 {}^{(\gamma )}{}_{ (\beta )},\nonumber \\
 &&{}\nonumber \\
 {}^4R^A{}_{BCD}&=&{}^4E^A_{(\gamma )}\, {}^4E^{(\delta )} _B\,
 {}^4\Omega_{CD}{}^{(\gamma )}{}_{(\delta )},\nonumber \\
 &&{}\nonumber \\
 {}^3\Omega_{rs(a)}&=&{1\over 2}\epsilon_{(a)(b)(c)}\,
 {}^3\Omega_{rs(b)(c)} = {1\over 2}\, \epsilon_{(a)(b)(c)}\,
 {}^3e_{(b)t}\, {}^3e^w_{(c)}\, {}^3R^t{}_{wrs} =\nonumber \\
 &=& \partial_r\, {}^3\omega_{s(a)}-\partial_s\, {}^3\omega_{r(a)} -\epsilon
 _{(a)(b)(c)}\, {}^3\omega_{r(b)}\, {}^3\omega_{s(c)},\nonumber \\
 &&{}\nonumber \\
 {}^3R^r{}_{stw}&=& \epsilon_{(a)(b)(c)}\, {}^3e^r_{(a)}\,
 \delta_{(b)(n)}\, {}^3e^{(n)}_s\, {}^3\Omega_{tw(c)},\qquad
 {}^3R_{rsuv} = \epsilon_{(a)(b)(c)}\, {}^3e_{(a)r}\,
 {}^3e_{(b)s}\, {}^3\Omega _{uv(c)}.\nonumber \\
 &&{}
 \label{2.22}
 \eea

\bigskip

The first Bianchi identity
${}^3R^t{}_{rsu}+{}^3R^t{}_{sur}+{}^3R^t{} _{urs}\equiv 0$ implies
the cyclic identity ${}^3\Omega_{rs(a)}\, {}^3e^s_{(a)} \equiv 0$.

\subsection{The Ashtekar Variables in the York Canonical Basis and their PM Limit}

As shown in Ref. \cite{8,10}, the canonical basis ${}^3e_{(a)r}$,
$\pi^r_{(a)}$, formed by the cotriads on $\Sigma_{\tau}$ and by
their conjugate momenta (see Eq.(\ref{2.8}) of paper I) can be
replaced by the following canonical basis of Ashtekar's variables
($\gamma$ is the Immirzi parameter; we use the conventions of
Ref.\cite{11}; $Q_a = e^{\sum_{\bar a}\, \gamma_{\bar aa}\, R_{\bar
a}}$)

\bea
 {}^3{\cal E}^r_{(a)} &=&  {}^3e\, {}^3e^r_{(a)} =
 {\tilde \phi}^{2/3}\, \sum_b\, R_{(a)(b)}(\alpha_{(e)})\,
 V_{ra}(\theta^i)\, e^{- \sum_{\bar a}\, \gamma_{\bar aa}\,
 R_{\bar a}},\nonumber \\
 &&{}\nonumber \\
 {}^3A_{(\gamma)(a)r} &=& {}^3\omega_{r(a)} + \gamma\,
 {}^3e^s_{(a)}\, {}^3K_{rs},
 \label{2.23}
 \eea

\noindent with ${}^3\omega_{r(a)}$ of Eqs.(\ref{2.20}) and with
${}^3K_{rs}$ of Eq.(2.3) of paper II. This formalism is usually
defined in the Schwinger time gauges $\varphi_{(a)}(\tau, \vec
\sigma) \approx 0$ of ADM tetrad gravity.\medskip

In Ref.\cite{11} it is shown that we have $\{ A_{(\gamma)(a)r}(\tau,
\vec \sigma ), A_{(\gamma)(b)s}(\tau ,{\vec \sigma}_1)\} = 0$ due to
the results $\{ {}^3K_{(a)r}(\tau ,\vec \sigma ),
{}^3\omega_{s(b)}(\tau ,{\vec \sigma}_1)\} = \{ {}^3K_{(b)s}(\tau
,\vec \sigma ), {}^3\omega_{r(a)}(\tau ,{\vec \sigma}_1)\} = 0$
(${}^3K_{(a)r} = {}^3K_{rs}\, {}^3e^s_{(a)}$), which are a
consequence of the fact that ${}^3e^r_{(a)}\, \delta\,
{}^3\omega_{r(a)}$ is a pure divergence (this implies
${}^3\omega_{r(a)}(\tau ,\vec \sigma ) = [{}^3e_{(b)r}\,
{}^3\omega_{(b)(a)}](\tau ,\vec \sigma ) =
  {{\delta}\over {\delta\, {}^3e^r_{(a)}(\tau ,\vec \sigma )}}\,
\int d^3\sigma_1\, \sum_b\, {}^3\omega_{(b)(b)}(\tau ,{\vec
\sigma}_1)$). We also have $\{ {}^3A_{(\gamma)(a)r}(\tau, \vec
\sigma), {}^3{\cal E}^s_{(b)}(\tau, {\vec \sigma}_1) \} = \gamma\,
\delta^s_r\, \delta_{(a)(b)}\, \delta^3(\vec \sigma - {\vec
\sigma}_1)$.
\bigskip

The SO(3) connection $A_{(\gamma)(a)r}$ is considered as a SU(2)
connection with field strength ${}^3F_{(\gamma)(a)rs} =
\partial_r\, {}^3A_{(\gamma)(a)s} - \partial_s\, {}^3A_{(\gamma)(a)r}
+ \epsilon_{(a)(b)(c)}\, {}^3A_{(\gamma)(b)r}\,
{}^3A_{(\gamma)(c)s}$. Instead the true SO(3) connection, associated
with the O(3) subgroup of the Lorentz group O(3,1), is ${1\over 2}\,
\epsilon_{(a)(b)(c)}\, \Big[R(\alpha_{(e)})\, \partial_r\,
R^T(\alpha_{(e)})\Big]_{(b)(c)} + R_{(b)(m)}(\alpha_{(e)})\,
{}^3{\bar \omega}_{r(m)(n)}\, R^T_{(m)(c)}(\alpha_{(e)})\Big)$.
\medskip

Instead the densitized triad ${}^3{\cal E}^r_{(a)}$ is considered an
analogue of an electric field.

\medskip

In the Ashstekar formalism the non-abelianized rotation constraint
${}^3M_{(a)}(\tau, \vec \sigma) \approx 0$ of Ref.\cite{3}, whose
Abelianized form in the York canonical basis is
$\pi^{(\alpha)}_{(a)} = - \sum_b\, {}^3M_{(b)}\,
A_{(b)(a)}(\alpha_{(e)}) \approx 0$, is replaced by the Gauss law
constraint $G_{(a)} = \sum_r\, \partial_r\, {}^3{\cal E}^r_{(a)} +
\epsilon_{(a)(b)(c)}\, {}^3A_{(\gamma)(b)r}\, {}^3{\cal E}^r_{(c)}
\approx 0$. The super-Hamiltonian constraint ${\cal H}(\tau, \vec
\sigma) \approx 0$ takes the form  $({}^3e)^{-2}\, \sum_{ars}\,
\Big[{}^3F_{(\gamma)(a)rs} - (1 + \gamma^2)\, \sum_{bc}\,
\epsilon_{(a)(b)(c)}\, {}^3K_{r(b)}\, {}^3K_{s(c)}\Big]\,
\sum_{mn}\, \epsilon_{(a)(m)(n)}\, {}^3{\cal E}^r_{(m)}\, {}^3{\cal
E}^s_{(n)} + \gamma^{-1}\, (1 + \gamma^2)\, \sum_{ar}\, G_{(a)}\,
\partial_r\, \Big[({}^3e)^{-2}\, {}^3{\cal E}^r_{(a)}\Big] \approx 0$,
while the super-momentum constraints become $\gamma^{-1}\, \sum_a\,
\Big[\sum_{s}\, {}^3F_{(\gamma)(a)rs}\, {}^3{\cal E}^s_{(a)} - (1 +
\gamma^2)\, {}^3K_{r(a)}\, G_{(a)} \Big] \approx 0$.

\bigskip

The linearized Ashtekar variables in gauges near the 3-orthogonal
ones  (with $\alpha_{(a)}(\tau, \vec \sigma) \not= 0$,
$\varphi_{(a)}(\tau, \vec \sigma) \not= 0$, $\theta^i(\tau, \vec
\sigma) = \theta^i_{(1)}(\tau, \vec \sigma) = O(\zeta) \not= 0$) are

\bea
 {}^3{\cal E}_{(a)}^r &=&\sum_b\, R_{(a)(b)}(\alpha_{(e)})\,
 \Big[(1 - \Gamma_r^{(1)} + 4\, \phi_{(1)})\, \delta_{rb} -
 \sum_i\, \epsilon_{rbi}\, \theta^i_{(1)}\Big] + O(\zeta^2),
 \nonumber \\
 {}^3A_{(\gamma)(a)r} &=& {1\over 2}\, \sum_{bc}\,
 \epsilon_{(a)(b)(c)}\, \Big(\Big[R(\alpha_{(e)})\,
 \partial_r\, R^T(\alpha_{(e)})\Big]_{(b)(c)} +\nonumber \\
 &+& \sum_{mn}\, R_{(b)(m)}(\alpha_{(e)})\, R_{(c)(n)}(\alpha_{(e)})\,
 \Big[(\delta_{rm}\, \partial_n - \delta_{rn}\, \partial_m)\,
 (\Gamma_r^{(1)} + 2\, \phi_{(1)}) - \sum_i\, \epsilon_{mni}\,
 \partial_r\, \theta^i_{(1)}\Big] +\nonumber \\
 &+& \gamma\, \sum_b\, R_{(a)(b)}(\alpha_{(e)})\, \Big[
 \sigma_{(1)(r)(b)}{|}_{r \not= b} + \delta_{rb}\, \Big({1\over 3}\,
 {}^3K - \partial_{\tau}\, \Gamma_r^{(1)} + \partial_r\, {\bar
 n}_{(1)(r)} - \sum_a\, \partial_a\, {\bar n}_{(1)(a)}\Big)
 \Big]\Big).\nonumber \\
 &&{}
 \label{2.24}
 \eea

\noindent After having solved the super-Hamiltonian and
super-momentum constraints with no fixation of the gauge one has to
replace $\phi_{(1)}$ and $\sigma_{(1)(r)(s)}{|}_{r \not= s}$ with
their expressions given in Eqs.(\ref{2.2}) and (\ref{2.5}). The
expression in the 3-orthogonal Schwinger time-gauges is obtained by
putting $\alpha_{(a)}(\tau, \vec \sigma) = 0$, $\varphi_{(a)}(\tau,
\vec \sigma) = 0$, $\theta^i(\tau, \vec \sigma) =
\theta^i_{(1)}(\tau, \vec \sigma) = 0$.

\medskip

With these results we can find the PM expression of\hfill\break
 1) the holonomy  along a closed loop $\Gamma$, i.e.  $P\,
e^{\int_{\Gamma}\, A} = \sum_{n=0}^{\infty}\, {}_{1
> s_n > .. > s_1 > 0}\,\, \int\, ... \int\, A(\Gamma(s_1)) .. A(\Gamma(s_n))\,
ds_1\, .. ds_n$, where $A[\Gamma] = \int_{\Gamma}\, A = \int_o^1\,
ds\, A_{(\gamma) (c) a}(x(s))\, {{d x^a(s)}\over {ds}}\, \tau_{(c)}$
($\tau_{(c)}$ are Pauli matrices);\hfill\break
 2) the flux of the electric field across a surface S, i.e. $\int_S\, d^2\sigma\, n_r\,
{}^3e^r_{(a)} = E_{(a)}(S)$.

\vfill\eject

\section{The PM Space-Time and its Instantaneous 3-Spaces}

In this Section we illustrate the general properties of  PM Einstein
space-times and how their properties depend on the choice of the
York time (selecting a member of our family of 3-orthogonal gauges)
through the non-local function ${}^3{\cal K}_{(1)} = {1\over
{\triangle}}\, {}^3K_{(1)}$. After describing the proper time of a
time-like observer and the properties of the non-Euclidean 3-spaces,
we compare the results in 3-orthogonal gauges with the IAU
conventions in harmonic gauges for the Solar System. Then we study
the PM time-like geodesics.

\subsection{The PM Proper Time of a Time-like Observer}

Given a time-like observer located in $(\tau, \sigma^r)$ (not too
near to the particles), the evaluation of the observer proper time
is done with the line element $\sgn\, ds^2{|}_{(\tau, \sigma^r)} =
\sgn\, {}^4g_{\tau\tau}(\tau, \sigma^r)\, d\tau^2 = d\, {\cal
T}^2_{(\tau, \sigma^r)}$. Therefore from Eqs.(\ref{2.14}) we get

\bea
 d\, {\cal T}_{(\tau, \sigma^r)} &=& \sqrt{\sgn\,
 {}^4g_{\tau\tau}(\tau, \sigma^r)}\, d\tau = \sqrt{1 + 2\,
 n_{(1)}(\tau, \sigma^r)}\, d\tau =\nonumber \\
 &=& \Big[1 - {G\over {c^3}}\, \sum_i\, {{\eta_i\, \sqrt{m_i^2c^2 +
 {\vec \kappa}_i^2(\tau)}}\over {|\vec \sigma - {\vec
 \eta}_i(\tau)|}}\, (1 + {{{\vec \kappa}_i^2(\tau)}\over {m_i^2c^2 +
 {\vec \kappa}_i^2(\tau)}}) -\nonumber \\
 &-& \partial_{\tau}\, {}^3{\cal K}_{(1)}(\tau, \vec \sigma)\Big]\, d\tau.
 \nonumber \\
 &&{}
 \label{3.1}
 \eea

As a consequence, the proper time depends on the $\tau$-derivative
of the non-local York time ${}^3{\cal K}_{(1)}$ at the position of
the observer in the 3-space $\Sigma_{\tau}$.

\subsection{The Instantaneous PM 3-Spaces $\Sigma_{\tau}$}

\subsubsection{The Spatial 3-Distance on the Instantaneous 3-Space
$\Sigma_{\tau}$}

Let us consider two points on the instantaneous 3-space
$\Sigma_{\tau}$ (whose intrinsic 3-curvature will be given in
Eq.(\ref{3.6})) with radar 3-coordinates $\sigma^r_o$ and
$\sigma^r_1$. They will be joined by a unique 3-geodesic
$\xi^r(\tau, s) = \sigma^r_0 + (\sigma^r_1 - \sigma^r_o)\, s +
\xi^r_{(1)}(\tau, s)$, $\xi^r(0) = \sigma^r_o$, $\xi^r(1) =
\sigma^r_1$, $\xi^r_{(1)}(0) = \xi^r_{(1)}(1) = 0$, solution of the
geodesic equation ${{d^2\, \xi^r(\tau, s)}\over {ds^2}} = -
\sum_{uv}\, {}^3\Gamma^r_{(1)uv}(\tau, \vec \xi(\tau, s))\, {{d
\xi^u(\tau, s)}\over {ds}}\, {{d \xi^v(\tau, s)}\over {ds}}$ with
the 3-Christoffel symbol given in Eq.(\ref{2.15}).\bigskip

At order $O(\zeta)$ we get the following solution for the 3-geodesic

\bea
 \xi^r(\tau, s) &=& \sigma^r_o + (\sigma^r_1 - \sigma^r_o)\, s +
 \nonumber \\
 &+& \sum_{uv}\, (\sigma^u_1 - \sigma^u_o)\, (\sigma^v_1 - \sigma^v_o)\,
 \Big(\int_o^1 - \int^s_o\Big)\, ds_1\, \int_o^{s_1} ds_2\,
 {}^3\Gamma^r_{(1)uv}(\tau, {\vec \sigma}_o + ({\vec \sigma}_1 -
 {\vec \sigma}_o)\, s_2).\nonumber \\
 &&{}
 \label{3.2}
 \eea

 \bigskip

Since Eqs.(\ref{2.14}) implies that at the first order the line
3-element joining the two points is

\bea
 d{\cal S}(\tau) &=& \sqrt{- \sgn\, \sum_{rs}\, {}^4g_{(1)rs}(\tau,
 \vec \xi(\tau, s))\, {{d
 \xi^r(\tau, s)}\over {ds}}\, {{d \xi^s(\tau, s)}\over {ds}}}\, ds =\nonumber \\
 &=& \sqrt{\Big({{d \vec \xi(\tau, s)}\over {ds}}\Big)^2 + 2\, \sum_r\,
 (\sigma_1^r - \sigma^r_o)^2\, (2\, \phi_{(1)} + \Gamma_r^{(1)})(\tau,
 \vec \xi(\tau, s))}\, ds,
 \label{3.3}
 \eea

 \noindent the geodesic 3-distance between the two points is ($d_{Euclidean}({\vec \sigma}_0,
{\vec \sigma}_1)  = |{\vec \sigma}_1 - {\vec \sigma}_o| =
\sqrt{\sum_r\, (\sigma^r_1 - \sigma^r_o)^2}$ is the Euclidean
distance with respect to the flat asymptotic 3-metric)

\bea
 d({\vec \sigma}_o, {\vec \sigma}_1)(\tau) &=& \int_o^1 \, d{\cal S}(\tau) =
 d_{Euclidean}({\vec \sigma}_0, {\vec \sigma}_1) +   \nonumber \\
  &+&\sum_r\, {{\sigma_1^r - \sigma_o^r}\over {|{\vec \sigma}_1 -
  {\vec \sigma}_o|}}\, \int_o^1 ds\, \Big( (\sigma_1^r - \sigma_o^r)\,
  (2\, \phi_{(1)} + \Gamma_r^{(1)})\, (\tau, {\vec \sigma}_o +
  ({\vec \sigma}_1 - {\vec \sigma}_o)\, s) -\nonumber \\
  &-&\sum_s\, (\sigma^s_1 - \sigma^s_o)\, \int_0^s ds_1\,
 \Big[2\, (\sigma^r_1 - \sigma^r_o)\, \partial_s\, (2\, \phi_{(1)}
 + \Gamma_r^{(1)}) -\nonumber \\
 &-& (\sigma^s_1 - \sigma^s_o)\, \partial_r\,
 (2\, \phi_{(1)} + \Gamma_s^{(1)})\Big]
 (\tau, {\vec \sigma}_o +  ({\vec \sigma}_1 - {\vec \sigma}_o)\, s_1)
 \Big).
 \label{3.4}
 \eea

As expected it does not depend upon the inertial gauge variable
${}^3{\cal K}_{(1)}$ \footnote{Instead a space-like 4-geodesic
depends on it. Indeed the extrinsic curvature tensor ${}^3K_{rs}$ is
a measure, at a point in the 3-space $\Sigma_{\tau}$, of the
curvature of a space-time geodesic tangent  to the 3-geodesic
(\ref{3.2}) at that point, see Refs.\cite{12}}.\medskip

Let us remark that in general a 3-geodesic of the 3-metric
${}^3g_{(1)rs} = - \sgn\, {}^4g_{(1)rs}$ on the 3-space
$\Sigma_{\tau}$ is not a space-like geodesics of the 4-metric
${}^4g_{(1)AB}$.

\subsubsection{The Extrinsic 3-Curvature Tensor}

From Eqs.(\ref{1.8}) and by using $\sum_{\bar a}\, \gamma_{\bar
aa}\, \gamma_{\bar ab} = \delta_{ab} - {1\over 3}$, we get that the
extrinsic curvature tensor of our 3-spaces in our family of
3-orthogonal gauges is the following first order quantity

\bea
 &&{}^3K_{(1)rs}(\tau, \vec \sigma)\, =\, \sigma_{(1)(r)(s)}{|}_{r \not= s}(\tau, \vec \sigma) +
 \delta_{rs}\, \Big({1\over 3}\, {}^3K_{(1)} - \partial_{\tau}\, \Gamma_r^{(1)}
 + \partial_r\, {\bar n}_{(1)(r)} - \sum_a\, \partial_a\, {\bar n}_{(1)(a)}\Big)(\tau, \vec \sigma),
 \nonumber \\
 &&{}
 \label{3.5}
 \eea

\noindent with ${\bar n}_{(1)(r)}$ and $\sigma_{(1)(r)(s)}{|}_{r
\not= s}$ given in Eqs.(\ref{2.4}) and (\ref{2.5}), respectively,
and with $\Gamma_r^{(1)}$ given by Eq.(\ref{2.6}). Therefore, our
(dynamically determined) 3-spaces have a first order deviation from
Euclidean 3-spaces, embedded in the asymptotically flat space-time,
determined by both instantaneous inertial matter effects and
retarded tidal ones. Moreover the inertial gauge variable
${}^3K_{(1)}$ (non existing in Newtonian gravity) is the free
numerical function labeling the members of the family of
3-orthogonal gauges. The extrinsic curvature tensor depends on  the
local (${}^3K_{(1)}$) and also on the non-local one (${}^3{\cal
K}_{(1)}$) through the shift function.

\subsubsection{The Intrinsic 3-Curvature Tensor}

The 3-Riemann tensor is given in Eq.(\ref{2.16}). The 3-Ricci tensor
and the 3-curvature scalar are

\bea
 {}^3R_{(1)rs} &=&\sum_u\, {}^3R_{(1)urus}
 = - \delta_{rs}\, \triangle\, (\Gamma^{(1)}_r + 2\, \phi_{(1)})
 + \partial_r\, \partial_s\, (\Gamma^{(1)}_r + \Gamma^{(1)}_s - 2\, \phi_{(1)})
 =\nonumber \\
 &=&{{4\pi\, G}\over {c^3}}\, (\delta_{rs} + {{\partial_r\, \partial_s}
 \over {\triangle}})\, {\cal M}_{(1)}^{(UV)} -\nonumber \\
 &-&\delta_{rs}\, \triangle\, (\Gamma_r^{(1)} + {1\over 2}\, \sum_c\,
 {{\partial_c^2}\over {\triangle}}\, \Gamma_c^{(1)}) + \partial_r\,
 \partial_s\, (\Gamma_r^{(1)} + \Gamma_s^{(1)} - {1\over 2}\, \sum_c\,
 {{\partial_c^2}\over {\triangle}}\, \Gamma_c^{(1)}) =\nonumber \\
 &=&{{4\pi\, G}\over {c^3}}\, (\delta_{rs} + {{\partial_r\, \partial_s}
 \over {\triangle}})\, {\cal M}_{(1)}^{(UV)} + {1\over 2}\, \triangle\,
 {}^4h_{(1)rs}^{TT},\nonumber \\
 &&{}\nonumber \\
 {}^3R_{(1)} &=& \sum_r\, {}^3R_{(1)rr} = - 8\, \triangle\, \phi_{(1)} + 2\, \sum_a\,
 \partial_a^2\, \Gamma^{(1)}_a = {{16\pi\, G}\over {c^3}}\, {\cal M}_{(1)}^{(UV)}.
 \label{3.6}
 \eea

We see that, apart from distributional contributions from the
particles,  the intrinsic 3-curvature ${}^3R_{(1)}$ of these
non-Euclidean 3-spaces is determined only by the tidal variables,
i.e. by the PM GW's propagating inside these 3-spaces.

\subsection{Comparison with the Barycentric Celestial Reference
System (BCRS) of IAU2000 in the Harmonic Gauge used for the Solar
System.}

In Refs.\cite{13} there is the 4-metric chosen in the astronomical
conventions IAU2000 to describe the Solar System in the Barycentric
Celestial Reference System (BCRS) centered in its barycenter by
using a PN approximation of Einstein's equations in a special system
of harmonic 4-coordinates $x^{\mu}_B$. The barycenter world-line (a
time-like geodesic of the PN 4-metric ${}^4g_{B\mu\nu}(x_B)$) is the
time axis $x^{\mu}_{B(B)}(\tau_B) = \Big(x^o_B(\tau_B); 0^i\Big)$,
where $\tau_B$ is the proper time of a standard clock in the solar
system barycenter, ($(d\tau_B)^2 = \sgn\, g_{Boo}(x_{B(B)})\,
(dx^o_B)^2$). It is approximately a straight line if we neglect
galactic and extra-galactic influences. Through each point of this
world-line we consider the hyper-surfaces $x^o_B = c\, t_B = const.$
as instantaneous 3-spaces $\Sigma_{x^o_B}$  with {\it rectangular}
3-coordinates (practically they are the quasi-Euclidean 3-spaces of
a quasi-inertial frame of Minkowski space-time, even if they do not
correspond to Einstein's 1/2 clock synchronization convention). In
each point of the barycenter world-line there is a {\it tetrad} with
the time-like 4-vector given by the barycenter 4-velocity and with
the 3 mutually orthogonal {\it kinematically non-rotating} spatial
axes (no systematic rotation with respect to certain fixed stars
(radio sources) in the instantaneous 3-spaces $t_B = const.$). This
is a {\it global} reference system, with the following PN solution
of Einstein's equations for the 4-metric ${}^4g_{B\mu\nu}(x_B)$ (the
potentials $w_B$ and $w_{BI}$ are static and of order $G$, so that
$w_B^2 = O(G^2)$)

\bea
 {}^4g_{Boo}(x_B) &=&\sgn\, \Big[N_B^2 - {}^3g^{ij}_B\, N_{Bi}\, N_{Bj}\Big](x_B) =
 = \sgn\, \Big[1 - {{2\, w_B}\over {c^2}} -
 {{2\, w^2_B}\over {c^4}} + O(c^{-5})\Big](x_B),\nonumber \\
 {}^4g_{Boi}(x_B) &=& - \sgn\, N_{Bi}(x_B) = - \sgn\, \Big[{{4\, w_{Bi}}\over {c^3}} +
 O(c^{-5})\Big](x_B),\nonumber \\
 {}^4g_{Bij}(x_B) &=& - \sgn\, {}^3g_{Bij} = - \sgn\,
 \Big[(1 + {{2\, w_B}\over {c^2}})\, \delta_{ij}  + O(c^{-4})\Big](x_B).
 \label{3.7}
 \eea
\bigskip

Eqs.(\ref{3.7}) imply an extrinsic curvature tensor ${}^3K_{Bij} =
{1\over {2N_B}}\, (N_{Bi|j} + N_{Bj|i} - \partial_o\, {}^3g_{Bij})$
of order $O(c^{-2})$, but the 3-sub-manifolds $x^o_B = const.$ of
space-time (the harmonic 3-spaces) are not specified: one has to
solve the inverse problem of finding the 3-sub-manifolds with the
given extrinsic curvature tensor.

\bigskip

By comparison let us consider the N particles in non-harmonic
3-orthogonal gauges as the Sun and the planets of the Solar System.
Let us neglect gravitational waves  (so that the 3-spaces have
negligible intrinsic 3-curvature except for a distributional
singularity at the particle locations, [see Eqs.(\ref{3.6})], where
our approximation breaks down). Then by using Eqs.(\ref{2.2}),
(\ref{2.3}), (\ref{2.4}), the non-relativistic limit of the 4-metric
(\ref{2.14}) in radar 4-coordinates (see the embedding in the
Introduction to get world 4-coordinates like the ones of BCRS) has
the following form

\begin{eqnarray*}
 {}^4g_{(1)\tau\tau}(\tau, \vec \sigma) &=& \sgn\, \Big[1 - {{2\, w}\over {c^2}} -
 {{2\, \tilde w}\over {c^4}} - 2\, \partial_{\tau}\, {}^3{\cal
 K}_{(1)}
 + O(c^{-5})\Big](\tau, \vec \sigma),\nonumber \\
 {}^4g_{(1)\tau r}(\tau, \vec \sigma) &=& - \sgn\, \Big({{4\, w_r}\over {c^3}} +
 \partial_r\, {}^3{\cal K}_{(1)} + O(c^{-5})\Big)(\tau, \vec \sigma), \nonumber \\
 {}^4g_{(1)rs}(\tau, \vec \sigma) &=& - \sgn\, \delta_{rs}\, \Big[1 + {{2\, w}\over
 {c^2}} + O(c^{-4})\Big](\tau, \vec \sigma),\end{eqnarray*}

 \bea
 &&w(\tau, \vec \sigma) = \sum_i\, w_i(\tau, \vec \sigma),\qquad
 w_i(\tau, \vec \sigma) = \eta_i\, {{G\,
 m_i}\over {|\vec \sigma - {\vec \eta}_i(\tau)|}},
 \qquad \tilde w(\tau, \vec \sigma) = \sum_i\,
 {{3\, {\vec \kappa}_i^2(\tau)}\over {2\, m_i^2\,
 c^2}}\, w_i(\tau, \vec \sigma),\nonumber \\
 &&w_r(\tau, \vec \sigma) = - {G\over 2}\, \sum_i\, {{\eta_i}\over
 {|\vec \sigma - {\vec \eta}_i(\tau)|}}\, \Big(\kappa_{ir}(\tau) +
 {{(\sigma^r - \eta^r_i(\tau))\, {\vec \kappa}_i(\tau) \cdot (\vec \sigma
 - {\vec \eta}_i(\tau))}\over {|\vec \sigma - {\vec
 \eta}_i(\tau)|^2}}\Big).\nonumber \\
 &&{}
 \label{3.8}
 \eea

\bigskip
Also in this 3-orthogonal gauge we can get quasi-static  potentials
(ignoring the motion of the sources and assuming that
$\partial_{\tau}\, {}^3{\cal K}_{(1)}$ and $\partial_r\, {}^3{\cal
K}_{(1)}$ are slowing varying functions of $\tau$) and the same
pattern as in Eq.(\ref{3.7}) till the order $1/c^3$ included. The
main difference is that $\tilde w \not= w^2 = O(G^2)$. Here $w$ is
the Newton potential and $w_r$ the gravito-magnetic one.
\medskip

If we choose the special 3-orthogonal gauge ${}^3{\cal
K}_{(1)}(\tau, \vec \sigma) = 0$ we recover agreement with the Solar
System conventions. Let us remark that the instantaneous 3-spaces
are not hyper-planes due to Eq.(\ref{3.5}), giving the non-vanishing
extrinsic curvature tensor ${}^3K_{(1)rs} = O(c^{-3})$ even if
${}^3K_{(1)} = 0$.\medskip

See Ref.\cite{14} for the status of knowledge on the possibility of
the presence of dark matter or of modifications of gravity in the
Solar System for explaining effects like the Pioneer anomaly (to be
mimicked by means of ${}^3{\cal K}_{(1)}(\tau, \vec \sigma)$ if
needed). Further restrictions on ${}^3{\cal K}_{(1)}$ near the Earth
will come from the gravito-magnetic Lense-Thirring (or
frame-dragging) effect (see Refs.\cite{15}, Ref.\cite{16} for Lageos
and Ref.\cite{17} for Gravity Probe B) when the experimental errors
will become acceptable.\medskip

Like in the case of the IAU 4-metric and by assuming that the
dependence on the inertial gauge variable ${}^3{\cal K}_{(1)}$ is
negligible inside the Solar System, by using the 4-metric
(\ref{3.8}) one could reproduce the standard general relativistic
effects like the perihelion precession and the deflection of light
rays by the Sun \footnote{For them a 4-metric approximating the
static spherically symmetric Schwartzschild solution is enough: see
for instance Ref.\cite{18}.} also in 3-orthogonal gauges. See
Ref.\cite{19} for the derivation of the Shapiro time delay and for
the gravitational redshift induced by the geo-potential (by using
its multipolar description).  With only one body (the Sun) in the
limit of spherical symmetry one can find the perihelion advance of
planets with the standard method of using the geodesic equation for
test particles (see Refs.\cite{15,18,20}).

\subsection{PM Time-like Geodesics}

Let now us consider a time-like geodesic $y^{\mu}(s) =
z^{\mu}(\sigma^A(s)) = x^{\mu}_o + \epsilon^{\mu}_A\, \sigma^A(s)$
(we use the natural adapted embedding of the Introduction) with
affine parameter $s$ and with radar 4-coordinates $\sigma^A(s)=\Big(
\tau (s); \sigma^u(s)\Big)$ to be used as the trajectory of a planet
or of a star. The tangent to the geodesic is $u^{\mu}(s) = {{d
y^{\mu}(s)}\over {ds}} = \epsilon^{\mu}_A\, p^A(s)$ with $p^A(s) =
{{d \sigma^A(s)}\over {ds}}$.\medskip

At the first order the parametrization of the geodesic (with
4-velocity $p^A(s)$) and the geodesic equation are

\bea
 \sigma^A(s) &=& \sigma^A_o(s) + \sigma^A_{(1)}(s) + O(\zeta^2),\qquad \sigma^A_o(s) =
 a^A + b^A\, s,\nonumber \\
  &&{}\nonumber \\
 p^A(s) &=& {{d\, \sigma^A(s)}\over {ds}} = b^A +
 {{\sigma^A_{(1)}(\sigma_o(s))}\over {ds}},\nonumber \\
 &&{}\nonumber \\
 {{d^2 \sigma^A(s)}\over {ds^2}} &=& -
 {}^4\Gamma^A_{(1)BC}(\sigma (s))\,
 {{d\sigma^B(s)}\over {ds}}\, {{d\sigma^C(s)}\over
 {ds}} = - {}^4\Gamma^A_{(1)BC}(\sigma_o(s))\, b^B\, b^C,
 \label{3.9}
 \eea

\noindent where $\sigma^{\alpha}_o(s) = a^{\alpha} + b^{\alpha}\, s$
is the flat Minkowski geodesic (with respect to the asymptotic flat
4-metric). The Christoffel symbols are given in Eq.(\ref{2.15}).
\medskip

The solution of the geodetic equation is

\beq
 \sigma^A(s) = a^A + b^A\, s - b^B\, b^C\, \int_0^s ds_1\,
 \int_0^{s_1} ds_2\, {}^4\Gamma^A_{(1)BC}(a + b\, s_2).
 \label{3.10}
 \eeq
\medskip

As Cauchy data at $s = 0$ we take the position $y^{\mu}(0) =
x^{\mu}_o + \epsilon^{\mu}_A\, a^A$ with $a^A = \sigma^A(0) =
\sigma^A_o$ and the tangent $u^{\mu}(0) = \epsilon^{\mu}_A\,
p^A(0)$.\medskip

For a time-like geodesics the tangent in the origin satisfies
$\sgn\, u^2(0) = 1$, i.e. $\sgn\, {}^4g_{(1)AB}(\sigma(0))\,
p^A(0)\, p^B(0) = 1$, if the parameter $s$ is the proper time. If
$u^i(0) = {\cal U}^i$, then we have $u^{\mu}(0) = (\sqrt{1 + {\vec
{\cal U}}^2}; {\cal U}^i)$, ${\vec {\cal U}}^2 = \sum_r\, ({\cal
U}^r)^2$. Therefore, with $b^r = {\cal U}^r$ and with the 4-metric
of Eq.(\ref{2.14}), for future-oriented geodesics the condition
$\sgn\, u^2(0) = 1$ leads to the following result for $b^A$

\bea
 b^{\tau} &=& \sqrt{1 + {\vec {\cal U}}^2} + d_{(1)}(\sigma_o),\nonumber \\
 &&{}\nonumber \\
 d_{(1)}(\sigma_o) &=& - \sqrt{1 + {\vec {\cal U}}^2}\,
 \Big[2\, n_{(1)}(\sigma_o) - {1\over 2}\,
 \sum_r\, {\cal U}^r\, {\bar n}_{(1)(r)}(\sigma_o) +\nonumber \\
 &+& \sum_r\, ({\cal U}^r)^2\,
 (\Gamma_r^{(1)} + 2\, \phi_{(1)})(\sigma_o)\Big].\nonumber \\
 &&{}\nonumber \\
 \Rightarrow&& b^A = b^A_{(o)} + \delta^{A\tau}\, d_{(1)}(\sigma_o),\qquad
 b^A_{(o)} = (\sqrt{1 + {\vec {\cal U}}^2}; {\cal U}^r).
 \label{3.11}
 \eea

 \medskip

Therefore, with these Cauchy data and by using Eqs.(\ref{2.15}), the
geodesic and its tangent  take the form

  \begin{eqnarray*}
 \tau (s) &=&\sigma^{\tau}(s) = \tau_o + \Big(\sqrt{1 +
 {\vec {\cal U}}^2} + d_{(1)}(\sigma_o)\Big)\, s -\nonumber \\
 &-& \int_0^s ds_1\, \int_0^{s_1} ds_2\,
 \Big((1 + {\vec {\cal U}}^2)\, \partial_{\tau}\,
 n_{(1)} + 2\, \sqrt{1 + {\vec {\cal U}}^2}\, \sum_u\, {\cal U}^u\,
 \partial_u\, n_{(1)} +\nonumber \\
 &+& \sum_{uv}\, {\cal U}^u\, {\cal U}^v\,
 \Big[ - {1\over 2}\, (\partial_u\,
 {\bar n}_{(1)(v)} + \partial_v\, {\bar n}_{(1)(u)}) + \delta_{uv}\,
 \partial_{\tau}\, (\Gamma_r^{(1)} + 2\, \phi_{(1)})
 \Big]\Big)(\sigma_o + {\cal U}\, s_2) =\nonumber \\
 &{\buildrel {def} \over =}& \tau_{({}^3K = 0)}(s) + \tau_{({}^3K)}(s),
 \nonumber \\
 &&{}\nonumber \\
 &&\tau_{({}^3K)}(s) = 2\, \sqrt{1 + {\vec {\cal U}}^2}\,
 \partial_{\tau}\, {}^3{\cal K}_{(1)}(\sigma_o) - {1\over 2}\,
 \sum_r\, {\cal U}^r\, \partial_r\, {}^3{\cal K}_{(1)}(\sigma_o)
 -\nonumber \\
 &-& \int_0^s ds_1\, \int_0^{s_1} ds_2\,
 \Big(- (1 + {\vec {\cal U}}^2)\,
 \partial_{\tau}^2\, {}^3{\cal K}_{(1)} -\nonumber \\
 &-& 2\, \sqrt{1 + {\vec {\cal U}}^2}\, \sum_u\, {\cal U}^u\,
 \partial_u\, \partial_{\tau}\, {}^3{\cal K}_{(1)}
 - \sum_{uv}\, {\cal U}^u\, {\cal U}^v\, \partial_u\,
 \partial_v\, {}^3{\cal K}_{(1)}
 \Big)(\sigma_o + {\cal U}\, s_2),
 \end{eqnarray*}

 \begin{eqnarray*}
 \sigma^r(s) &=& \sigma^r(0) + {\cal U}^r\, s - \int_0^s ds_1\,
 \int_0^{s_1} ds_2\, \Big((1 + {\vec {\cal U}}^2)\, (\partial_r\,
 n_{(1)} + \partial_{\tau}\, {\bar n}_{(1)(r)}) +\nonumber \\
 &+& 2\, \sqrt{1 + {\vec {\cal U}}^2}\, \sum_u\, {\cal U}^u\,
 \Big[ \delta_{ur}\, \partial_{\tau}\,
 (\Gamma_r^{(1)} + 2\, \phi_{(1)}) - {1\over 2}\, (\partial_r\,
 {\bar n}_{(1)(u)} - \partial_u\, {\bar n}_{(1)(r)}) \Big] +\nonumber \\
 &+& \sum_{uv}\, {\cal U}^u\, {\cal U}^v\, \Big[ 2\, \delta_{ru}\,
 \partial_v\, (\Gamma_r^{(1)} + 2\, \phi_{(1)}) - \delta_{uv}\,
 \partial_r\, (\Gamma_u^{(1)} + 2\, \phi_{(1)})\Big]\Big)(\sigma(0)
 + {\cal U}\, s_2) =\nonumber \\
 &{\buildrel {def}\over =}& \sigma^r_{({}^3K =0)}(s) +
 \sigma^r_{({}^3K)}(s) = \sigma_{({}^3K = 0)}(s),\nonumber \\
 &&{}\nonumber \\
 &&\sigma^r_{({}^3K)}(s) = 0,
 \end{eqnarray*}

 \bea
  p^A(s) &=& b^A_{(o)} + p^A_{(1)}(s),\nonumber \\
  &&{}\nonumber \\
  p^{\tau}(s) &=& \sqrt{1 +
 {\vec {\cal U}}^2} + d_{(1)}(\sigma_o) -\nonumber \\
 &-& \int_0^{s} ds_2\, \Big((1 + {\vec {\cal U}}^2)\, \partial_{\tau}\,
 n_{(1)} + 2\, \sqrt{1 + {\vec {\cal U}}^2}\, \sum_u\, {\cal U}^u\,
 \partial_u\, n_{(1)} +\nonumber \\
 &+& \sum_{uv}\, {\cal U}^u\, {\cal U}^v\,
 \Big[ - {1\over 2}\, (\partial_u\,
 {\bar n}_{(1)(v)} + \partial_v\, {\bar n}_{(1)(u)}) + \delta_{uv}\,
 \partial_{\tau}\, (\Gamma_r^{(1)} + 2\, \phi_{(1)})
 \Big]\Big)(\sigma_o + {\cal U}\, s_2),\nonumber \\
  p^r(s) &=& {\cal U}^r -
 \int_0^{s} ds_2\, \Big((1 + {\vec {\cal U}}^2)\, (\partial_r\,
 n_{(1)} + \partial_{\tau}\, {\bar n}_{(1)(r)}) +\nonumber \\
 &+& 2\, \sqrt{1 + {\vec {\cal U}}^2}\, \sum_u\, {\cal U}^u\,
 \Big[ \delta_{ur}\, \partial_{\tau}\,
 (\Gamma_r^{(1)} + 2\, \phi_{(1)}) - {1\over 2}\, (\partial_r\,
 {\bar n}_{(1)(u)} - \partial_u\, {\bar n}_{(1)(r)}) \Big] +\nonumber \\
 &+& \sum_{uv}\, {\cal U}^u\, {\cal U}^v\, \Big[ 2\, \delta_{ru}\,
 \partial_v\, (\Gamma_r^{(1)} + 2\, \phi_{(1)}) - \delta_{uv}\,
 \partial_r\, (\Gamma_u^{(1)} + 2\, \phi_{(1)})\Big]\Big)(\sigma(0)
 + {\cal U}\, s_2).\nonumber \\
 &&{}
  \label{3.12}
  \eea

\medskip

This is the {\it trajectory of a massive test particle}.\medskip

By using Eqs.(\ref{2.2}) - (\ref{2.4}), it turns out that all the
dependence of the geodesic upon the non-local York time is contained
in the function $\tau_{({}^3K)}(s)$, which contributes with ${{d\,
\tau_{({}^3K)}(s)}\over {ds}}$ to the component $p^{\tau}(s)$ of the
tangent.

\bigskip

Once the time-like geodesic $\sigma^A(s)$ starting at $\sigma^A_o =
\sigma^A(s = 0)$ and arriving at $\sigma^A_1 = \sigma^A(s = 1)$ is
known in terms the 4-metric of Eq.(\ref{2.14}) and denoted
$\gamma_{01}$, we can evaluate the HPM expression of the Synge world
function (see Refs. \cite{21,22,23}), i.e. of the two-point function
(for a space-like 4-geodesics it has the opposite sign)

\bea
 \Omega(\sigma_o, \sigma_1) &=& {1\over 2}\,
 \int_{(\Gamma_{01})\, 0}^1 ds\, \sgn\, {}^4g_{AB}(\sigma^D(s))\, {{
 d\, \sigma^A(s)}\over {ds}}\, {{d\, \sigma^B(s)}\over {ds}} =
 \nonumber \\
 &=&{1\over 2}\, \int_{(\Gamma_{01})\, 0}^1 ds\, \Big[ \sqrt{1 + {\vec
 {\cal U}}^2}\, (\sqrt{1 + {\vec {\cal U}}^2}\, + 2\,
 p^{\tau}_{(1)}(\sigma(s))) -\nonumber \\
 &-& \sum_r\, {\cal U}^r\, ({\cal U}^r + 2\,
 p_{(1)}^r(\sigma(s))) + (1 + {\vec {\cal U}}^2)\, (1 + 2\,
 n_{(1)}(\sigma(s))) - \nonumber \\
 &-& 2\, \sqrt{1 + {\vec {\cal U}}^2}\,
 \sum_r\, {\cal U}^r\, {\bar n}_{(1)(r)} - 2\, ({\cal U}^r)^2\,
 (\Gamma_r^{(1)} + 2\, \phi_{(1)})(\sigma(s))\Big].
 \label{3.13}
 \eea

\medskip

This is a 4-scalar in both points (the simplest case of bi-tensors
\cite{22}) defined in terms of the 4-geodesic distance between them.
Its gradients with respect to the end points give the vectors
tangent to the 4-geodesic at the end points.

\vfill\eject

\section{PM Null Geodesics, the Red-Shift, the Geodesic Deviation Equation and the
PM Luminosity Distance}

In this Section we study the null geodesics, the red-shift, the
geodesic deviation equation and the luminosity distance in PM
space-times.

\subsection{The PM Null Geodesics and the Red-Shift}

Let us now consider a null geodesic $y^{\mu}(s) =
z^{\mu}(\sigma^A(s)) = x^{\mu}_o + \epsilon^{\mu}_A\, \sigma^A(s)$
through the point $y^{\mu}_o = y^{\mu}(0) = x^{\mu}_o +
\epsilon^{\mu}_A\, \sigma^A(0)$  with $\sigma^A(0) = \sigma^A_o =
(\tau_o; {\vec \sigma}_o)$. It will have the form (\ref{4.2}) with
$a^A = \sigma^A_o$.\medskip

However now the tangent vector $u^{\mu}(s) = \epsilon^{\mu}_A\,
p^A(s)$, with $p^A(s) = {{d \sigma^A(s)}\over {ds}} = b^A - b^B\,
b^C\, \int_0^s ds_2\, {}^4\Gamma^A_{(1)BC}(\sigma_o + b\, s_2)$, is
a null vector, $\sgn\, {}^4g_{(1)AB}(\sigma(s))\, p^A(s)\, p^B(s) =
0$. Therefore we must require the initial condition $\sgn\,
{}^4g_{(1)AB}(\sigma_o)\, p^A(0)\, p^B(0) = \sgn\,
{}^4g_{(1)AB}(\sigma_o)\, b^A\, b^B + O(\zeta^2) = 0$ on $b^A =
(b^{\tau}; b^r)$.\medskip

By using Eq.(\ref{2.14})  we get that to each given value of $b^r$
there are two values of $b^{\tau}$ determined by the following
equation

\bea
 &&[1 + 2\, n_{(1)}(\sigma_o)]\, (b^{\tau})^2 - 2\, b^{\tau}\,
 \sum_r\, b^r\, {\bar n}_{(1)(r)}(\sigma_o) -
  [{\vec b}^2 + 2\, \sum_r\, (b^r)^2\,
 (\Gamma_r^{(1)} + 2\, \phi_{(1)})(\sigma_o)] = 0,\nonumber \\
 &&{}\nonumber \\
 &&\Downarrow\nonumber \\
 &&{}\nonumber \\
 b^{\tau} &=& \pm \sqrt{{\vec b}^2} + c_{(1)\pm}(\sigma_o),\nonumber \\
 &&{}\nonumber \\
 &&c_{(1)\pm}(\sigma_o) = \mp \sqrt{{\vec b}^2}\, [2\, n_{(1)}(\sigma_o) +
 \sum_r\, (b^r)^2\, (\Gamma_r^{(1)} + 2\, \phi_{(1)})(\sigma_o)] +
 {1\over 2}\, \sum_r\, b^r\, {\bar n}_{(1)(r)}(\sigma_o),\nonumber \\
 &&b^A = b^A_{(o)\pm} + \delta^{A\tau}\, c_{(1)\pm}(\sigma_o),\qquad
 b^A_{(o)\pm} = (\pm \sqrt{{\vec b}^2}; b^r).
 \label{4.1}
 \eea

\medskip

Therefore we get the following form of a future-oriented null
geodesic emanating from $\sigma_o^A$ with tangent $b^A = b^A_{(o)+}
+ \delta^{A\tau}\, c_{(1)+}(\sigma_o)$

\begin{eqnarray*}
 \sigma^A(s) &=& \sigma_o^A + (b^A_{(o)+} + \delta^{A\tau}\, c_{(1)+})\, s
 - b^B_{(o)+}\, b^C_{(o)+}\, \int_0^s ds_1\, \int_0^{s_1} ds_2\,
 {}^4\Gamma^A_{(1)BC}(\sigma_o + b_{(o)+}\, s_2),
 \end{eqnarray*}

\begin{eqnarray*}
 \tau(s) &=& \tau_o + (\sqrt{{\vec b}^2} + c_{(1)+}(\sigma_o))\, s
 -\nonumber \\
 &-&  \int_0^s ds_1\, \int_0^{s_1} ds_2\,
 \Big({\vec b}^2\, \partial_{\tau}\,
 n_{(1)} + 2\, \sqrt{{\vec b}^2}\, \sum_u\, {\cal U}^u\,
 \partial_u\, n_{(1)} +\nonumber \\
 &+& \sum_{uv}\, b^u\, b^v\,
 \Big[ - {1\over 2}\, (\partial_u\,
 {\bar n}_{(1)(v)} + \partial_v\, {\bar n}_{(1)(u)}) + \delta_{uv}\,
 \partial_{\tau}\, (\Gamma_r^{(1)} + 2\, \phi_{(1)})
 \Big]\Big)(\sigma(0) + b_{(o)+}\, s_2) =\nonumber \\
 &{\buildrel {def}\over =}& \tau_{({}^3K = 0)}(s) + \tau_{({}^3K)}(s),
 \nonumber \\
 &&{}\nonumber \\
 &&\tau_{({}^3K)}(s) = - \int_0^s ds_1\, \int_0^{s_1} ds_2\,
 \Big(- {\vec b}^2\,
 \partial_{\tau}^2\, {}^3{\cal K}_{(1)} -\nonumber \\
 &-& 2\, \sqrt{{\vec b}^2}\, \sum_u\, b^u\,
 \partial_u\, \partial_{\tau}\, {}^3{\cal K}_{(1)}
 - \sum_{uv}\, b^u\, b^v\, \partial_u\,
 \partial_v\, {}^3{\cal K}_{(1)}
 \Big)(\sigma_o + b_{(o)+}\, s_2),
 \end{eqnarray*}

\bea
 \sigma^r(s) &=& \sigma^r_o + b^r\, s -  \int_0^s ds_1\,
 \int_0^{s_1} ds_2\, \Big({\vec b}^2\, (\partial_r\,
 n_{(1)} + \partial_{\tau}\, {\bar n}_{(1)(r)}) +\nonumber \\
 &+& 2\, \sqrt{{\vec b}^2}\, \sum_u\, b^u\,
 \Big[ \delta_{ur}\, \partial_{\tau}\,
 (\Gamma_r^{(1)} + 2\, \phi_{(1)}) - {1\over 2}\, (\partial_r\,
 {\bar n}_{(1)(u)} - \partial_u\, {\bar n}_{(1)(r)}) \Big] +\nonumber \\
 &+& \sum_{uv}\, b^u\, b^v\, \Big[ 2\, \delta_{ru}\,
 \partial_v\, (\Gamma_r^{(1)} + 2\, \phi_{(1)}) - \delta_{uv}\,
 \partial_r\, (\Gamma_u^{(1)} + 2\, \phi_{(1)})\Big]\Big)(\sigma(0)
 + b_{(o)+}\, s_2) =\nonumber \\
 &{\buildrel {def}\over =}& \sigma^r_{({}^3K = 0)}(s) +
 \sigma^r_{({}^3K)}(s) = \sigma^r_{({}^3K = 0)}(s),\nonumber \\
 &&{}\nonumber \\
 &&\sigma^r_{({}^3K)}(s) = 0.
 \label{4.2}
 \eea

 This is the trajectory of a {\it ray of light}.
 \medskip

 The tangent to the null geodesic is

\begin{eqnarray*}
 p^A(s) &=& b^A_{(o)} + p^A_{(1)}(s) =\nonumber \\
 &=& b^A_{(o)+} +
 \delta^{A \tau}\, c_{(1)+}(\sigma_o) - b^B_{(o)+}\, b^C_{(o)+}\,
 \int_0^s ds_2\, {}^4\Gamma^A_{(1)BC}(\sigma_o + b_{(o)+}\, s_2),
 \nonumber \\
 p^A(0) &=&b^A_{(o)+} + \delta^{A \tau}\, c_{(1)+}(\sigma_o),
 \end{eqnarray*}

 \begin{eqnarray*}
 p^{\tau}(s) &=& \sqrt{{\vec b}^2} + c_{(1)+}(\sigma_o) - \int_0^s ds_2\,
 \Big({\vec b}^2\, \partial_{\tau}\,
 n_{(1)} + 2\, \sqrt{{\vec b}^2}\, \sum_u\, {\cal U}^u\,
 \partial_u\, n_{(1)} +\nonumber \\
 &+& \sum_{uv}\, b^u\, b^v\,
 \Big[ - {1\over 2}\, (\partial_u\,
 {\bar n}_{(1)(v)} + \partial_v\, {\bar n}_{(1)(u)}) + \delta_{uv}\,
 \partial_{\tau}\, (\Gamma_r^{(1)} + 2\, \phi_{(1)})
 \Big]\Big)(\sigma(0) + b_{(o)+}\, s_2) =\nonumber \\
 &{\buildrel {def}\over =}& \sqrt{{\vec b}^2} + c_{(1)+}(\sigma_o)
 + p^{\tau}_{(1)({}^3K = 0)}(s) + p^{\tau}_{(1)({}^3K)}(s),
 \nonumber \\
 &&{}\nonumber \\
 &&p^{\tau}_{(1)({}^3K)}(s) =
 \int_0^s ds_2\, \Big( {\vec b}^2\,
 \partial_{\tau}^2\, {}^3{\cal K}_{(1)} +\nonumber \\
 &+& 2\, \sqrt{{\vec b}^2}\, \sum_u\, b^u\,
 \partial_u\, \partial_{\tau}\, {}^3{\cal K}_{(1)}
 + \sum_{uv}\, b^u\, b^v\, \partial_u\,
 \partial_v\, {}^3{\cal K}_{(1)}
 \Big)(\sigma_o + b_{(o)+}\, s_2),
 \end{eqnarray*}

\bea
 p^r(s) &=& b^r - \int_0^s ds_2\, \Big({\vec b}^2\, (\partial_r\,
 n_{(1)} + \partial_{\tau}\, {\bar n}_{(1)(r)}) +\nonumber \\
 &+& 2\, \sqrt{{\vec b}^2}\, \sum_u\, b^u\,
 \Big[ \delta_{ur}\, \partial_{\tau}\,
 (\Gamma_r^{(1)} + 2\, \phi_{(1)}) - {1\over 2}\, (\partial_r\,
 {\bar n}_{(1)(u)} - \partial_u\, {\bar n}_{(1)(r)}) \Big] +\nonumber \\
 &+& \sum_{uv}\, b^u\, b^v\, \Big[ 2\, \delta_{ru}\,
 \partial_v\, (\Gamma_r^{(1)} + 2\, \phi_{(1)}) - \delta_{uv}\,
 \partial_r\, (\Gamma_u^{(1)} + 2\, \phi_{(1)})\Big]\Big)(\sigma(0)
 + b_{(o)+}\, s_2) =\nonumber \\
 &{\buildrel {def}\over =}& b^r + p^r_{(1)({}^3K = 0)}(s),
 \label{4.3}
 \eea

\noindent  with $\sgn\, {}^4g_{(1)AB}(\sigma(s))\, p^A(s)\, p^B(s) =
0 + O(\zeta^2)$.

 \bigskip

 The point $\sigma^A_1 = \sigma^A(s = 1)$ satisfies the equation

 \bea
  (\tau_1 - \tau_o)^2 &-& ({\vec \sigma}_1 - {\vec \sigma}_2)^2
  = 2\, \sqrt{{\vec b}^2}\, c_{(1)+}(\sigma_o)
  -\nonumber \\
  &-& 2\, b^B_{(o)+}\, b^C_{(o)+}\, \int_0^1 ds_1\, \int_0^{s_1}
  ds_2\, \Big[\sqrt{{\vec b}^2}\, {}^4\Gamma^{\tau}_{(1)BC} - \sum_r\,
  b^r\, {}^4\Gamma^r_{(1)BC}\Big](\sigma_o + b_{(o)+}\,
  s_2),\nonumber \\
  &&{}
  \label{4.4}
  \eea

\noindent which gives an idea of the first order deviation of the
null geodesic from the flat one joining the same two points
$\sigma^A_o$ and $\sigma^A_1$ on the Minkowski light-cone $\sgn\,
{}^4\eta_{AB}\, (\sigma^A_1 - \sigma^A_o)\, (\sigma^B_1 -
\sigma^B_o) = (\sigma_1 - \sigma_o)^2 = (\tau_1 - \tau_o)^2 -
\sum_r\, (\sigma_1^r - \sigma_o^r)^2 = 0$. See Ref. \cite{24} for
the use of a similar equations in the IAU conventions for the
definition of the radial velocity of stars.

\medskip

Let us remark that the already introduced Synge world function
$\Omega(\sigma_o, \sigma_1)$ of Eq.(\ref{4.5}) vanishes when
evaluated along a null 4-geodesics joining the two points: therefore
$\Omega(\sigma_o, \sigma) = 0$ is the equation of the null cone at
the point $\sigma^A_o$. If one solves the equation $\Omega(\sigma_o,
\sigma_1) = 0$ in $\tau_1$, one can find the emission time transfer
function for an electromagnetic signal emitted at $\tau_o$ in
$\sigma_o^r$ and absorbed in $\sigma^r_1$ and then study {\it time
delays} \cite{23} and their dependence upon the York time.

\bigskip

By using the embedding given in the Introduction we get the
following expressions for the end points and the tangent vector

\bea
 y^{\mu}(s) &=& x^{\mu}_o + \epsilon_A^{\mu}\, \sigma^A(s) =
 x^{\mu}_{2({\vec \sigma}(s))}(\tau_s = \tau(s)),\qquad
 y^{\mu}_o = y^{\mu}(0) = x^{\mu}_o + \epsilon_A^{\mu}\, \sigma^A_o =
 x^{\mu}_{1({\vec \sigma}_o)}(\tau_o), \nonumber \\
 &&{}\nonumber \\
 k^{\mu}(s) &=& {{d\, y^{\mu}(s)}\over {ds}} = \epsilon^{\mu}_A\,
 p^A(s).
 \label{4.5}
 \eea
\bigskip

With the PM null geodesics one can study the light deflection from a
massive body and the Shapiro time delay (see for instance
Ref.\cite{23,25}): in both cases the main ${}^3{\cal
K}_{(1)}$-dependence comes from the lapse function $n_{(1)}$.

\subsubsection{The PM Red-Shift}

If $v_1^{\mu}(0) = {{{\dot x}^{\mu}_1(\tau_o)}\over {\sqrt{\sgn\,
{\dot x}^2_1(\tau_o)}}}$ is the unit 4-velocity of the object
emitting the ray of light at $\tau_o$ and $v_2^{\mu}(s) = {{{\dot
x}^{\mu}_2(\tau_s)}\over {\sqrt{\sgn\, {\dot x}^2_2(\tau_s)}}}$ of
the observer detecting it at $\tau_s = \sigma^{\tau}(s)$, the
emitted frequency $\omega(0)$, the absorbed frequency $\omega(s)$
and the red-shift $z(s)$  (see Ref.\cite{25}) have the following PM
expressions

\begin{eqnarray*}
 \omega(0) &=& c\, k^{\mu}(0)\, v_{1 \mu}(0) = c\, v_{1 \mu}(0)\,
 \epsilon^{\mu}_A\, p^A(0) = c\, v_{1\mu}(0)\, \epsilon^{\mu}_A\,
 (b^A_{(o)+} + \delta^{A\tau}\, c_{(1)+}(\sigma_o)),\nonumber \\
 \omega(s) &=& c\, k^{\mu}(s)\, v_{2 \mu}(s) = c\, v_{2 \mu}(s)\,
 \epsilon^{\mu}_A\, p^A(s),\nonumber \\
 &&{}\nonumber \\
 {1\over {1 + z(s)}} &=& {{\omega(s)}\over {\omega(0)}} =
 {{v_{2\mu}(s)\, \epsilon^{\mu}_A\, p^A(s)}\over
 {v_{1\mu}(0)\, \epsilon^{\mu}_A\, p^A(0)}},
 \end{eqnarray*}

\bea
 z(s) &=& 1 - {{v_{1\mu}(0)\, \Big(\epsilon^{\mu}_{\tau}\, \sqrt{{\vec b}^2}
 + \epsilon^{\mu}_r\, b^r\Big)}\over {v_{2\mu}(s)\, \Big(\epsilon^{\mu}_{\tau}\,
 \sqrt{{\vec b}^2} + \epsilon^{\mu}_r\, b^r\Big)}}\times
 \Big[1 + {{v_{1\mu}(o)\, \epsilon^{\mu}_{\tau}\, c_{(1)+}(\sigma_o)}\over
 {v_{1\mu}(0)\, \Big(\epsilon^{\mu}_{\tau}\, \sqrt{{\vec b}^2}
 + \epsilon^{\mu}_r\, b^r\Big)}} -\nonumber \\
 &-& {{v_{2\mu}(s)\, \Big(\epsilon^{\mu}_{\tau}\, \Big[c_{(1)+}(\sigma_o) +
 p^{\tau}_{(1)({}^3K=0)}(s) + p^{\tau}_{(1)({}^3K)}(s)\Big] +
 \epsilon^{\mu}_r\, p^r_{(1)({}^3K=0)}(s) \Big)}\over {v_{2\mu}(s)\,
 \Big(\epsilon^{\mu}_{\tau}\, \sqrt{{\vec b}^2}
 + \epsilon^{\mu}_r\, b^r\Big)}}\Big].\nonumber \\
 &&{}
 \label{4.6}
 \eea

This equation allows to find the dependence of the red-shift $z(s)$
upon the non-local York time ${}^3{\cal K}(\sigma(s))$.

\subsection{The PM Geodesic Deviation Equation along a PM Null Geodesic and the
PM Luminosity Distance}

In the inertial frames of Minkowski space-time the flat null
geodesics joining $x_1^{\mu}$ to $x_2^{\mu}$ with $(x_1 - x_2)^2 =
0$ is $x^{\mu}(p) = x^{\mu}_1 + (x^{\mu}_2 - x^{\mu}_1)\, p$,
$k^{\mu} = {{d x^{\mu}(p)}\over {dp}} = x^{\mu}_2 - x^{\mu}_1$: this
implies $|x^o_1 - x^o_2| =  \sqrt{({\vec x}_1 - {\vec x}_2)^2} =
d_{Euclidean}(1,2)$, where $d_{Euclidean}$ is the Euclidean spatial
distance between the two points in the instantaneous inertial
3-spaces.
\bigskip

In curved space-time we have to solve the equation for the null
geodesics (see the previous Subsection and the Appendix of the first
paper in Refs.\cite{19}). However in astrophysics one uses the {\it
luminosity distance} \cite{25} between the emission point on a star
and the absorption point on the Earth. We have to find the relation
of the luminosity distance with the dynamical spatial distance
between the star and the Earth in the dynamical instantaneous PM
3-spaces.

\subsubsection{The PM Geodesic Deviation Equation}

As shown in Ref.\cite{25}, to find the luminosity distance between a
point (a star) emitting a ray of light (eikonal approximation) and a
point (the Earth) where the ray of light (propagating along a null
geodesics) is absorbed, we must solve the geodesics deviation
equation for nearby null geodesics with the same emission point and
propagate the resulting deviation vector to the absorption point.

\bigskip

Let the emitting star $S$ have the world-line $y_S^{\mu}(\tau(s_S) )
= x^{\mu}_o + \epsilon^{\mu}_A\, \sigma^A_S(s_S)$  (a time-like
geodesic with parameter $s_S$ if the star is considered a test
particle) with the unit time-like 4-velocity $v_S^{\mu}(\tau(s_S) )
= \epsilon^{\mu}_A\, u^A_S(s_S) = \epsilon^{\mu}_A\,
{{\sigma^A_S(s_S)}\over {ds_S}}$ ($s_S$ is the proper time). Let
$s_S = 0$, with $\sigma^A_S(0) = \sigma^A_o$, be the proper time of
the emission point.\medskip

Let $y^{\mu}(s) = \epsilon^{\mu}_A\, \sigma^A(s)$, $\sigma^A(s) =
\sigma_o^A + b^A\, s + \sigma_{(1)}^A(s)$ be the null geodesic
(\ref{4.2}) followed by the emitted ray of light, whose tangent
vector $k^{\mu}(s) = \epsilon^{\mu}_A\, p^A(s)$, $p^A(s) = b^A +
p^A_{(1)}(s)$ ($b^A = (\sqrt{{\vec b}^2}; b^r)$), is given in
Eq.(\ref{4.3}).
\medskip

At the emission point we have the unit time-like vector
$u_S^A(\sigma_o)$ and the null vector $p^A(0) = b^A +
\delta^{A\tau}\, c_{(1)}(\sigma_o)$ satisfying $\sgn\,
{}^4g_{(1)AB}(\sigma_o)\, u^A_S(\sigma_o)\, u^B_S(\sigma_o) = 1$ and
$\sgn\, {}^4g_{(1)AB}(\sigma_o)\, p^A(0)\, p^B(0) = 0$
\footnote{With the 4-metric (\ref{2.14}),  we have $\sgn\,
{}^4g_{(1)AB}\, (A_{(o)}^A + A_{(1)}^A)\, (B^B_{(o)} + B^B_{(1)}) =
A^{\tau}_{(o)}\, B^{\tau}_{(o)} - \sum_r\, A^r_{(o)}\, B^r_{(o)} +
2\, A^{\tau}_{(o)}\, B^{\tau}_{(o)}\, n_{(1)} + A^{\tau}_{(o)}\,
B^{\tau}_{(1)} + A^{\tau}_{(1)}\, B^{\tau}_{(o)} - \sum_r\,
(A^{\tau}_{(o)}\, B^r_{(o)} + A^r_{(o)}\, B^{\tau}_{(o)})\, {\bar
n}_{(1)(r)} - \sum_r\, (A^r_{(o)}\, B^r_{(1)} + A^r_{(1)}\,
B^r_{(o)}) - 2\, \sum_r\, A^r_{(o)}\, B^r_{(o)} (\Gamma_r^{(1)} +
2\, \phi_{(1)})$.}, respectively. To form a (non-orthogonal) frame
at $\sigma_o^A$ we must add two space-like vectors
$E^A_{S(\lambda)}(\sigma_o)$, $\lambda = 1,2$ satisfying $\sgn\,
{}^4g_{(1)AB}(\sigma_o)\, u^A_S(\sigma_o)\, E^B_{S(\lambda)} =
\sgn\, {}^4g_{(1)AB}(\sigma_o)\, p^A(0)\, E^B_{S(\lambda)}= 0$ and
$\sgn\, {}^4g_{(1)AB}(\sigma_o)\, E^A_{S(\lambda)}\,
E^B_{S(\lambda_1)} = - \delta_{\lambda\lambda_1}$ (they span a
2-plane orthogonal the star velocity and to the tangent to the ray
of light at the emission point).\medskip

A set of four vectors satisfying these conditions is
(${}^4\eta_{AB}\, b^A\, b^B = 0$, ${}^4\eta_{AB}\, b^A\,
E^B_{(o)S(\lambda)} = 0$, $\sgn\, {}^4\eta_{AB}\, b^A\, u^B_{(o)S} =
1$, $\sgn\, {}^4g_{(1)AB}(\sigma_o)\, p^A(0)\, u^B_S(\sigma_o) = 1 +
(c_{(1)} + n_{(1)} - {\bar n}_{(1)(3)})(\sigma_o) =
{{\omega_S(\sigma_o)}\over c}$ with $\omega_S(\sigma_o)$ the
emission frequency)

\begin{eqnarray*}
 u^A_S(\sigma_o) &=& u^A_{(o)S} - \delta^{A\tau}\,
 n_{(1)}(\sigma_o),\qquad u^A_{(o)S} = (1; 0,0,0),\nonumber \\
  p^A(0) &=& b^A + \delta^{A\tau}\, c_{(1)}(\sigma_o),\qquad b^A = (1; 0,0,1),
  \nonumber \\
 E^A_{S(\lambda)}(\sigma_o) &=& E^A_{(o)S(\lambda)} +
 E^A_{(1)S(\lambda)}(\sigma_o),
 \end{eqnarray*}

 \bea
 &&E^A_{(o)S(\lambda)} = \Big(0; e^1_{(o)S(\lambda)},
 e^2_{(o)S(\lambda)}, 0\Big),\nonumber \\
 &&E^A_{(1)S(\lambda)}(\sigma_o) =\Big(\sum_{s \not= 3}\, {\bar
 n}_{(1)(s)}(\sigma_o)\, e^s_{(o)S(\lambda)}; -(\Gamma_1^{(1)} +
 2\, \phi_{(1)})(\sigma_o)\, e^1_{(o)S(\lambda)},\nonumber \\
 &&\qquad  - (\Gamma_2^{(1)} + 2\, \phi_{(1)})(\sigma_o)\, e^2_{(o)S(\lambda)},
 0 \Big),\nonumber \\
 &&\sum_{r \not= 3}\, e^r_{(o)S(\lambda)}\,
 e^r_{(o)S(\lambda_1)} = \delta_{\lambda\lambda_1},\qquad
 e^3_{(o)S(\lambda)} = 0.\nonumber \\
 &&{}
 \label{4.7}
 \eea
\medskip

Let the absorbing Earth $E$ have the world-line $y_E^{\mu}(\tau(s_E)
) = x^{\mu}_o + \epsilon^{\mu}_A\, \sigma^A_E(s_E)$  (a time-like
geodesic with parameter $s_E$ if the Earth is considered a test
particle) with the unit time-like 4-velocity $v_E^{\mu}(\tau(s_E) )
= \epsilon^{\mu}_A\, u^A_E(s_E) = \epsilon^{\mu}_A\,
{{\sigma^A_E(s_E)}\over {ds_E}}$ ($s_E$ is the proper time). Let
$s_E = s_1$, with $\sigma^A_S(s_1) = \sigma^A_1$, be the proper time
of the absorption point.\medskip

At the absorption point $s = s_1$ we have the unit time-like vector
$u_E^A(\sigma_1)$ and the null vector $p^A(s_1) = b^A +
p^A_{(1)}(s_1)$, with $p^A_{(1)}(s_1)$ given in Eq.(\ref{4.1}),
satisfying $\sgn\, {}^4g_{(1)AB}(\sigma_1)\, u^A_E(\sigma_1)\,
u^B_E(\sigma_1) = 1$ and $\sgn\, {}^4g_{(1)AB}(\sigma_1)\,
p^A(s_1)\, p^B(s_1) = 0$, respectively.\medskip

To form a (non-orthogonal) frame at $\sigma_1^A$ we must add two
space-like vectors $F^A_{E(\lambda)}(\sigma_1)$, $\lambda = 1,2$
satisfying $\sgn\, {}^4g_{(1)AB}(\sigma_1)\, u^A_E(\sigma_1)\,
F^B_{E(\lambda)} = \sgn\, {}^4g_{(1)AB}(\sigma_1)\, p^A(s_1)\,
F^B_{E(\lambda)}= 0$ and $\sgn\, {}^4g_{(1)AB}(\sigma_1)\,
F^A_{E(\lambda)}\, F^B_{E(\lambda_1)} = - \delta_{\lambda\lambda_1}$
(they span a 2-plane orthogonal the Earth velocity and to the
tangent to the ray of light at the absorption point).\medskip

A set of four vectors satisfying these conditions is
(${}^4\eta_{AB}\, b^A\, b^B = 0$, ${}^4\eta_{AB}\, b^A\,
F^B_{(o)E(\lambda)} = 0$, $\sgn\, {}^4g_{(1)AB}(\sigma_1)\,
p^A(s_1)\, u^B_E(\sigma_1) = 1 + (p^{\tau}_{(1)} + n_{(1)} - {\bar
n}_{(1)(3)})(\sigma_1) = {{\omega_E(\sigma_1)}\over c}$ with
$\omega_E(\sigma_1)$ the absorption frequency)

\begin{eqnarray*}
 u^A_E(\sigma_1) &=& u^A_{(o)E} - \delta^{A\tau}\,n_{(1)}(\sigma_1),
 \qquad u^A_{(o)E} = (1; 0,0,0), \nonumber \\
 p^A(s_1) &=& b^A + p^A_{(1)}(s_1),\qquad b^A = (1;
 0,0,1),\nonumber \\
  F^A_{E(\lambda)}(\sigma_1) &=&F^A_{(o)E(\lambda)} +
  F^A_{(1)E(\lambda)}(\sigma_1),
  \end{eqnarray*}

  \bea
  &&F^A_{(o)E(\lambda)} = \Big(0; f^1_{(o)E(\lambda)},
  f^2_{(o)E(\lambda)}, 0\Big),\nonumber \\
  &&F^A_{(1)E(\lambda)}(\sigma_1) = \Big(\sum_{s \not= 3}\, {\bar
 n}_{(1)(s)}(\sigma_1)\, f^s_{(o)E(\lambda)}; - (\Gamma_1^{(1)} +
 2\, \phi_{(1)})(\sigma_1)\, f^1_{(o)E(\lambda)}, \nonumber \\
 &&- (\Gamma_2^{(1)} + 2\, \phi_{(1)})(\sigma_1)\, f^2_{(o)E(\lambda)},
 - \sum_{s \not= 3}\, p^s_{(1)}(s_1)\, f^s_{(o)E(\lambda)} \Big),
 \nonumber \\
 &&\sum_{r \not= 3}\, f^r_{(o)E(\lambda)}\, f^r_{(o)E(\lambda_1)}
 = \delta_{\lambda\lambda_1},\qquad f^3_{(o)E(\lambda)} = 0.
 \nonumber \\
 &&{}
 \label{4.8}
 \eea

\medskip

We can choose $e^r_{(o)S(\lambda)} = f^r_{(o)E(\lambda)} =
g^r_{(o)(\lambda)}$ with $\sum_{r=1,2}\, g^r_{(o)(\lambda)}\,
g^r_{(o)(\lambda_1)} = \delta_{\lambda\lambda_1}$ and
$g^3_{(o)(\lambda} = 0$.

\bigskip

As shown in Ref.\cite{25} the deviation vector $Y^{\mu}(y(s)) =
\epsilon^{\mu}_A\, Y^A(\sigma(s))$, with $Y^A(\sigma_o) = 0$, along
the null geodesic connecting $\sigma^A_o$ and $\sigma^A_1$ has the
following properties:\medskip

\noindent A) it vanishes at $\sigma^A_o$;\medskip

\noindent B) its covariant derivative along the tangent to the null
geodesic

\beq
 {{D\, Y^A(\sigma(s))}\over {ds}} = p^B(s)\, \Big[\partial_B\,
 Y^A(\sigma(s)) + {}^4\Gamma^A_{BC}(\sigma(s))\, Y^C(\sigma(s))\Big],
 \label{4.9}
 \eeq

\noindent  is orthogonal to the star velocity  $u^A_S(\sigma_o)$ and
to the tangent $p^a(0)$ to the ray of light at the emission point
$\sigma^A_o$;\medskip

\noindent C) its covariant differential along the tangent to the
null geodesic is also orthogonal to the Earth velocity
$u^A_E(\sigma_1)$ and to the tangent $p^A(s_1)$ to the ray of light
at the absorption point $\sigma^A_1$.\medskip

Therefore we have

\bea
 Y^A(\sigma_o) &=& 0,\nonumber \\
 &&{}\nonumber \\
 {{D\, Y^A(\sigma(s))}\over {ds}}{|}_{\sigma_o} &=&
 \sum_{\lambda = 1,2}\, A_{(\lambda)}\,
 E^A_{S(\lambda)}(\sigma_o),\nonumber \\
 &&{}\nonumber \\
  {{D\, Y^A(\sigma(s))}\over {ds}}{|}_{\sigma_1} &=&
 \sum_{\lambda = 1,2}\, B_{(\lambda)}\,
 F^A_{E(\lambda)}(\sigma_1),
 \label{4.10}
 \eea

 \medskip

The deviation vector is solution of the geodesic deviation equation

\bea
 {{D^2\, Y^A(\sigma(s))}\over {ds^2}} &=&p^B(s)\, \Big(\partial_B\,
 \Big[p^C(s)\, \Big(\partial_C\, Y^A(\sigma(s)) +
 {}^4\Gamma^A_{CD}(\sigma(s))\, Y^D(\sigma(s))\Big)\Big] +\nonumber \\
 &+& {}^4\Gamma^A_{BE}(\sigma(s))\, \Big[p^C(s)\, \Big(\partial_C\,
 Y^E(\sigma(s)) + {}^4\Gamma^E_{CD}(\sigma(s))\, Y^D(\sigma(s))
 \Big)\Big]\Big) =\nonumber \\
 &&{}\nonumber \\
 &=& {}^4g^{AB}(\sigma(s))\, {}^4R_{BCDE}(\sigma(s))\, p^C(s)\,
 p^D(s)\, Y^E(\sigma(s)),
 \label{4.11}
 \eea

\noindent with the initial data $Y^A(\sigma_o) = 0$ and ${{D\,
Y^A(\sigma(s))}\over {ds}}{|}_{\sigma_o} = \sum_{\lambda = 1,2}\,
A_{(\lambda)}\, E^A_{S(\lambda)}(\sigma_o)$.\medskip

Its solution, evaluated at the absorption point $\sigma^A_1$, can be
put in the form \cite{25}

 \bea
 Y^A(\sigma_1) &=& J^A{}_B(E,S)\, {c\over {\omega_S(\sigma_o)}}\,
 {{D\, Y^B(\sigma(s))}\over {ds}}{|}_{s = 0},\nonumber \\
 &=& \sum_{\lambda\lambda_1}\, F^A_{E(\lambda_1)}(\sigma_1)\,
 {\cal J}_{\lambda_1\lambda}(E,S)\, E_{S(\lambda)B}(\sigma_o)\,
 {{A_{(\lambda)}}\over {\omega_S(\sigma_o)/c}},\nonumber \\
 &&{}\nonumber \\
 with&& J^A{}_B(E,S) =  \sum_{\lambda_1\lambda}\, F^A_{E(\lambda_1)}(\sigma_1)\,
 {\cal J}_{\lambda_1\lambda}\, E_{S(\lambda) B}(\sigma_o),\nonumber \\
 &&{}\nonumber \\
 &&\sgn\, {}^4g_{(1)AC}(\sigma_1)\, Y^A(\sigma_1)\, F^C_{E(\lambda_1)}(\sigma_1)
 = - \sum_{\lambda}\, {\cal J}_{\lambda_1\lambda}(E,S)\,
 {{A_{(\lambda)}}\over {\omega_S(\sigma_o)/c}},\nonumber \\
 &&{}
 \label{4.12}
 \eea

\noindent where $\omega_S(\sigma_o)$ is the emission circular
frequency of the light-ray. The {\it Jacobi map}
$J^{\mu}{}_{\nu}(E,S)$ maps vectors at $S$ into vectors at $E$

 \subsubsection{The PM Luminosity Distance}

The {\it luminosity distance} is \cite{25}

\beq
 d_{lum}(S,E) = (1 + z)\, \sqrt{|det\, {\cal J}|} = {{\omega_S(\sigma_o)}\over
 {\omega_E(\sigma_1)}}\, \sqrt{|det\, {\cal J}|},
 \label{4.13}
 \eeq

\noindent where $z$ is the {\it red-shift} of the source as seen by
the observer: $1 + z = \omega_S(\sigma_o)/\omega_E(\sigma_1)$, with
$\omega_E(\sigma_1)$ the absorption frequency. The {\it corrected
luminosity distance} is $D_{lum}(S,E) = \sqrt{|det\, {\cal J}|}$.
\medskip

In the inertial frames of Minkowski space-time one gets
$d_{lum}(S,E) = (1 + z)\, d_{Euclidean}(S,E)$, namely the corrected
luminosity distance is the Euclidean spatial distance.

\bigskip

In the weak field approximation, by using  $\sigma^A(s) =
\sigma_{(o)}(s) + \sigma^A_{(1)}(s)$ we get

\bea
 Y^A(\sigma(s)) &=& Y^A(\sigma_{(o)}(s) + \sigma_{(1)}(s)) =
 \nonumber \\
 &=& Y^A(\sigma_{(o)}(s)) + {{\partial\, Y^A(\sigma_{(o)}(s))}\over
 {\partial\, \sigma^E}}\, \sigma^E_{(1)}(s) + O(\zeta^2) =\nonumber \\
 &=&  Y^A_{(o)}(\sigma_{(o)}(s)) + Y^A_{(1)}(\sigma_{(o)}(s)) +
 {{\partial\, Y_{(o)}^A(\sigma_{(o)}(s))}\over
 {\partial\, \sigma^E_{(o)}}}\, \sigma^E_{(1)}(s) + O(\zeta^2),
 \nonumber \\
 &&{}\nonumber \\
  \partial_B\, Y^A(\sigma(s)) &=& {{\partial\, Y_{(o)}^A(\sigma_{(o)}(s))}\over
 {\partial\, \sigma^B_{(o)}}} + {{\partial\, Y_{(1)}^A(\sigma_{(o)}(s))}\over
 {\partial\, \sigma^B_{(o)}}} +\nonumber \\
 &+& {{\partial^2\, Y_{(o)}^A(\sigma_{(o)}(s))}\over {\partial\,
 \sigma^B_{(o)}\, \partial\, \sigma^E_{(o)}}}\, \sigma^E_{(1)}(s)
 + O(\zeta^2),\nonumber \\
  \partial_C\, \partial_B\, Y^A(\sigma(s)) &=& {{\partial^2\,
  Y_{(o)}^A(\sigma_{(o)}(s))}\over {\partial\, \sigma^C_{(o)}\,
  \partial\, \sigma^B_{(o)}}} + {{\partial^2\, Y_{(1)}^A(\sigma_{(o)}(s))}\over
 {\partial\, \sigma^C_{(o)}\, \partial\, \sigma^B_{(o)}}} +\nonumber \\
 &+& {{\partial^3\, Y_{(o)}^A(\sigma_{(o)}(s))}\over {\partial\, \sigma^C_{(o)}\,
 \partial\, \sigma^B_{(o)}\, \partial\, \sigma^E_{(o)}}}\, \sigma^E_{(1)}(s)
 + O(\zeta^2).
 \label{4.14}
 \eea
\medskip

By using $p^A(s) = b^A + p^A_{(1)}(s)$,  as implied by
Eq.(\ref{4.2}), Eq.(\ref{4.9}) becomes

\bea
 {{D\, Y^A(\sigma(s))}\over {ds}} &=& b^B\, \Big[\partial_B\,
 Y^A_{(o)}(\sigma_{(o)}(s)) + \partial_B\, Y^A_{(1)}(\sigma_{(o)}(s))
 +\nonumber \\
 &+& \partial_E\, \partial_B\, Y^A_{(o)}(\sigma_{(o)}(s))\,
 \sigma_{(1)}^E + {}^4\Gamma^A_{(1)BC}(\sigma_{(o)}(s))\,
 Y^C_{(o)}(\sigma_{(o)}(s))\Big] +\nonumber \\
 &+& p_{(1)}^B(s)\, \partial_B\, Y^A_{(o)}(\sigma_{(o)}(s))
 + O(\zeta^2) =\nonumber \\
 &=& {{D\, Y_{(o)}^A(\sigma_{(o)}(s))}\over {ds}} +
 {{D\, Y_{(1)}^A(\sigma_{(o)}(s))}\over {ds}}.
 \label{4.15}
 \eea

As a consequence,  the geodesic deviation equation (\ref{4.11})
becomes

\bea
 &&b^B\, b^c\, \Big(\partial_B\, \partial_C\, Y^A_{(o)}(\sigma_{(o)}(s))
 + \partial_B\, \partial_C\, Y^A_{(1)}(\sigma_{(o)}(s)) +
 \partial_B\, \partial_C\, \partial_E\, Y^A_{(o)}(\sigma_{(o)}(s))\,
 \sigma^E_{(1)}(s) +\nonumber \\
 &&+ \partial_B\, {}^4\Gamma^A_{(1)CE}(\sigma_{(o)}(s))\,
 Y^E_{(o)}(\sigma_{(o)}(s)) + 2\,
 {}^4\Gamma^A_{(1)BE}(\sigma_{(o)}(s))\, \partial_C\,
 Y^E_{(o)}(\sigma_{(o)}(s))\Big) +\nonumber \\
 &&+ b^B\, p^C_{(1)}(s)\, \partial_B\, \partial_C\, Y^A_{(o)}(\sigma_{(o)}(s))
 =\nonumber \\
 &&{}\nonumber \\
 &&= {}^4\eta^{AD}\, {}^4R_{(1)DBCE}(\sigma_{(o)}(s))\, b^B\, b^C\,
 Y^E_{(o)}(\sigma_{(o)}(s)),
 \label{4.16}
 \eea

\noindent with the Christoffel symbols and the Riemann tensor of
Eqs. (\ref{2.15}) and (\ref{2.16}).\medskip

Therefore we have to solve the following two equations (the
dependence upon $\sigma^E_{(1)}(s)$ is eliminated by the first
equation)

\bea
 b^B\, b^C\, \partial_B\, \partial_C\, Y^A_{(o)}(\sigma_{(o)}(s)) &=& 0,\nonumber \\
 &&{}\nonumber \\
 b^B\, b^C\, \partial_B\, \partial_C\,
 Y^A_{(1)}(\sigma_{(o)}(s)) &=& b^B\, b^C\, \Big(\Big[
 {}^4\eta^{AD}\, {}^4R_{(1)DBCE}(\sigma_{(o)}(s)) -\nonumber \\
 &-& \partial_B\, {}^4\Gamma^A_{(1)CE}(\sigma_{(o)}(s))\Big]\,
 Y^E_{(o)}(\sigma_{(o)}(s)) -\nonumber \\
 &-& 2\, {}^4\Gamma^A_{(1)BE}(\sigma_{(o)}(s))\, \partial_C\,
 Y^E_{(o)}(\sigma_{(o)}(s)) \Big) -\nonumber \\
 &-& b^B\, p^C_{(1)}(s)\, \partial_B\, \partial_C\,
 Y^A_{(o)}(\sigma_{(o)}(s)).
 \label{4.17}
 \eea
\bigskip

From Eqs.(\ref{4.10}) the initial conditions are

\bea
 Y^A_{(o)}(\sigma_o) &=& Y^A_{(1)}(\sigma_o) = 0,
 \nonumber \\
 \Big({{D\, Y_{(o)}^A(\sigma_{(o)}(s))}\over {ds}}\Big){|}_{s = 0} &=&
 \Big(b^B\, \partial_B\,
 Y^A_{(o)}(\sigma_{(o)}(s))\Big){|}_{s = } = \sum_{\lambda =1,2}\, A_{(\lambda)}\,
 E^A_{(o)S(\lambda)},\nonumber \\
 \Big({{D\, Y_{(1)}^A(\sigma_{(o)}(s))}\over {ds}}\Big){|}_{s = 0}
 &=& \Big(b^B\, \partial_B\, Y^A_{(1)}(\sigma_{(o)}(s))\Big){|}_{s = 0}
 + c_{(1)}(\sigma_o)\, \partial_{\tau_o}\, Y^A_{(o)}(\sigma_o) =
 \sum_{\lambda =1,2}\, A_{(\lambda)}\,
 E^A_{(1)S(\lambda)}.\nonumber \\
 &&{}
 \label{4.18}
 \eea

Since at the zero order we have ${}^4\eta_{AB}\, b^A\, b^B = 0$,
${}^4\eta_{AB}\, b^A\, E^B_{(o)S(\lambda)} = 0$ and $\sgn\,
{}^4\eta_{AB}\, u^A_{(o)S}\, b^B = 1$, due to Eqs.(\ref{4.7}), the
solution of the first equation, satisfying the initial conditions
(\ref{4.18}), is

\bea
 Y^A_{(o)}(\sigma_{(o)}(s)) &=& \Big(\sgn\, {}^4\eta_{BC}\, u^B_{(o)S}\,
 (\sigma^C_{(o)}(s) - \sigma^C_o)\Big)\, \sum_{\lambda =1,2}\, A_{(\lambda)}\,
 E^A_{(o)S(\lambda)} =\nonumber \\
 &=& \Big(\tau_{(o)}(s) - \tau_o\Big)\, \sum_{\lambda =1,2}\, A_{(\lambda)}\,
 E^A_{(o)S(\lambda)},\nonumber \\
 &&{}\nonumber \\
 {{D\, Y_{(o)}^A(\sigma_{(o)}(s))}\over {ds}} &=& b^B\, \partial_B\,
 Y^A_{(o)}(\sigma_{(o)}(s)) = \sum_{\lambda =1,2}\, A_{(\lambda)}\,
 E^A_{(o)S(\lambda)},\qquad independently\, from\, s.\nonumber \\
 &&{}
 \label{4.19}
 \eea

Let us remark that $Y^A_{(o)}(\sigma_1)$ is proportional to $\tau_1
- \tau_o = \sqrt{({\vec \sigma}_1 - {\vec \sigma}_o)^2} =
d_{Euclidean}(1,0)$ as expected at the zero order in Minkowski
space-time.

\bigskip

Then the second of equations (\ref{4.17}) and its initial conditions
(\ref{4.18}) become

\bea
 b^B\, b^C\, \partial_B\, \partial_C\, Y^A_{(1)}(\sigma_{(o)}(s))
 &=& \Big(\sgn\, {}^4\eta_{UV}\, u^U_{(o)S}\,
 (\sigma^V_{(o)}(s) - \sigma^V_o)\Big)\, b^B\, b^C\,
 \Big[{}^4\eta^{AD}\, {}^4R_{(1)DBCE}(\sigma_{(o)}(s))\,
   -\nonumber \\
 &-& \partial_B\, {}^4\Gamma^A_{(1)CE}(\sigma_{(o)}(s))\Big]\,
 \sum_{\lambda =1,2}\, A_{(\lambda)}\, E^E_{(o)S(\lambda)},
 \nonumber \\
 &&{}\nonumber \\
 Y^A_{(1)}(\sigma_o) &=& 0,\nonumber \\
 \Big(b^B\, \partial_B\, Y^A_{(1)}(\sigma_{(o)}(s))\Big){|}_{s = 0} &=& \sum_{\lambda = 1,2}\,
 A_{(\lambda)}\, \Big(E^A_{(1)S(\lambda)}(\sigma_o) -
 c_{(1)}(\sigma_o)\, E^A_{(o)S(\lambda)}\Big),
 \label{4.20}
 \eea

\noindent with $E^A_{(1)S(\lambda)}(\sigma_o)$ given in
Eq.(\ref{4.7}).\medskip

Since we have $\sigma_{(o)}(s) = \sigma^A_o + b^A\, s$, we get
$b^B\, {{\partial\, Y^A_{(1)}(\sigma_{(o)}(s))}\over {\partial\,
\sigma_{(o)}^B}} = {d\over {ds}}\, Y^A_{(1)}(\sigma_{(o)}(s))$ and
$\Big(b^B\, \partial_B\, Y^A_{(1)}(\sigma_{(o)}(s))\Big){|}_{s = 0}
= {{d\, Y^A_{(1)}(\sigma_{(o)}(s))}\over {ds}}{|}_{s = 0}$ .
\medskip

Therefore the solution of Eq.(\ref{4.17}) with the given initial
data is

\bea
 Y^A_{(1)}(\sigma_{(o)}(s)) &=& \Big[\sum_{\lambda = 1,2}\,
 A_{(\lambda)}\, \Big(E^A_{(1)S(\lambda)}(\sigma_o) -
 c_{(1)}(\sigma_o)\, E^A_{(o)S(\lambda)}\Big)\Big]\, s +\nonumber \\
 &+& \int_0^s ds_1\, \int_0^{s_1} ds_2\, \Big[
 \Big(\sgn\, {}^4\eta_{BC}\, u^B_{(o)S}\,
 (\sigma^C_{(o)}(s_2) - \sigma^C_o)\Big)\nonumber \\
 &&b^B\, b^C\, \Big({}^4\eta^{AD}\, {}^4R_{(1)DBCE}(\sigma_{(o)}(s_2)) -
 \partial_B\, {}^4\Gamma^A_{(1)CE}(\sigma_{(o)}(s_2))\Big) \Big]\nonumber \\
 &&\sum_{\lambda =1,2}\, A_{(\lambda)}\,
 E^E_{(o)S(\lambda)}.
 \label{4.21}
 \eea

\bigskip

By using Eqs. (\ref{4.14}), (\ref{4.19}) and (\ref{4.20}) the last
line of Eq.(\ref{4.12}) becomes \footnote{We also use $b^A =
(1;0,0,1)$, $\sgn\, {}^4g_{(1)FA(\sigma_1)}\,
F^F_{E(\lambda_1)}(\sigma_1)\, E^A_{(1)S(\lambda)}(\sigma_o) = -
\sum_{r=1,2}\, [1 + (\Gamma_r^{(1)} + 2\, \phi_{(1)})(\sigma_1)]\,
g^r_{(o)(\lambda_1)}\, g^r_{(o)(\lambda)}$, $\sgn\,
{}^4g_{(1)FA}(\sigma_1)\, F^F_{(o)E(\lambda_1)}\,
E^A_{(1)S(\lambda)}(\sigma_o) = \sum_{r=1,2}\, (\Gamma_r^{(1)} + 2\,
\phi_{(1)})(\sigma_o)\, g^r_{(o)(\lambda_1)}\, g^r_{(o)(\lambda)}$
and $\omega_S(\sigma_o)/c = 1 + (c_{(1)} + n_{(1)} - {\bar
n}_{(1)(3)})(\sigma_o) = 1 - (n_{(1)} + \Gamma_3^{(1)} + 2\,
\phi_{(1)} + {1\over 2}\, {\bar n}_{(1)(3)})(\sigma_o)$ (we used
Eq.(\ref{4.1}) for $c_{(1)}(\sigma_o)$).}

\begin{eqnarray*}
 &&- \sgn\, {}^4g_{(1)AC}(\sigma_1)\, Y^A(\sigma_1)\, F^C_{E(\lambda_1)}(\sigma_1)
 =\nonumber \\
 &&= - \sgn\, {}^4g_{(1)AC}(\sigma_1)\, \Big[Y^A_{(o)}(\sigma_1) +
 Y^A_{(1)}(\sigma_1) + \partial_E\, Y^A_{(o)}(\sigma_1)\Big]\, F^C_{E(\lambda_1)}(\sigma_1)
 =\nonumber \\
 &=&\sum_{\lambda}\, {\cal J}_{\lambda_1\lambda}(E,S)\, {{A_{(\lambda)}}\over
 {\omega_S(\sigma_o)}},\nonumber \\
 &&{}\nonumber \\
 && {\cal J}(E,S)_{\lambda_1\lambda}  = \Big((\tau_1 - \tau_o)\,
 [1 + 2\, \phi_{(1)}(\sigma_1) + (c_{(1)} + n_{(1)} -
 {\bar n}_{(1)(3)})(\sigma_o)] +\nonumber \\
 &&+ \tau_{(1)}(\sigma_1) - (c_{(1)} + 2\, \phi_{(1)})(\sigma_o)\, s_1
 \Big)\, \delta_{\lambda_1\lambda} +\nonumber \\
 &&+\sum_{r=1,2}\, \Big((\tau_1 - \tau_o)\, \Gamma_r^{(1)}(\sigma_1) -
 \Gamma_r^{(1)}(\sigma_o)\, s_1\Big)\, g^r_{(o)(\lambda_1)}\, g^r_{(o)(\lambda)}
 -\nonumber \\
 &&- \int_0^{s_1} ds_2\, \int_0^{s_2} ds_3\, (\tau_{(o)}(s_3) -
 \tau_o)\, {\cal W}_{(1)\lambda_1\lambda}(
 \sigma_{(o)}(s_3)) = {\cal J}(E,S)_{(o)\lambda_1\lambda}
 + {\cal J}(E,S)_{(1)\lambda_1\lambda},
 \end{eqnarray*}

\begin{eqnarray*}
 {\cal J}(S,E)_{(o)\lambda_1\lambda} &=& (\tau_1 - \tau_o)\,
 \delta_{\lambda_1\lambda} = d_{Euclidean}(S,E)\,
 \delta_{\lambda_1\lambda},\nonumber \\
 {\cal J}(S,E)_{(1)\lambda_1\lambda} &=& \Big(\tau_{(1)}(\sigma_1)
 - (2\, n_{(1)} + \Gamma_3^{(1)} - {1\over 2}\,
 {\bar n}_{(1)(3)})(\sigma_o)\, s_1 +\nonumber \\
 &+&d_{Euclidean}(S,E)\, (n_{(1)} + \Gamma_3^{(1)} +
 2\, \phi_{(1)} - {1\over 2}\, {\bar n}_{(1)(3)})\Big)\,
 \delta_{\lambda_1\lambda} +\nonumber \\
 &+&\sum_{r=1,2}\, \Big(d_{Euclidean}(S,E)\, \Gamma_r^{(1)}(\sigma_1)
 - \Gamma_r^{(1)}(\sigma_o)\, s_1\Big)\, g^r_{(o)(\lambda_1)}\,
 g^r_{(o)(\lambda)} -\nonumber \\
 &-&\int_0^{s_1} ds_2\, \int_0^{s_2} ds_3\, (\tau_{(o)}(s_3) -
 \tau_o)\, {\cal W}_{(1)\lambda_1\lambda}(\sigma_{(o)}(s_3)),
 \end{eqnarray*}

 \bea
 {\cal W}_{(1)\lambda_1\lambda}(\sigma_{(o)}(s_3)) &=&
 \sgn\, {}^4\eta_{FA}(\sigma_1)\, F^F_{(o)E(\lambda_1)}\, b^B\,
 b^C\, \Big[{}^4\eta^{AD}\, {}^4R_{(1)DBCK}(\sigma_{(o)}(s_3))
 -\nonumber \\
 &&- \partial_B\, {}^4\Gamma^A_{(1)CK}(\sigma_{(o)}(s_3))\Big]\,
 E^K_{(o)S(\lambda)} =\nonumber \\
 &=&\sum_{r,s = 1,2}\, g^r_{(o)(\lambda_1)}\, \Big[\sgn\, \Big(
 {}^4R_{(1)r3\tau s} - {}^4R_{(1)\tau r\tau s} + {}^4R_{(1)r33s}
 - {}^4R_{(1)\tau r3s}\Big) +\nonumber \\
 &+&(\partial_{\tau} + \partial_3)\, \Big({}^4\Gamma^r_{(1)\tau s} +
 {}^4\Gamma^r_{(1)3s}\Big) \Big]\, g^s_{(o)(\lambda)}.
 \label{4.22}
 \eea

By using $\tau_1 - \tau_o = d_{Euclidean}(S,E)$  we get ${\cal
J}_{(o)\lambda_1\lambda}(S,E) = d_{Euclidean}(S,E)\,
\delta_{\lambda_1\lambda}$. As a consequence we get the following
expression of the corrected luminosity distance

\begin{eqnarray*}
 D_{lum}(S,E) &=& {{d_{lum}(S,E)}\over {1 + z(s_1)}} =
 \sqrt{|det\, {\cal J}(S,E)|}
 =\nonumber \\
 &&{}\nonumber \\
 &=&\sqrt{{\cal J}(S,E)_{(o)11}\, {\cal J}(S,E)_{(o)22}
 + {\cal J}(S,E)_{(o)11}\, {\cal J}(S,E)_{(1)22} +
 {\cal J}(S,E)_{(1)11}\, {\cal J}(S,E)_{(o)22} }=\nonumber \\
 &&{}\nonumber \\
 &=& d_{Euclidean}(S,E)\, \sqrt{1 + {{{\cal J}(S,E)_{(1)11} +
 {\cal J}(S,E)_{(1)22}}\over {d_{Euclidean}(S,E)}} + O(\zeta^2)}
 =\nonumber \\
 &=& d_{Euclidean}(S,E)\, \sqrt{1 + {{\sum_{\lambda = 1,2}\, \Big(
 {\cal J}(S,E)_{(1)\lambda\lambda({}^3K=0)} +
 {\cal J}(S,E)_{(1)\lambda\lambda({}^3K)}\Big)}\over
 {d_{Euclidean}(S,E)}} + O(\zeta^2)},
 \end{eqnarray*}

\begin{eqnarray*}
 \sum_{\lambda = 1,2}\, {\cal J}(S,E)_{(1)\lambda\lambda} &=& 2\, \Big(\tau_{(1)}(\sigma_1)
 - (2\, n_{(1)} + \Gamma_3^{(1)} - {1\over 2}\,
 {\bar n}_{(1)(3)})(\sigma_o)\, s_1 +\nonumber \\
 &+&d_{Euclidean}(S,E)\, (n_{(1)} + \Gamma_3^{(1)} +
 2\, \phi_{(1)} - {1\over 2}\, {\bar n}_{(1)(3)})(\sigma_1)\Big)
  +\nonumber \\
 &+&\sum_{\lambda, r=1,2}\, \Big(d_{Euclidean}(S,E)\, \Gamma_r^{(1)}(\sigma_1)
 - \Gamma_r^{(1)}(\sigma_o)\, s_1\Big)\, \Big(g^r_{(o)(\lambda)}\Big)^2 -\nonumber \\
 &-&\int_0^{s_1} ds_2\, \int_0^{s_2} ds_3\, (\tau_{(o)}(s_3) -
 \tau_o)\, \sum_{\lambda = 1,2}\, {\cal W}_{(1)\lambda\lambda}(
 \sigma_{(o)}(s_3)),
 \end{eqnarray*}

\bea
 \sum_{\lambda = 1,2}\, {\cal J}(S,E)_{(1)\lambda\lambda({}^3K)} &=&
 s_1\, \Big((4\, \partial_{\tau} + \partial_3)\,
 {}^3{\cal K}_{(1)}\Big)(\sigma_o) - {1\over 2}\, d_{Euclidean}(S,E)\,
 \Big((2\, \partial_{\tau} + \partial_3)\,
 {}^3{\cal K}_{(1)}\Big)(\sigma_1) -\nonumber \\
 &-&\int_0^{s_1} ds_2\, \int_0^{s_2} ds_3\, (\tau_{(o)}(s_3) -
 \tau_o)\, \sum_{\lambda,r,s = 1,2}\, g^r_{(o)(\lambda)}\,
 g^s_{(o)(\lambda)}\, \Big(\partial_r\, \partial_s\,
 {}^3{\cal K}_{(1)}\Big)(\sigma_{(o)}(s_3)),\nonumber \\
 &&{}
 \label{4.23}
 \eea

\noindent where Eqs.(\ref{2.15}) and (\ref{2.16}) have been used to
find the dependence upon the non-local York time ${}^3{\cal
K}_{(1)}$.
\bigskip

Let us remark that Eq. (\ref{4.6}) implies that the frequency
$\omega(0)$ of the light emitted from the star is $\omega(0) = (1 +
z(s_1))\, \omega(s_1)$, where $\omega(s_1)$ is the frequency
absorbed on the Earth. Since $\omega(0) = c\, v_{S \mu}(0)\,
\epsilon^{\mu}_A\, p^A(0) = v_{rec}(S,E)$ \footnote{Due to the use
of proper time $c\, v^{\mu}_S$ has the dimension of an ordinary
velocity with respect to $t = \tau/c$.} is also the radial (i.e.
along the line of sight) recessional velocity of the star, we have
that the recessional velocity is proportional to the red-shift (i.e.
it is a red-shift-velocity $c\, z$). On the other hand,  for small
deviations from the Euclidean distance, Eq.(\ref{4.23}) can be
written as

\beq
 D_{lum}(S,E) \approx d_{Euclidean}(S,E) + {1\over 2}\, \sum_{\lambda
 =1,2}\, {\cal J}(S,E)_{(1)\lambda\lambda} = \alpha + \beta\, (1 +
 z(s_1)),
 \label{4.24}
 \eeq

\noindent because the term $- (2\, n_{(1)} + \Gamma_3^{(1)} -
{1\over 2}\, {\bar n}_{(1)(3)})(\sigma_o)\, s_1 $ contains
$\omega(0)$, i.e. a linear dependence on the red-shift.\medskip

These two results imply that the recessional velocity of the star is
proportional to its luminosity distance from the Earth
($V_{rec}(S,E) = A\, z(s_1) + B$) at least for small distances. This
is in accord with the Hubble old  redshift-distance relation which
is formalized in the Hubble law (velocity-distance relation) when
the standard cosmological model is used (see for instance
Ref.\cite{26} on these topics). Again these results have a
dependence on the trace of the non-local York time, which could play
a role in giving a different interpretation of the data from
super-novae, which are used as a support for dark energy \cite{54}.

\vfill\eject

\section{The PM Equations of Motion of the Particles and their PN Expansion}

In this Section we study the HPM equations of motion for the
particles. We formulate the problem of the definition of the center
of mass and relative variables first in general relativity and then
in PM space-times, using the two-body problem as an example. Then we
study the PN expansion of the equations of motion and we consider
the 1PN limit of PM binaries for vanishing York time.

\subsection{The PM Equations of Motion for the Particles}

From Eqs.(5.2) and (5.3) of paper II, by using Eqs.(\ref{2.1}), we
get the following expression for the momenta and the equations of
motion of the particles ($\eta^r_i(\tau), {\dot \eta}^r_i(\tau) =
O(1)$, $m_i = M\, O(\zeta)$, with $M$ the ultraviolet cutoff)

\begin{eqnarray*}
 {{\kappa_{ir}(\tau)}\over {m_i\, c}} = {{{\dot \eta}^r_i(\tau)}\over
 {\sqrt{1 - {\dot {\vec \eta}}_i(\tau)}}} + {M\over
 {m_i}}\, O(\zeta),
 \end{eqnarray*}

\bea
 \eta_i\, {d\over {d\tau}}\,&& \Big[{{ {\dot
 \eta}^r_i(\tau)}\over {\sqrt{1 - {\dot {\vec \eta}}_i^2(\tau)
 }}}\, \Big(1 + 2\, (\Gamma_r^{(1)} + 2\,
 \phi_{(1)}) -\nonumber \\
 &&- {{n_{(1)} - \sum_c\, {\dot \eta}_i^c(\tau)\, [{\bar
 n}_{(1)(c)} + (\Gamma_c^{(1)} + 2\, \phi_{(1)})\, {\dot
 \eta}^c_i(\tau)]}\over {1 - {\dot {\vec \eta}}_i^2(\tau)}}
 \Big) +\nonumber \\
 &+& {{ {\bar n}_{(1)(r)}}\over {\sqrt{1 -
 {\dot {\vec \eta}}_i^2(\tau)}}}\,
 \Big]{|}_{{\vec \kappa}_i = {{m_ic\, {\dot {\vec \eta}}_i}\over
 {\sqrt{1 - {\dot {\vec \eta}}_i^2}}}}(\tau ,{\vec
 \eta}_i(\tau))\, \cir\nonumber \\
 &&{}\nonumber \\
 &&\cir\, \eta_i\,  {1\over {\sqrt{1 -
 {\dot {\vec \eta}}_i^2(\tau)}}}\, \Big[\sum_a\, {\dot
 \eta}^a_i(\tau)\, \Big({{\partial\, {\bar n}_{(1)(a)}}\over
 {\partial\, \eta^r_i}} +
  {{\partial\, (\Gamma_a^{(1)} + 2\, \phi_{(1)})}\over
 {\partial\, \eta_i^r}}\, {\dot
 \eta}^a_i(\tau)\Big) -\nonumber \\
 &-& {{\partial\, n_{(1)}}\over {\partial\,
 \eta^r_i}}\Big]{|}_{{\vec \kappa}_i = {{m_ic\, {\dot {\vec \eta}}_i}\over
 {\sqrt{1 - {\dot {\vec \eta}}_i^2}}}} (\tau ,{\vec \eta}_i(\tau)),
 \nonumber \\
 &&{}\nonumber \\
 \Rightarrow&&  {\ddot \eta}^r_i(\tau)\, \cir\, O(\zeta).
 \label{5.1}
 \eea

The last line is a consequence of the ultraviolet cutoff, which
allows the definition of the HPM linearization.

\medskip

Eqs(\ref{5.1}), being implied by Hamilton equations derived from a
standard relativistic particle Lagrangian (see the action (3.1) in
paper I), are equal to the geodesic equations for point-like scalar
particles notwithstanding these particles are dynamical and not test
objects (for spinning particles this is not true due to
spin-curvature couplings, see for instance Ref.\cite{27}).\bigskip

Eqs.(\ref{5.1}) may be rewritten by putting all the terms involving
the accelerations at the first member. Since
Eqs.(\ref{2.2})-(\ref{2.4}) and ${\vec \kappa}_i = {{m_ic\, {\dot
{\vec \eta}}_i}\over {\sqrt{1 - {\dot {\vec \eta}}_i^2}}} + M\,
O(\zeta)$ imply that the functions $f_{(1)}(\tau, \vec \sigma) =
\phi_{(1)}(\tau, \vec \sigma), n_{(1)}(\tau, \vec \sigma), {\bar
n}_{(1)(r)}(\tau, \vec \sigma)$ depend on $\eta^r_k(\tau)$ and
${\dot \eta}^r_k(\tau)$ with $k = 1,..,N$, for each of these
functions we have ${d\over {d\tau}}\, f_{(1)}(\tau, \vec \sigma) =
\sum_k^{1..N}\, \sum_s\, \Big({\dot \eta}^s_k(\tau)\, {{\partial\,
f_{(1)}(\tau, \vec \sigma)}\over {\partial\, \eta^s_k}} + {\ddot
\eta}^s_k(\tau)\, {{\partial\, f_{(1)}(\tau, \vec \sigma)}\over
{\partial\, {\dot \eta}^s_k}}\Big)$. Due to the result ${\ddot
\eta}^r_i(\tau) \, \cir\, O(\zeta)$, we get ${d\over {d\tau}}\,
f_{(1)}(\tau, \vec \sigma) = \sum_k^{1..N}\, \sum_s\, {\dot
\eta}^s_k(\tau)\, {{\partial\, f_{(1)}(\tau, \vec \sigma)}\over
{\partial\, \eta^s_k}} + O(\zeta^2)$. Therefore, the terms involving
the accelerations have the following expression

\bea
  &&\eta_i\, \Big( {{{\ddot \eta}^r_i(\tau)}\over {\sqrt{1 - {\dot {\vec \eta}}^2_i(\tau)}}}\,
 \Big[1 + 2\, (\Gamma_r^{(1)} + 2\, \phi_{(1)}) - {{n_{(1)} - \sum_c\,
 {\dot \eta}^c_i(\tau)\, {\bar n}_{(1)(c)} - \sum_c\, ({\dot \eta}^c_i(\tau))^2\,
 (\Gamma_c^{(1)} + 2\, \phi_{(1)})}\over
 {1 - {\dot {\vec \eta}}^2_i(\tau)}}\Big] +\nonumber \\
 &+& {{{\dot {\vec \eta}}_i(\tau) \cdot {\ddot {\vec \eta}}_i(\tau)}\over
 {(1 - {\dot {\vec \eta}}^2_i(\tau))^{3/2}}}\, {\bar n}_{(1)(r)} +
 {{{\dot \eta}^r_i(\tau)}\over {(1 - {\dot {\vec \eta}}^2_i(\tau))^{3/2}}}\,
 \Big[\sum_c\, {\ddot \eta}^c_i(\tau)\, \Big({\bar n}_{(1)(c)} + 2\,
 {\dot \eta}^c_i(\tau)\, (\Gamma_c^{(1)} + 2\, \phi_{(1)})\Big) +\nonumber \\
 &+& {\dot {\vec \eta}}_i(\tau) \cdot {\ddot {\vec \eta}}_i(\tau)\, \Big(1 +
 2\, (\Gamma_r^{(1)} + 2\, \phi_{(1)}) - 3\, {{n_{(1)} - \sum_c\,
 {\dot \eta}^c_i(\tau)\, {\bar n}_{(1)(c)} - \sum_c\, ({\dot \eta}^c_i(\tau))^2\,
 (\Gamma_c^{(1)} + 2\, \phi_{(1)})}\over {1 - {\dot {\vec \eta}}^2_i(\tau)}}
 \Big] +\nonumber \\
 &+&{1\over {\sqrt{1 - {\dot {\vec \eta}}^2_i(\tau)}}}\, \sum_u\,
 \sum_{j \not= i}\, {\ddot \eta}^u_j(\tau)\, \Big[{{\partial\, {\bar n}_{(1)(r)}}
 \over {\partial\, {\dot \eta}^u_j}} + {\dot \eta}^r_i(\tau)\,
 \Big(2\, {{\partial\, (\Gamma_r^{(1)} + 2\, \phi_{(1)})}\over
 {\partial\, {\dot \eta}^u_j}} -\nonumber \\
 &-&{1\over {1 - {\dot {\vec \eta}}^2_i(\tau)}}\, \Big[{{\partial\,
 n_{(1)}}\over {\partial\, {\dot \eta}^u_j}} + \sum_c\, {\dot
 \eta}^c_i(\tau)\, {{\partial\, {\bar n}_{(1)(c)}}\over {\partial\,
 {\dot \eta}^u_j}} + \sum_c\, ({\dot \eta}^c_i(\tau))^2\,
 {{\partial\, (\Gamma_c^{(1)} + 2\, \phi_{(1)})}\over {\partial\, {\dot \eta}^u_j}}
  \Big]\Big)\Big]\Big)(\tau ,{\vec \eta}_i(\tau))
 =\nonumber \\
 &&{}\nonumber \\
  &=& {{\eta_i}\over {\sqrt{1 - {\dot {\vec \eta}}^2_i(\tau)}}}\,
  \Big( {\ddot \eta}^r_i(\tau) + {{{\dot \eta}^r_i(\tau)\, {\dot
  {\vec \eta}}_i(\tau) \cdot {\ddot {\vec \eta}}_i(\tau)}\over {1 -
  {\dot {\vec \eta}}^2_i(\tau)}}
 \Big) + O(\zeta^2).
 \label{5.2}
 \eea

\medskip

As a consequence, after having rewritten the lapse and shift
functions in the form $n_{(1)} = {\check n}_{(1)} -
\partial_{\tau}\, {}^3{\cal K}_{(1)}$, ${\bar n}_{(1)(r)} = {\check
{\bar n}}_{(1)(r)} + \partial_r\, {}^3{\cal K}_{(1)}$, to display
their dependence on the inertial gauge variable ${}^3{\cal K}_{(1)}
= {1\over {\triangle}}\, {}^3K_{(1)}$, we get the following form of
the PM equations of motion of the particles

 \begin{eqnarray*}
  && {{\eta_i}\over {\sqrt{1 - {\dot {\vec \eta}}^2_i(\tau)}}}\,
  \Big( {\ddot \eta}^r_i(\tau) + {{{\dot \eta}^r_i(\tau)\, {\dot
  {\vec \eta}}_i(\tau) \cdot {\ddot {\vec \eta}}_i(\tau)}\over {1 -
  {\dot {\vec \eta}}^2_i(\tau)}}
 \Big) \cir
 \end{eqnarray*}

\bea
 &&\cir\,  {{\eta_i}\over {\sqrt{1 - {\dot {\vec \eta}}^2_i(\tau)}}}\,
 \Big(  - {{\partial\, {\check n}_{(1)}(\tau,
 {\vec \eta}_i(\tau))}\over {\partial\, \eta^r_i}} +
  {{{\dot \eta}^r_i(\tau)}\over {1 - {\dot {\vec
 \eta}}^2_i(\tau)}}\, \sum_u\, \Big[{\dot \eta}^u_i(\tau)\, {{\partial\,
 {\check n}_{(1)}}\over {\partial\,
 \eta^u_i}} + \sum_{j \not= i}\, {\dot \eta}^u_j(\tau)\, {{\partial\,
 {\check n}_{(1)} }\over {\partial\,
 \eta^u_j}} \Big](\tau, {\vec \eta}_i(\tau)) +\nonumber \\
 &+& \Big(\sum_u\, {\dot \eta}^u_i(\tau)\, \Big[{{\partial\, {\check
 {\bar n}}_{(1)(u)}}\over {\partial\, \eta^r_i}} - {{\partial\, {\check
 {\bar n}}_{(1)(r)}}\over {\partial\, \eta^u_i}}\Big] -
  \sum_{j \not= i}\, \sum_u\, {\dot \eta}^u_j(\tau)\, {{\partial\, {\check
 {\bar n}}_{(1)(r)}}\over {\partial\, \eta^u_j}} -\nonumber \\
 &-&{{{\dot \eta}^r_i(\tau)}\over {1 - {\dot {\vec \eta}}^2_i(\tau)}}\,
 \sum_u\, {\dot \eta}_i^u(\tau)\, \sum_s\, \Big[{\dot
 \eta}_i^s(\tau)\, {{\partial\, {\check {\bar n}}_{(1)(u)}}\over
 {\partial\, \eta^s_i}} + \sum_{j \not= i}\, {\dot \eta}_j^s(\tau)\,
 {{\partial\, {\check {\bar n}}_{(1)(u)}}\over {\partial\, \eta^s_j}}
 \Big]\Big)(\tau, {\vec \eta}_i(\tau)) +\nonumber \\
 &+& \Big(\sum_u\, ({\dot \eta}^u_i(\tau))^2\, {{\partial\, (\Gamma_u^{(1)} +
 2\, \phi_{(1)})}\over {\partial\, \eta^r_i}} -\nonumber \\
 &-& {\dot \eta}^r_i(\tau)\, \sum_u\, \Big[ {\dot \eta}^u_i(\tau)\,
 \Big(2\, {{\partial\, (\Gamma_r^{(1)} + 2\, \phi_{(1)})}\over {\partial\, \eta^u_i}}
 + \sum_c\, {{({\dot \eta}^c_i(\tau))^2}\over {1 - {\dot {\vec \eta}}^2_i(\tau)}}\,
 {{\partial\, (\Gamma_c^{(1)} + 2\, \phi_{(1)})}\over {\partial\, \eta^u_i}}
 \Big) +\nonumber \\
 &+& \sum_{j \not= i}\, {\dot \eta}^u_j(\tau)\,
 \Big(2\, {{\partial\, (\Gamma_r^{(1)} + 2\, \phi_{(1)})}\over {\partial\, \eta^u_j}}
 + \sum_c\, {{({\dot \eta}^c_i(\tau))^2}\over {1 - {\dot {\vec \eta}}^2_i(\tau)}}\,
 {{\partial\, (\Gamma_c^{(1)} + 2\, \phi_{(1)})}\over {\partial\, \eta^u_j}}
 \Big) \Big]\Big)(\tau, {\vec \eta}_i(\tau)) -\nonumber \\
 &-&  {{{\dot \eta}^r_i(\tau)}\over {1 - {\dot {\vec \eta}}^2_i(\tau)}}
 \Big[\partial_{\tau}^2{|}_{{\vec \eta}_i}\, {}^3{\cal K}_{(1)} +
 2\, \sum_s\, {\dot \eta}_i^s(\tau)\, {{\partial\, \partial_{\tau}{|}_{{\vec
 \eta}_i}\, {}^3{\cal K}_{(1)} }\over {\partial\, \eta_i^s}} +
 \sum_{su}\, {\dot \eta}_i^s(\tau)\,
 {\dot \eta}_i^u(\tau)\, {{\partial^2\, {}^3{\cal K}_{(1)}}\over
 {\partial\, \eta_i^u\, \partial\, \eta_i^s}}
 \Big](\tau, {\vec \eta}_i(\tau))
  \Big) + O(\zeta^2) =\nonumber \\
 &{\buildrel {def}\over =}& \eta_i\, {{{\cal F}^r_i(\tau |{\vec \eta}_i(\tau)| {\vec
 \eta}_{k \not= i}(\tau))}\over {m_i }} + O(\zeta^2).
 \label{5.3}
 \eea
\medskip

Since Eqs.(\ref{5.3}) imply $\eta_i\, {\dot {\vec \eta}}_i(\tau)
\cdot {\ddot {\vec \eta}}_i(\tau)\, \cir\,\eta_i\, m_i^{-1}\, (1 -
{\dot {\vec \eta}}_i^2(\tau))^{3/2}\, {\dot {\vec \eta}}_i(\tau)
\cdot {\vec {\cal F}}_i(\tau |{\vec \eta}_i(\tau)| {\vec \eta}_{k
\not= i}(\tau))$, the final form of the equation of motion of the
particles is

\bea
 m_i\, \eta_i\, {\ddot \eta}_i^r(\tau) &\cir& \eta_i\, \sqrt{1 -
 {\dot {\vec \eta}}_i^2(\tau)}\, \Big({\cal F}^r_i - {\dot
 \eta}_i^r(\tau)\, {\dot {\vec \eta}}_i(\tau) \cdot {\vec {\cal
 F}}_i\Big)(\tau |{\vec \eta}_i(\tau)| {\vec
 \eta}_{k \not= i}(\tau)) =\nonumber \\
 &{\buildrel {def}\over =}& \eta_i\, F^r_i(\tau |{\vec \eta}_i(\tau)| {\vec
 \eta}_{k \not= i}(\tau)).
 \label{5.4}
 \eea

\medskip

The second member of Eqs.(\ref{5.4}) defines the effective force
$F_i^r$ acting on particle $i$. It contains:\medskip

a) the contribution of the lapse function ${\check n}_{(1)}$, which
generalizes the Newton force;\hfill\break

b) the contribution of the shift functions ${\check {\bar
n}}_{(1)(r)}$, which gives the gravito-magnetic effects
\footnote{Since we have ${\dot {\vec \eta}}_i(\tau = ct) = {{{\vec
v}_i(t)}\over c}$ (see Eq.(\ref{5.5})), the term $\sum_a\,
{{v_i^a}\over c}\, \Big({{\partial\, {\check {\bar
n}}_{(1)(a)}}\over {\partial\, \eta^r}} - {{\partial\, {\check {\bar
n}}_{(1)(r)}}\over {\partial\, \eta^a}}\Big)$ in Eqs.(\ref{5.3}) is
proportional to ${{\vec v}\over c}\, \times {\vec B}_G$, where
${\vec B}_G$ is the gravito-magnetic field. It is of order
$({{v}\over c})^2$ as shown in Eq.(\ref{5.6}).};\hfill\break

c) the retarded contribution of GW's, described by the functions
$\Gamma_r^{(1)}$ of Eqs.(\ref{2.6}), whose contribution at the order
$O(\zeta)$ is given in Eq.(\ref{2.7});\hfill\break

d) the contribution of the volume element $\phi_{(1)}$ (${\tilde
\phi} = 1 + 6\, \phi_{(1)} + O(\zeta^2)$), always summed to the
GW's, giving forces of Newton type;\hfill\break

e) the contribution of the inertial gauge variable (the non-local
York time) ${}^3{\cal K}_{(1)} = {1\over {\triangle}}\,
{}^3K_{(1)}$.\hfill\break

\medskip

Let us remark that in  the gravitational case the regularization
with Grassmann-valued signs of the particle energies leads to
equations of motion for the particles of the type $\eta_i\, {\ddot
\eta}_i(\tau) \cir\eta_i....$ with $\eta_i^2 = 0$ with instantaneous
action-at-a-distance effects coming from the lapse $n_{(1)}$ and
shift ${\bar n}_{(1)(r)}$ functions and from the volume 3-element
${\tilde \phi}_{(1)} = 1 + 6\, \phi_{(1)}$. However the retardation
present in the solution (\ref{2.6}) for the GW's is not eliminable
and formally the equations of motion of the particles are of
integro-differential type for the N functions $\eta^r_i(\tau)$, $i =
1,..,N$. However, since ${\ddot \eta}_i^r(\tau) = O(\zeta)$, the
retardation effects in the GW's are pushed to higher HPM order as
shown in Eqs.(\ref{2.7}), so that at the lowest order we have
coupled differential equations for the particles. This shows that
our semi-classical approximation, obtained with our Grassmann
regularization, of a unspecified "quantum gravity" theory does not
take into account only a "one-graviton exchange diagram" but also
more complex structures already present at the tree level (namely
they are not radiative corrections) but showing up only at higher
HPM orders.

\medskip

By comparison in electro-magnetism the coupled equations of motion
for the charged particles and the transverse electro-magnetic field
in the radiation gauge, containing the Coulomb potential, allow to
find the Lienard-Wiechert solution \cite{28} for the transverse
vector potential with the no-incoming radiation condition. The
regularization with Grassmann-valued electric charges \cite{7}
implies equations of motion of the type ${\ddot \eta}_i(\tau) \cir
Q_i......$ with $Q^2_i = 0$, so that we get the elimination of
retardation effects, i.e. $Q_i\, {\dot \eta}_i(\tau)(\tau - |\vec
\sigma|) \cir Q_i\, {\dot \eta}_i(\tau)$. In this way the difference
between retarded and advanced (or symmetric) Lienard-Wiechert
solutions is killed and it is possible to identify the hidden common
action-at-a-distance part of such solutions and to express the
resulting semi-classical Lienard-Wiechert transverse
electro-magnetic fields in terms of the canonical variables ${\vec
\eta}_i(\tau)$, ${\vec \kappa}_i(\tau)$ of the particles. This
implies that the electro-magnetic retardation effects are to be
described as radiative corrections to the one-photon exchange
diagram of QED, which is replaced by a Cauchy problem with a well
defined action-at-a-distance potential. As a consequence, the final
equations for the particles are second order coupled ordinary
differential equations. To reduce the original phase space
containing the charged particles and the electro-magnetic field in
the radiation gauge, one added the second class constraints
identifying the transverse electro-magnetic field with the
Lienard-Wiechert solution and one evaluated the Dirac brackets. It
turned out that the resulting reduced phase space containing only
particles has a canonical basis spanned by new particle variables
${\hat \eta}^r_i$, ${\hat \kappa}_{ir}$ [interpretable as the old
ones $\eta^r_i$, $\kappa_{ir}$, (no more canonical with respect to
the Dirac brackets) dressed with a Coulomb cloud] with a mutual
action-at-a-distance interaction governed by the sum of the {\it
Coulomb and Darwin} potentials. In the rest-frame instant form of
dynamics \cite{4,29,30,31},one can find the expression of the
internal Poincare' generators:  $p^o = M c$, $p^r \approx 0$,
$j^{rs}$, $j^{\tau r} \approx 0$ (the potentials appear in the
energy $p^o$ and in the Lorentz boosts $J^{\tau r}$). Then, after
having gone from the canonical basis ${\hat \eta}^r_i$, ${\hat
\kappa}_{ir}$, to a canonical basis containing internal
center-of-mass variables $\vec \eta = {{\sum_i\, m_i\, {\vec {\hat
\eta}}_i}\over {\sum_i\, m_i}}$, $\vec p = \sum_i\, {\vec {\hat
\kappa}}_i$ and relative ones ${\vec \rho}_a$, ${\vec \pi}_a$, $a =
1,..,N-1$, (see Eqs. (2.1), (2.2) of Ref.\cite{30}), the rest-frame
conditions $p^r \approx 0$, $j^{\tau r} \approx 0$, eliminated the
collective variables: $\vec p \approx 0$, $\vec \eta \approx \vec
f({\vec \rho}_a, {\vec \pi}_a)$. As a consequence, in the reduced
phase space there are second order equations of motion only for the
relative variables ${\vec \rho}_a$, ${\vec \pi}_a$. \bigskip

In the HPM gravitational case the analogue of the Hamiltonian
action-at-a-distance Lienard-Wiechert transverse electromagnetic
fields are the action-at-a-distance fields $\phi_{(1)}(\tau, \vec
\sigma)$, $n_{(1)}(\tau, \vec \sigma)$, ${\bar n}_{(1)(r)}(\tau,
\vec \sigma)$, $\sigma_{(1)(a)(a)}{|}_{a \not= b}(\tau, \vec
\sigma)$, of Eqs.(\ref{2.2})-(\ref{2.5}) and the tidal fields
$R_{\bar a}(\tau, \vec \sigma)$ of Eqs.(\ref{2.7}). To find a
reduced phase space containing only particles (how it was done in
the electro-magnetic case), we have to add the second class
constraints which identify the tidal fields $R_{\bar a}(\tau, \vec
\sigma)$ and $\Pi_{\bar a}(\tau, \vec \sigma)$ with the HPM solution
of Eqs.(\ref{2.7}) and ((\ref{2.9}) in our family of 3-orthogonal
gauges. To get a set of second class constraints for the elimination
of the gravitational field we must add to the existing first class
constraints: 1) $\pi_{\tilde \phi}(\tau ,\vec \sigma) - {{c^3}\over
{12\pi\, G}}\, {}^3K_{(1)}(\tau ,\vec \sigma) \approx 0$ to the
super-Hamiltonian constraint written in the form $\tilde \phi(\tau
,\vec \sigma) - [1 + 6\, \phi_{(1)}(\tau ,\vec \sigma)] \approx 0$;
2) $\theta^i(\tau ,\vec \sigma) \approx 0$ to the super-momentum
constraints written in the form $\sigma_{(a)(b)}{|}_{a \not= b}(\tau
,\vec \sigma) - \sigma_{(1)(a)(b)}{|}_{a \not= b}(\tau ,\vec \sigma)
\approx 0$; 3) $n(\tau ,\vec \sigma) - n_{(1)}(\tau, \vec \sigma)
\approx 0$ to $\pi_n(\tau ,\vec \sigma) \approx 0$; 4) ${\bar
n}_{(a)}(\tau ,\vec \sigma) - {\bar n}_{(1)(a)}(\tau ,\vec \sigma)
\approx 0$ to $\pi_{{\bar n}_{(a)}}(\tau ,\vec \sigma) \approx 0$;
5) $R_{\bar a}(\tau ,\vec \sigma) - R_{(1)\bar a}(\tau ,\vec \sigma)
\approx 0$ and $\Pi_{\bar a}(\tau ,\vec \sigma) - \Pi_{(1)\bar
a}(\tau ,\vec \sigma) \approx 0$. Differently from the
electro-magnetic case, the evaluation of the Dirac brackets for the
reduced phase space containing only particles at the lowest HPM
order is trivial, because all the linearized solutions are sums of
terms proportional to $G\, m_i$, $i = 1,..,N$ and the gauge variable
${}^3{\cal K}_{(1)}(\tau, \vec \sigma)$ is a numerical function.
Therefore in the evaluation of the Dirac brackets of the variables
$\eta^r_i(\tau)$, $\kappa_{ir}(\tau)$, all the extra terms added to
the ordinary Poisson bracket are quadratic in $[G m_i\, G m_j]_{j
\not= i}$ and can be discarded being of order $O(\zeta^2)$ due to
the ultraviolet cutoff $m_i = M\, O(\zeta)$. As a consequence the
variables $\eta^r_i$, $\kappa_{ir}$, are also a canonical basis of
the Dirac brackets at the lowest order: the analogue of the
electro-magnetic Coulomb dressing is pushed to higher HPM order.

\subsection{The Center-of-Mass Problem in General Relativity and in
the HPM Linearization.}

As we have seen Eqs.(\ref{5.4}), with the gravitational waves
$\Gamma^{(1)}_r(\tau, \vec \sigma)$ given in Eqs.(\ref{2.7}), are
the final equations of motion for the particles in a reduced phase
space containing only particles described by the canonical variables
$\eta^r_i(\tau)$, $\kappa_{ir}(\tau)$. The forces appearing in
Eqs.(\ref{5.4}) are the gravitational analogue of the
electro-magnetic mutual interaction produced by the Coulomb and
Darwin potentials.

\medskip

As said in papers I and II, the instantaneous (non-Euclidean)
3-spaces $\Sigma_{\tau}$ are non-inertial rest frames of the
3-universe (an isolated system including the gravitational field in
the chosen family of space-times) due to the rest-frame conditions
${\hat P}^r_{ADM} \approx 0$. This remains true when the
gravitational field is expressed in terms of the particles by means
of the solution (\ref{2.7}). The gauge fixings to these three first
class constraints, eliminating the collective 3-variable of the
3-universe, are ${\hat J}^{\tau r}_{ADM}\, {\buildrel {def}\over =}\
{\hat K}^r_{ADM} \approx 0$. If we introduce the definition ${\hat
K}^r_{ADM}\, {\buildrel {def}\over =}\, - ({1\over c}\, {\hat
E}_{ADM})\, R^r_{ADM}$, with $R^r_{ADM}$ being the gravitational
analogue of the (neither covariant nor canonical) M$\o$ller 3-center
of energy, the conditions ${\hat K}^r_{ADM} \approx 0$ imply
$R^r_{ADM} \approx 0$. Therefore the 3-center of energy is put in
the origin of the 3-coordinates on $\Sigma_{\tau}$ (so that there is
only an external decoupled center of mass of the 3-universe which
can be built in terms of the ADM Poincare' generators and of the
embedding of the 3-space into space-time as shown in
Refs.\cite{4,29,30,31}) and is carried by the reference time-like
observer using the radar 4-coordinates. In the inertial rest frames
of special relativity \cite{4,29,30,31} \footnote{In these papers
there is a complete clarification of the center-of-mass problem in
special relativity.} this implies that the reference observer has to
be identified with the covariant non-canonical Fokker-Pryce center
of inertia of the isolated system. This is true also in the
non-inertial rest frames of special relativity as shown in
Ref.\cite{4}. As a consequence the isolated system can be seen as an
external (unobservable) decoupled (canonical non-covariant)
Newton-Wigner center of mass carrying a pole-dipole structure (the
relative motion of the components of the isolated system inside the
3-space $\Sigma_{\tau}$) identified by the rest mass $M$ and the
rest spin $S^r$ of the system. This is also true in the
gravitational case with $Mc^2$ replaced by ${\hat E}_{ADM}$ and with
the rest spin $S^r$ replaced by ${\hat J}^{rs}_{ADM}$. The reference
observer defining the radar 4-coordinates should be replaced also in
this case by an ADM Fokker-Pryce center of inertia (a non-geodesic
time-like observer corresponding to the the asymptotic inertial
observers existing in our class of space-times, whose spatial axes
are determined by the fixed stars of star catalogues).

\medskip

This is our way out from the the problem of the center of mass in
general relativity and of its world-line, a still  open problem in
generic space-times as can be seen from Refs. \cite{32,33} (and
Ref.\cite{34} for the PN approach). Usually, by means of some
supplementary condition, the center of mass is associated to the
monopole of a multipolar expansion of the energy-momentum of a small
body (see Ref.\cite{35} for the special relativistic case).

\medskip

Another open problem in general relativity is the replacement of the
particle canonical variables $\eta^r_i(\tau)$, $\kappa_{ir}(\tau)$
($i=1,..,N$) with relative canonical variables after the elimination
of the internal center of mass. In the special relativistic
electro-magnetic case (see Eqs.(2.1) of Ref.\cite{30}) one replaces
them with a naif (Newton mechanics oriented) canonical basis
$\eta^r_{+}(\tau)$, $\kappa_{+ r}(\tau)$, $\rho^r_a(\tau)$, $\pi_{a
r}(\tau)$ ($a=1,..,N-1$) with the relative variables in the rest
frame being Wigner spin-1 3-vectors. The rest-frame conditions imply
$\kappa_{+ r}(\tau) \approx 0$ and their gauge fixings $K^r\,
{\buildrel {def}\over =}\, - Mc\, R^r_{+} \approx 0$ allow to
express $\eta^r_{+}(\tau)$ as a function only of the relative
variables, $\eta^r_{+} \approx f^r({\vec \rho}_a, {\vec \pi}_a)$, as
shown in Refs. \cite{4,29,30,31}. In these papers it is also shown
how to reconstruct the world-lines of the particles by using the
embedding of the 3-space $\Sigma_{\tau}$ into the space-time.

\medskip

However relative variables do not exist in the non-Euclidean
3-spaces of curved space-times, where flat objects like ${\vec
r}_{ij}(\tau) = {\vec \eta}_i(\tau) - {\vec \eta}_j(\tau)$ have to
be replaced with a quantity proportional to the tangent vector to
the space-like 3-geodesics joining the two particles in the
non-Euclidean 3-space $\Sigma_{\tau}$ (see Ref. \cite{36} for an
implementation of this idea). This problem is another reason why
extended objects tend to be replaced with point-like multipoles,
which, however, do not span a canonical basis of phase space.

\medskip

However, at the HPM order in our family of space-times we can rely
on special relativity in the inertial rest frame by using the
asymptotic inertial frame (with the asymptotic background Minkowski
metric) like it has been done in the solution of the constraints and
of the wave equation for the tidal variables \footnote{In the
solutions $|{\vec \eta}_i(\tau) - {\vec \eta}_j(\tau)|$ is the
Euclidean 3-distance between the two particles, which differs by
quantities of order $O(\zeta)$ from the real non-Euclidean
3-distance on $\Sigma_{\tau}$ as shown in Eq.(\ref{3.2}).}. This
allows to define HPM collective and relative canonical variables for
the particles. Now the collective variables are eliminated with Eqs.
(\ref{2.11}) and (\ref{2.13}), which at the lowest order become
$p^r_{(1)} \approx 0$ and $j^{\tau r}_{(1)}\, {\buildrel {def}\over
=}\, - M_{(1)}c\, R^r \approx 0$ like in special relativity. If we
take into account also the terms of order $O(\zeta^2)$ in
Eqs.(\ref{2.11}) and (\ref{2.13}), we can find the first HPM
deviation (depending also on tidal terms) from the special
relativistic solution and the HPM equations of motion for the
relative variables implied by Eqs.(\ref{5.4}).
\bigskip

As an example let us consider the two-body case in presence of GW's,
namely without using the retarded solution (\ref{2.7}). Instead of
defining the overall collective variable of the two particles and of
GW's (like it was done in Ref.\cite{30}, where the transverse
electro-magnetic field was replacing GW's), let us define the naive
(Newton mechanics oriented) collective and relative canonical
variables only of the particles (with masses $m_1$ and $m_2$; $M=
m_1 + m_2$ and $\mu = {{m_1 m_2}\over M}$ are the total and reduced
masses respectively; we choose positive energy particles so that we
can replace the Grassmann variables $\eta_1$ and $\eta_2$ with their
mean value $< \eta_i > = + 1$, after having done the regularization,
as said in footnote 17 of paper I)

\bea
 &&{\vec \eta}_{12}(\tau) = {{m_1\, {\vec \eta}_1(\tau) + m_2\,
 {\vec \eta}_2}\over M},\qquad  {\vec \rho}_{12}(\tau) = {\vec
 \eta}_1(\tau) - {\vec \eta}_2(\tau),\nonumber \\
 &&{\vec \kappa}_{12}(\tau) = {\vec \kappa}_1(\tau) + {\vec
 \kappa}_2(\tau),\qquad {\vec \pi}_{12}(\tau) = {{m_2\, {\vec \kappa}_1(\tau) -
 m_1\, {\vec \kappa}_2(\tau)}\over M},\nonumber \\
 &&{}\nonumber \\
 &&{\vec \eta}_1(\tau) = {\vec \eta}_{12}(\tau) + {{m_2}\over M}\,
 {\vec \rho}_{12}(\tau),\qquad {\vec \eta}_2(\tau) = {\vec \eta}_{12}(\tau) - {{m_1}\over M}\,
 {\vec \rho}_{12}(\tau),\nonumber \\
 &&{\vec \kappa}_1(\tau) = {{m_1}\over M}\, {\vec \kappa}_{12}(\tau)
 + {\vec \pi}_{12}(\tau),\qquad {\vec \kappa}_2(\tau) = {{m_2}\over M}\, {\vec \kappa}_{12}(\tau)
 - {\vec \pi}_{12}(\tau).
 \label{5.5}
 \eea

All these quantities are of the type $a = a_{(o)} + a_{(1)} +
a_{(2)} + O(\zeta^3)$: the coordinates start with $O(1)$ terms,
while the momenta start with $O(\zeta)$ terms due to the ultravilet
cutoff.\medskip

The rest-frame condition (\ref{2.11}) implies ${\vec
\kappa}_{(1)12}(\tau) \approx 0$ and a certain expression for ${\vec
\kappa}_{(2)12}(\tau)$ in terms of GW's, ${\vec
\eta}_{(o)12}(\tau)$, ${\vec \rho}_{(o)12}(\tau)$, ${\vec
\pi}_{(1)12}(\tau)$ and of the non-local York time.\medskip

Instead the condition (\ref{2.13}) determines ${\vec
\eta}_{(o)12}(\tau)$ in terms of ${\vec \rho}_{(o)12}(\tau)$ and
${\vec \pi}_{(1)12}(\tau)$

\bea
 {\vec \eta}_{(o)12}(\tau) &\approx & - A_{(o)}(\tau)\, \, {\vec
 \rho}_{(o)12}(\tau),\nonumber \\
 &&{}\nonumber \\
 &&A_{(o)}(\tau) = {{ {{m_2}\over M}\, \sqrt{m_1^2\, c^2 +
 {\vec \pi}^2_{(1)12}(\tau)} -  {{m_1}\over M}\, \sqrt{m_2^2\, c^2 +
 {\vec \pi}^2_{(1)12}(\tau)}}\over {\sqrt{m_1^2\, c^2 +
 {\vec \pi}^2_{(1)12}(\tau)} + \sqrt{m_2^2\, c^2 +
 {\vec \pi}^2_{(1)12}(\tau)} }}.
 \nonumber \\
  &&{}
 \label{5.6}
 \eea

\noindent It vanishes in the equal mass case. As a consequence
${\vec \kappa}_{(2)12}(\tau)$ depends only on relative
variables.\medskip

Then ${\vec \eta}_{(1)12}(\tau)$ is determined in terms of GW's, of
particle relative variables and of both the local and non-local York
times: ${\vec \eta}_{(1)12}(\tau) \approx {\vec
f}_{(1)}(\tau)[rel.var.]$. As a consequence the particle
3-coordinates depend only on relative variables

\bea
 {\vec \eta}_1(\tau) &=& {\vec \eta}_{(o)1}(\tau) + {\vec
 \eta}_{(1)1}(\tau) \approx \nonumber \\
 &\approx& \Big({{m_2}\over M} - A_{(o)}(\tau)\Big)\, {\vec
 \rho}_{(o)12}(\tau) + {{m_2}\over M}\, {\vec \rho}_{(1)12}(\tau) +
 {\vec f}_{(1)}(\tau)[rel.var.],\nonumber \\
 {\vec \eta}_2(\tau) &=& {\vec \eta}_{(o)2}(\tau) + {\vec
 \eta}_{(1)2}(\tau) \approx \nonumber \\
 &\approx& - \Big({{m_1}\over M} + A_{(o)}(\tau)\Big)\, {\vec
 \rho}_{(o)12}(\tau) - {{m_1}\over M}\, {\vec \rho}_{(1)12}(\tau) +
 {\vec f}_{(1)}(\tau)[rel.var.],
 \label{5.7}
 \eea

\noindent and the world-lines can be reconstructed by using the
embedding of 3-spaces into space-time: $x^{\mu}_i(\tau) =
z^{\mu}(\tau, {\vec \eta}_i(\tau))$  \footnote{We have the following
reconstruction of the particle world-lines in the preferred adapted
world 4-coordinate system defined in the Introduction (${\tilde
\eta}^r_i(t) = \eta^r_i(\tau = ct)$)\hfill\break
\begin{eqnarray*}
 x^{\mu}_i(\tau) = z^{\mu}(\tau, \vec \eta(\tau)) = {\tilde
 x}^{\mu}_i(t) = x^{\mu}_o + \epsilon^{\mu}_{\tau}\, \tau +
 \epsilon^{\mu}_r\, \eta^r_i(\tau)
 = x^{\mu}_o + \epsilon^{\mu}_{\tau}\, c\, t +
 \epsilon^{\mu}_r\, {\tilde \eta}_i^r(t).
 \end{eqnarray*}}.\medskip

At the lowest order the velocities ${\dot {\vec \eta}}_{(o)i}(\tau)$
depend on ${\dot {\vec \rho}}_{(o)12}(\tau)$, ${\vec
\rho}_{(o)12}(\tau)$, and on the function $A_{(o)}(\tau)$ and its
$\tau$-derivative.

The ADM energy and angular momentum of Eqs.(\ref{2.10}) and
(\ref{2.12}) become

\begin{eqnarray*}
 {\hat E}_{ADM} &\approx& c\, \Big(\sqrt{m_1^2\, c^2 +
 {\vec \pi}^2_{(1)12}(\tau)} + \sqrt{m_2^2\, c^2 +
 {\vec \pi}^2_{(1)12}(\tau)}\Big) -\nonumber \\
 &-& {G\over {c^3}}\, {{  {\vec \pi}^2_{(1)12}(\tau)}\over
 {|{\vec \rho}_{(o)12}(\tau)|}}\, \Big({{\sqrt{m_2^2\, c^2 +
 {\vec \pi}^2_{(1)12}(\tau)}}\over {\sqrt{m_1^2\, c^2 +
 {\vec \pi}^2_{(1)12}(\tau)}}} + {{\sqrt{m_1^2\, c^2 +
 {\vec \pi}^2_{(1)12}(\tau)} }\over {\sqrt{m_2^2\, c^2 +
 {\vec \pi}^2_{(1)12}(\tau)}}}\Big) -\nonumber \\
 &-& {1\over 2}\, {\vec \pi}^2_{(1)12}(\tau)\,
 \sum_c\, {{\partial_c^2}\over {\triangle}}\,
 \Big({{\Gamma_c^{(1)}(\tau, {\vec \eta}_1(\tau))}\over {\sqrt{m_1^2\, c^2 +
 {\vec \pi}^2_{(1)12}(\tau)}}} + {{\Gamma_c^{(1)}(\tau,
 {\vec \eta}_2(\tau))}\over {\sqrt{m_2^2\, c^2 +
 {\vec \pi}^2_{(1)12}(\tau)}}}\Big) -\nonumber \\
 &-&\sum_c\, (\pi^c_{(1)12}(\tau))^2\,
 \Big({{\Gamma_c^{(1)}(\tau, {\vec \eta}_1(\tau))}\over {\sqrt{m_1^2\, c^2 +
 {\vec \pi}^2_{(1)12}(\tau)}}} + {{\Gamma_c^{(1)}(\tau,
 {\vec \eta}_2(\tau))}\over {\sqrt{m_2^2\, c^2 +
 {\vec \pi}^2_{(1)12}(\tau)}}}\Big) -\nonumber \\
 &-& {G\over {c^2}}\, {{\sqrt{m_1^2\, c^2 +
 {\vec \pi}^2_{(1)12}(\tau)}\, \sqrt{m_2^2\, c^2 +
 {\vec \pi}^2_{(1)12}(\tau)}}\over {|{\vec \rho}_{(o)12}|}} -
 {G\over {c^2}}\, \sum_{rs}\, \pi^r_{(1)12}(\tau)\,
 \pi^s_{(1)12}(\tau)\, \Big(\frac{7}{2}\,{{\, \delta^{rs}}\over
 {|{\vec \rho}_{(o)12}(\tau)|}} +\nonumber \\
 &+& {1\over {2}}\,
 {{\rho^r_{(o)12}(\tau)\,\rho^s_{(o)12}(\tau)}\over
 {|{\vec \rho}_{(o)12}(\tau)|^3}}\Big) +\nonumber \\
 &+& {{c^4}\over {16\pi\, G}}\, \sum_{\bar a\bar b}\, \int d^3\sigma\,
  \Big[\partial_{\tau}\, R_{\bar a}\, M_{\bar a\bar b}\,
 \partial_{\tau}\, R_{\bar b} + \sum_a\, \partial_a\,
 R_{\bar a}\, M_{\bar a\bar b}\, \partial_a\, R_{\bar b}\Big] \,
 \Big)(\tau, \vec \sigma) + Mc\, O(\zeta^3),
 \end{eqnarray*}

 \begin{eqnarray*}
 {\hat J}^{rs}_{ADM} &\approx& \Big(\rho^r_{(o)12}(\tau) +
 \rho^r_{(1)12}(\tau)\Big)\, \pi^s_{(1)12}(\tau) - \Big(\rho^s_{(o)12}(\tau)
 + \rho^s_{(1)12}(\tau)\Big)\, \pi^r_{(1)12}(\tau) +\nonumber \\
 &+& \rho^r_{(o)12}(\tau)\, \Big(\pi^s_{(2)12}(\tau) - A_{(o)}(\tau)\,
 \kappa^s_{(2)12}(\tau)\Big) - \rho^s_{(o)12}(\tau)\, \Big(\pi^r_{(2)12}(\tau)
 - A_{(o)}(\tau)\, \kappa^r_{(2)12}(\tau)\Big) +\nonumber \\
 &+& \sqrt{m_1^2\, c^2 + {\vec \pi}^2_{(1)12}(\tau)}\, ({{m_2}\over M} - A_{(o)}(\tau))\,
 \Big(\rho^r_{(o)12}(\tau)\, {{\partial}\over {\partial\, \eta^s_{(o)1}}}
 - \rho^s_{(o)12}(\tau)\, {{\partial}\over {\partial\, \eta^r_{(o)1}}}\Big)\,
 {}^3{\cal K}_{(1)}(\tau, {\vec \eta}_{(o)1}(\tau))
 -\nonumber \\
 &-& \sqrt{m_2^2\, c^2 + {\vec \pi}^2_{(1)12}(\tau)}\, ({{m_1}\over M} + A_{(o)}(\tau))\,
 \Big(\rho^r_{(o)12}(\tau)\, {{\partial}\over {\partial\, \eta^s_{(o)2}}}
 - \rho^s_{(o)12}(\tau)\, {{\partial}\over {\partial\, \eta^r_{(o)2}}}\Big)\,
 {}^3{\cal K}_{(1)}(\tau, {\vec \eta}_{(o)2}(\tau)) -
 \end{eqnarray*}

\begin{eqnarray*}
 &-& 2\, \sum_u\, \pi^u_{(1)12}(\tau)\nonumber \\
 &&\Big[({{m_2}\over M} - A_{(o)}(\tau))\,
 \Big(\rho^r_{(o)12}(\tau)\, {{\partial}\over {\partial\, \eta^s_{(o)1}}}
 - \rho^s_{(o)12}(\tau)\, {{\partial}\over {\partial\, \eta^r_{(o)1}}}\Big)\,
 {{\partial_u}\over {\triangle}}\, \Big(\Gamma_u^{(1)} - {1\over 4}\, \sum_c\,
 {{\partial_c^2}\over {\triangle}}\, \Gamma_c^{(1)}\Big)(\tau, {\vec \eta}_{(o)1}(\tau))
 -\nonumber \\
 &-&({{m_1}\over M} + A_{(o)}(\tau))\,
  \Big(\rho^r_{(o)12}(\tau)\, {{\partial}\over {\partial\, \eta^s_{(o)2}}}
 - \rho^s_{(o)12}(\tau)\, {{\partial}\over {\partial\, \eta^r_{(o)2}}}\Big)\,
 {{\partial_u}\over {\triangle}}\, \Big(\Gamma_u^{(1)} - {1\over 4}\, \sum_c\,
 {{\partial_c^2}\over {\triangle}}\, \Gamma_c^{(1)}\Big)(\tau, {\vec \eta}_{(o)2}(\tau))
 \Big] +\nonumber \\
 &+&2\, \Big[\pi^r_{(1)12}(\tau)\, \sum_i\, {{\partial_s}\over
 {\triangle}}\, \Big(\Gamma_s^{(1)} - {1\over 4}\, \sum_c\,
 {{\partial_c^2}\over {\triangle}}\, \Gamma_c^{(1)}\Big)(\tau,
 {\vec \eta}_{(o)i}(\tau)) -\nonumber \\
 &-& \pi^s_{(1)12}(\tau)\, \sum_i\, {{\partial_r}\over
 {\triangle}}\, \Big(\Gamma_r^{(1)} - {1\over 4}\, \sum_c\,
 {{\partial_c^2}\over {\triangle}}\, \Gamma_c^{(1)}\Big)(\tau,
 {\vec \eta}_{(o)i}(\tau)) \Big] -
 \end{eqnarray*}

\bea
 &-& {{c^3}\over {8\pi\, G}}\, \int d^3\sigma\,
 \Big[\sum_{\bar a\bar b}\, (\sigma^r\, \partial_s
 - \sigma^s\, \partial_r)\, R_{\bar a}\, M_{\bar a\bar b}\,
 \partial_{\tau}\, R_{\bar b} + 2\, {}^3K_{(1)}\, \partial_r\,
 \partial_s\, (\Gamma^{(1)}_s - \Gamma_r^{(1)}) +\nonumber \\
 &+& 2\, (\partial_{\tau}\, \Gamma_r^{(1)} + \partial_{\tau}\, \Gamma_s^{(1)} -
 {1\over 2}\, \sum_c\, {{\partial_c^2}\over {\triangle}}\,
 \partial_{\tau}\, \Gamma_c^{(1)})\,
 {{\partial_r\, \partial_s}\over {\triangle}}\, (\Gamma_s^{(1)} - \Gamma_r^{(1)})
 \Big](\tau, \vec \sigma) + Mc\, L\, O(\zeta^3).\nonumber \\
 &&{}
  \label{5.8}
  \eea

The dependence of the rest-frame energy on the York time is pushed
to order $O(\zeta^3)$. Instead the rest-frame angular momentum (or
spin) depends on both local and non-local York time.\medskip

Let us now consider the equations of motion (\ref{5.4}). By using
Eqs.(\ref{5.5}) and the previous results, they can be written in the
following form

\bea
 M\, {\ddot \eta}^r_{12}(\tau) &\approx& M\, \Big[{{d^2}\over {d\tau^2}}\,
 \Big(A_{(o)}(\tau)\, \rho^r_{(o)12}(\tau) + f_{(1)}(\tau)[rel.var.]
 \Big)\Big] \cir \nonumber \\
 &\cir& F^r_{1}(\tau| {\vec \eta}_{(o)1}(\tau)| {\vec \eta}_{(o)2}(\tau))
 + F^r_{2}(\tau| {\vec \eta}_{(o)2}(\tau)| {\vec
 \eta}_{(o)1}(\tau)),\nonumber \\
 &&{}\nonumber \\
 \mu\, {\ddot \rho}^r_{12}(\tau) &\approx& \mu\, \Big({\ddot \rho}^r_{(o)12}(\tau)
 + {\ddot \rho}^r_{(1)12}(\tau)\Big) \cir {{m_2}\over M}\,
 F^r_{1}(\tau| {\vec \eta}_{(o)1}(\tau)| {\vec \eta}_{(o)2}(\tau))
 - {{m_1}\over M}\,  F^r_{2}(\tau| {\vec \eta}_{(o)2}(\tau)| {\vec
 \eta}_{(o)1}(\tau)).\nonumber \\
 &&{}
 \label{5.9}
 \eea

Since the forces (depending on ${\vec \rho}_{(o)12}(\tau)$ and
${\dot {\vec \eta}}_{(o)i}(\tau)$) are of order $O(\zeta)$, since
${d\over {d\tau}}\, A_{(o)}(\tau)$ is of order $O(\zeta)$ involving
${\dot \pi}^r_{(1)}(\tau)$ and since ${\ddot \rho}^r_{(1)12}(\tau)$
is of higher order with respect to ${\ddot \rho}^r_{(o)12}(\tau)$,
Eqs.(\ref{5.9}) imply the following equations of motion

\bea
 \mu\, {\ddot \rho}^r_{(o)12}(\tau) &\cir& {{m_2}\over M}\,
 F^r_{1}(\tau| {\vec \eta}_{(o)1}(\tau)| {\vec \eta}_{(o)2}(\tau))
 - {{m_1}\over M}\,  F^r_{2}(\tau| {\vec \eta}_{(o)2}(\tau)| {\vec
 \eta}_{(o)1}(\tau)),\nonumber \\
 &&{}\nonumber \\
 &&{}\nonumber \\
 M&& \Big[{\ddot \eta}^r_{(1)12}(\tau) - {{d^2}\over {d\tau^2}}\, \Big(
  A_{(o)}(\tau)\, \rho^r_{(o)12}(\tau) \Big)\Big] \cir\nonumber \\
 &\cir& F^r_{1}(\tau| {\vec \eta}_{(o)1}(\tau)| {\vec \eta}_{(o)2}(\tau))
 + F^r_{2}(\tau| {\vec \eta}_{(o)2}(\tau)| {\vec
 \eta}_{(o)1}(\tau)).\nonumber \\
 &&{}
 \label{5.10}
 \eea

While the first equation determines the relative motion, the second
equation is a consistency relation connecting the motion of the
particle 3-center of mass to GW's and expressing the fact that the
overall 3-center mass of the isolated system "particles plus GW's"
is in free motion.\medskip

If the previous discussion is formulated in the reduced phase space
containing only particles after the elimination of GW's, by forcing
them to coincide with the solution (\ref{2.7}), nothing changes
except that the second of Eqs.(\ref{5.10}) should become the time
derivative of the rest-frame condition ${\vec \eta}_{(1)12}(\tau)
\approx {\vec f}_{(1)}(\tau)[rel.var.]$.

\subsection{The PN Expansion at all Orders in the Slow Motion
Limit.}

Due to our ultraviolet cutoff $M$ we have been able to obtain a HPM
linearization without never making PN expansions. However, if all
the particles are contained in a compact set of radius $l_c$, we can
add the slow motion condition in the form $\sqrt{\epsilon} = {v\over
c} \approx \sqrt{{{R_{m_i}}\over {l_c}}}$, $i=1,..N$ ($R_{m_i} =
{{2\, G\, m_i}\over {c^2}}$ is the gravitational radius of particle
$i$) with $l_c \geq R_M$ and $\lambda >> l_c$ (see the
Introduction). In this case we can do the PN expansion of
Eqs.(\ref{5.4}).
\medskip

Since we have $\tau = c\, t$, we make the following change of
notation

\bea
 &&{\vec \eta}_i(\tau) = {\vec {\tilde \eta}}_i(t),\qquad {\vec
 v}_i(t) = {{d {\vec {\tilde \eta}}_i(t)}\over {dt}},\qquad {\vec
 a}_i(t) = {{d^2 {\vec {\tilde \eta}}_i(t)}\over {dt^2}},\nonumber \\
 &&\qquad {\dot {\vec \eta}}_i(\tau) = {{{\vec v}_i(t)}\over c},\qquad
 {\ddot {\vec \eta}}_i(\tau) = {{{\vec a}_i(t)}\over {c^2}}.
 \label{5.11}
 \eea

For the non-local York time we use the notation ${}^3{\tilde {\cal
K}}_{(1)}(t, \vec \sigma) = {}^3{\cal K}_{(1)}(\tau, \vec \sigma) =
{1\over {\triangle}}\, {}^3K_{(1)}(\tau, \vec \sigma) = {1\over
{\triangle}}\, {}^3{\tilde K}(t, \vec \sigma)$.

\bigskip

By using $(1 - x)^{-1/2} = \sum_{k=0}^{\infty}\, (-)^k\, {{(2k -
1)!!}\over {(2k)!!}}\, x^k$ (valid for $x^2 < 1$), Eqs.(\ref{2.7})
and (\ref{2.2})-(\ref{2.4}) can be written in the form (here
$A_{[(k)]} = O(\epsilon^{k/2} = ({v\over c})^k)$ is of order
${k\over 2}\, PN$)

\begin{eqnarray*}
 \eta_i\, \Gamma_r^{(1)}(\tau, {\vec \eta}_i(\tau)) &=& - {{2
 G}\over {c^2}}\, \sum_b\, {\tilde M}^{-1}_{rs}({\vec {\tilde
 \eta}}_i(t))\,\, \sum_{j \not= i}\, \eta_j\, m_j\, \sum_{uv}\,
 {\cal P}_{ssuv}({\vec {\tilde \eta}}_i(t))\, {{{{v_j^u(t)}\over c}\,
 {{v_j^v(t)}\over c}}\over {\sqrt{1 - ({{{\vec v}_j(t)}\over c})^2}}}
 \nonumber \\
 &&\Big[|{\vec {\tilde \eta}}_i(t) -
 {\vec {\tilde \eta}}_j(t)|^{- 1} + \sum_{m=1}^{\infty}\, {1\over {(2m)!}}\,
 \Big({{{\vec v}_j(t)}\over c} \cdot {{\partial}\over {\partial\,
 {\vec {\tilde \eta}}_i}}\Big)^{2m}\, |{\vec {\tilde \eta}}_i(t) -
 {\vec {\tilde \eta}}_j(t)|^{2m - 1}\Big] =\nonumber \\
 &=& \eta_i\, {\tilde \Gamma}^{(1)}_r(t, {\vec {\tilde \eta}}_i(t))
 = \eta_i\, \sum_{j \not= i}\, \eta_j\, {{G\, m_j}\over {c^2}}\,
 \sum_{k=1}^{\infty}\, {\hat \Gamma}^{(1)}_{ r j[(2k)]}(t,
 {\vec {\tilde \eta}}_i(t) |{\vec {\tilde \eta}}_j(t)),\nonumber \\
 &&{}\nonumber \\
 &&{\hat \Gamma}^{(1)}_{j r [(2k)]}(t, {\vec {\tilde \eta}}_i(t)
 |{\vec {\tilde \eta}}_j(t)) =  \sum_s\, {\tilde M}^{-1}_{rs}({\vec
 {\tilde \eta}}_i(t))\, \sum_{uv}\,
 {\cal P}_{ssuv}({\vec {\tilde \eta}}_i(t))\, {{v_j^u(t)}\over c}\,
 {{v_j^v(t)}\over c} \nonumber \\
 &&\sum_{h = 1}^k\, {{(-)^{h}(2h-3)!!}\over {(2h-2)!!\, [2 (k-h)]!}}\,
 \Big({{{\vec v}_j(t)}\over c}\Big)^{2\, (h-1)}\,
 \Big({{{\vec v}_j(t)}\over c} \cdot {{\partial}\over {\partial\,
 {\vec {\tilde \eta}}_i}}\Big)^{2 (k - h)}\, |{\vec {\tilde \eta}}_i(t) -
 {\vec {\tilde \eta}}_j(t)|^{2 (k - h) - 1},
 \end{eqnarray*}

\begin{eqnarray*}
 \eta_i\, \phi_{(1)}(\tau, {\vec \eta}_i(\tau)) &=&
 \eta_i\, \sum_{j \not= i}\, \eta_j\, {{G\, m_j}\over
 {c^2}}\, \Big({1\over {2\, |{\vec {\tilde \eta}}_i(t) -
 {\vec {\tilde \eta}}_j(t)|}} + \sum_{k=1}^{\infty}\,
 {\hat \phi}_{(1) j [(2k)]}(t, {\vec {\tilde \eta}}_i(t)
 | {\vec {\tilde \eta}}_j(t))\Big),\nonumber \\
 &&{}\nonumber \\
 &&{\hat \phi}_{(1) j [(2k)]}(t, {\vec {\tilde \eta}}_i(t)
 | {\vec {\tilde \eta}}_j(t)) = (-)^k\,{{(2k - 1)!!}\over {(2k)!!}}\,
 {{({{{\vec v}_j(t)}\over c})^{2k}}\over {2\, |{\vec {\tilde \eta}}_i(t) -
 {\vec {\tilde \eta}}_j(t)|}} -\nonumber \\
 &-& \int d^3\sigma\, {{\sum_r\, \partial_r^2\,
 {\hat \Gamma}^{(1)}_{j r [(2k)]}(t, \vec \sigma |{\vec
 {\tilde \eta}}_j(t))}\over {16\, |{\vec {\tilde \eta}}_i(t) -
 \vec \sigma|}},
 \end{eqnarray*}

 \begin{eqnarray*}
 \eta_i\, n_{(1)}(\tau, {\vec \eta}_i(\tau)) &=& \eta_i\, \Big[
 - \partial_{\tau}{|}_{{\vec \eta}_i}\, {}^3{\cal K}_{(1)}(\tau, {\vec \eta}_i(\tau))
 + {\check n}_{(1)}(\tau, {\vec \eta}_i(\tau))\Big] =\nonumber \\
 &=&\eta_i\, \Big[ - {1\over c}\,\partial_t{|}_{{\vec {\tilde \eta}}_i}\,
 {}^3{\tilde {\cal K}}_{(1)}(t, {\vec {\tilde \eta}}_i(t)) - \sum_{j \not= i}\,
 \eta_j\, {{G\, m_j}\over {c^2}}\, {1\over {|{\vec {\tilde \eta}}_i(t) -
 {\vec {\tilde \eta}}_j(t)|}} +\nonumber \\
 &+&\sum_{j \not= i}\,
 \eta_j\, {{G\, m_j}\over {c^2}}\, \sum_{k=1}^{\infty}\, {\hat
 n}_{(1) j [(2k)]}(t, {\vec {\tilde \eta}}_i(t) | {\vec {\tilde
 \eta}}_j(t))\Big],\nonumber \\
 &&{}\nonumber \\
 &&{\hat n}_{(1) j [(2)]}(t, {\vec {\tilde \eta}}_i(t) | {\vec {\tilde
 \eta}}_j(t)) = -{{{3\over 2}\,  ({{{\vec v}_j(t)}\over c})^2 }\over {|{\vec {\tilde \eta}}_i(t) -
 {\vec {\tilde \eta}}_j(t)|}},\nonumber \\
 &&{\hat n}_{(1) j [(2k)]}(t, {\vec {\tilde \eta}}_i(t) | {\vec {\tilde
 \eta}}_j(t)) =-\Big((-)^k\, {{(2k-1)!!}\over {(2k)!!}} +
 (-)^{k-1}\,{{(2k-3)!!}\over {(2k-2)!!}}\Big)\times\nonumber \\
 &&\qquad\qquad \times{{  ({{{\vec v}_i(t)}\over c})^{2k} }\over {|{\vec {\tilde \eta}}_i(t) -
 {\vec {\tilde \eta}}_j(t)|}},\qquad k \geq 2,
 \end{eqnarray*}

 \begin{eqnarray*}
 \eta_i\, {\bar n}_{(1)(r)}(\tau, {\vec \eta}_i(\tau)) &=& \eta_i\, \Big[
  \partial_r\, {}^3{\cal K}_{(1)}(\tau, {\vec \eta}_i(\tau)) +
  {\check {\bar n}}_{(1)(r)}(\tau, {\vec \eta}_i(\tau)\Big]
  =\nonumber \\
 &=&\eta_i\, \Big[ \partial_r\, {}^3{\tilde {\cal K}}_{(1)}(t,
 {\vec {\tilde \eta}}_i(t)) + \sum_{j \not= i}\,
 \eta_j\, {{G\, m_j}\over {c^2}}\,
 {\hat {\bar n}}_{(1)(r) j [(1)]}(t, {\vec {\tilde \eta}}_i(t) | {\vec {\tilde
 \eta}}_j(t) +\nonumber \\
 &+& \sum_{j \not= i}\, \eta_j\, {{G\, m_j}\over {c^2}}\, \sum_{k=1}^{\infty}\,
 \Big({\hat {\bar n}}_{(1)(r) j [(2k)]} + {\hat {\bar n}}_{(1)(r) j [(2k+1)]}(t,
 {\vec {\tilde \eta}}_i(t) | {\vec {\tilde \eta}}_j(t))\Big)\Big],
 \end{eqnarray*}

 \bea
 &&{\hat {\bar n}}_{(1)(r) j [(1)]}(t, {\vec {\tilde \eta}}_i(t) | {\vec {\tilde
 \eta}}_j(t)) = -{1\over {|{\vec {\tilde \eta}}_i(t) - {\vec {\tilde \eta}}_j(t)|  }}\,
 \Big(   \frac{7}{2}\,{{v^r_j(t)}\over c}  +\nonumber \\
 &-& \frac{1}{2}{{ ({\tilde \eta}^r_i(t) - {\tilde \eta}^r_j(t))\, {{{\vec v}_j(t)}\over
 c}\, \cdot ({\vec {\tilde \eta}}_i(t) - {\vec {\tilde \eta}}_j(t))}\over
 {|{\vec {\tilde \eta}}_i(t) - {\vec {\tilde \eta}}_j(t)|^2}}\Big),
 \nonumber \\
  &&{\hat {\bar n}}_{(1)(r) j [(2k+1)]}(t, {\vec {\tilde \eta}}_i(t) | {\vec {\tilde
 \eta}}_j(t)) = {{(2k - 1)!!}\over {(2k)!!}}\,
 {{(-)^{k+1}}\over {|{\vec {\tilde \eta}}_i(t) - {\vec {\tilde \eta}}_j(t)|  }}\,
 \Big(  \frac{7}{2} {{v^r_j(t)}\over c}  +\nonumber \\
 &-& \frac{1}{2}{{ ({\tilde \eta}^r_i(t) - {\tilde \eta}^r_j(t))\, {{{\vec v}_j(t)}\over
 c}\, \cdot ({\vec {\tilde \eta}}_i(t) - {\vec {\tilde \eta}}_j(t))}\over
 {|{\vec {\tilde \eta}}_i(t) - {\vec {\tilde \eta}}_j(t)|^2}}\Big)\,
 \Big({{{\vec v}_i(t)}\over c}\Big)^{2k},\nonumber \\
 &&{\hat {\bar n}}_{(1)(r) j [(2k)]}(t, {\vec {\tilde \eta}}_i(t) | {\vec {\tilde
 \eta}}_j(t)) = -\int {{d^3\sigma_1}\over {4\pi\, |{\vec {\tilde \eta}}_i(t) -
 {\vec \sigma}_1|}}\, {{\partial_{1r}\, \partial_t{|}_{{\vec {\tilde \eta}}_j}}\over c}\,
 \Big[2\, {\hat \Gamma}^{(1)}_{j r [(2k)]}(t, {\vec \sigma}_1
 |{\vec {\tilde \eta}}_j(t)) -\nonumber \\
 &-& \int d^3\sigma_2\, {{\sum_c\, \partial_{2c}^2\,
 {\hat \Gamma}^{(1)}_{j c [(2k)]}(t, {\vec \sigma}_2
 |{\vec {\tilde \eta}}_j(t)) }\over {8\pi\, |{\vec \sigma}_1
 - {\vec \sigma}_2|}}\Big].
 \label{5.12}
 \eea

\noindent All the quantities are even in ${v\over c}$ except the
shift functions which have both odd and even terms. As a
consequence, ${}^4g_{(1)\tau\tau}$ and ${}^4g_{(1)rs}$ are even in
${v\over c}$, but this is not true for ${}^4g_{(1)\tau r}$.

\bigskip

By using Eqs.(\ref{5.12}), Eqs.(\ref{5.4}) can be written in the
following form after having being multiplied by $c^2$ and $m_i$ (we
use $(1 - x)^{-1} = \sum_{h=0}^{\infty}\, x^h$, valid for $x < 1$)

 \begin{eqnarray}
  m_i\, \eta_i\, a^r_i(t)
  &\cir& \eta_i\, {\tilde F}^r_i(t |{\vec {\tilde \eta}}_i(t) |
 {\vec {\tilde \eta}}_j(t)) =\nonumber\\
 &&\nonumber\\
 &=& \eta_i\,m_i\,c^2 \Big[
-\frac{\partial\check{n}_{(1)}}{\partial\tilde{\eta}^r_i}+
2\frac{v^r_i}{c}\,\frac{\vec{v}_i}{c}\cdot\frac{\partial\check{n}_{(1)}}{\partial\vec{\tilde{\eta}}_i}+
\sum_{j\neq i}\,\frac{v^r_i}{c}\,\frac{\vec{v}_j}{c}\cdot\frac{\partial\check{n}_{(1)}}{\partial\vec{\tilde{\eta}}_j}+\nonumber\\
&&\nonumber\\
&&+\sum_u\Big(\,\frac{v^u_i}{c}\,\frac{\partial\check{\bar{n}}_{(1)(u)}}{\partial\tilde{\eta}^r_i}-
\frac{v^u_i}{c}\,\frac{\partial\check{\bar{n}}_{(1)(r)}}{\partial\tilde{\eta}^u_i}-
\sum_{j\neq i}\frac{v^u_j}{c}\,\frac{\partial\check{\bar{n}}_{(1)(r)}}
{\partial\tilde{\eta}^u_j}\Big)+\nonumber\\
&&\nonumber\\
&&+\frac{v^r_i}{c}\sum_{u,v}\frac{v^u_i}{c}\frac{v^v_i}{c}\,
\frac{\partial\check{\bar{n}}_{(1)(u)}}
{\partial\tilde{\eta}^v_i}+\nonumber\\
&&\nonumber\\
&&+\sum_s\,\Big(\frac{v_i^s}{c}\Big)^2\,\Big(\frac{\partial}{\partial\tilde{\eta}^r_i}
-2\frac{v^r_i}{c}\,\frac{\vec{v}_i}{c}\cdot\frac{\partial}{\partial\vec{\tilde{\eta}}_i}\Big)
(\tilde{\Gamma}^{(1)}_s+2\phi_{(1)})-\nonumber\\
&&\nonumber\\
&&-2\,\frac{v^r_i}{c}\,\Big(1-\frac{v_i^2}{c^2}\Big)\,\Big(\frac{\vec{v}_i}{c}
\cdot\frac{\partial}{\partial\vec{\tilde{\eta}}_i}+\sum_{j\neq i}\,\frac{\vec{v}_j}{c}
\cdot\frac{\partial}{\partial\vec{\tilde{\eta}}_j}
\Big)(\tilde{\Gamma}^{(1)}_s+2\phi_{(1)})-\nonumber\\
&&\nonumber\\
&&-\frac{v^r_i}{c}\,\sum_s\,\Big(\frac{v_i^s}{c}\Big)^2\,\sum_{j\neq i}\,
\frac{\vec{v}_j}{c}\cdot\frac{\partial}{\partial\vec{\tilde{\eta}}_j}
(\tilde{\Gamma}^{(1)}_s+2\phi_{(1)})
-\nonumber\\
&&\nonumber\\
&&-\frac{v^r_i}{c}\,\frac{1}{c^2}
\Big(\partial^2_t{|}_{{\vec {\tilde \eta}}_i}\, {}^3{\tilde {\cal K}}_{(1)}
 + 2\, \sum_u\, v_i^u(t)\, {{\partial\, \partial_t{|}_{{\vec {\tilde \eta}}_i}\,
 {}^3{\tilde {\cal K}}_{(1)} }\over {\partial\, {\tilde \eta}_i^u}}
 +\nonumber \\
 &&\qquad+\sum_{uv}\, v_i^u(t)\, v^v_i(t)\, {{\partial^2\, {}^3{\tilde
 {\cal K}}_{(1)}}\over {\partial\, {\tilde \eta}_i^u\, \partial\, {\tilde \eta}_i^v}}
 \Big)\,\,\Big](t, {\vec {\tilde \eta}}_i(t))=\nonumber\\
 &=&-\eta_i\, \sum_{j \not= i}\, \eta_j\, G\, m_i\, m_j\,
 {{{\tilde \eta}_i^r(t) - {\tilde \eta}_j^r(t)}\over {|{\vec {\tilde \eta}}_i(t) -
 {\vec {\tilde \eta}}_j(t)|^3}}\nonumber\\
 &&\nonumber\\
 &&- \eta_i\,  {{v_i^r(t)}\over c}\,
 \Big[ \partial^2_t{|}_{{\vec {\tilde \eta}}_i}\, {}^3{\tilde {\cal K}}_{(1)}
 + 2\, \sum_u\, v_i^u(t)\, {{\partial\, \partial_t{|}_{{\vec {\tilde \eta}}_i}\,
 {}^3{\tilde {\cal K}}_{(1)} }\over {\partial\, {\tilde \eta}_i^u}}
 +\nonumber \\
 &&\qquad +\sum_{uv}\, v_i^u(t)\, v^v_i(t)\, {{\partial^2\, {}^3{\tilde
 {\cal K}}_{(1)}}\over {\partial\, {\tilde \eta}_i^u\, \partial\, {\tilde \eta}_i^v}}
 \Big] (t, {\vec {\tilde \eta}}_i(t))+\nonumber\\
 &&+\,m_i\,\sum_{j\neq i}\,\eta_i\eta_j\,\sum_{k=1}^\infty
 \Big(\tilde{F}^r_{ij[(2k)]}+\tilde{F}^r_{ij[(2k+1)]}\Big)
 \nonumber
\end{eqnarray}

where

\begin{eqnarray}
\tilde{F}^r_{ij[(2)]}&=&-\frac{\partial\hat{n}_{(1)j[(2)]}}{\partial\tilde{\eta}^r_i}+
2\frac{v^r_i}{c}\,\frac{\vec{v}_i}{c}\cdot\frac{\partial\hat{n}_{(1)j[(0)]}}{\partial\vec{\tilde{\eta}}_i}+
\sum_{j\neq i}\,\frac{v^r_i}{c}\,\frac{\vec{v}_j}{c}\cdot
\frac{\partial\hat{n}_{(1)j[(0)]}}{\partial\vec{\tilde{\eta}}_j}+\nonumber\\
&&\nonumber\\
&&+\sum_u\Big(\,\frac{v^u_i}{c}\,\frac{\partial\hat{\bar{n}}_{(1)(u)j[(1)]}}{\partial\tilde{\eta}^r_i}-
\frac{v^u_i}{c}\,\frac{\partial\hat{\bar{n}}_{(1)(r)j[(1)]}}{\partial\tilde{\eta}^u_i}-
\frac{v^u_j}{c}\,\frac{\partial\hat{\bar{n}}_{(1)(r)j[(1)]}}{\partial\tilde{\eta}^u_j}\Big)+\nonumber\\
&&\nonumber\\
&&+2\,\sum_s\,\Big(\frac{v_i^s}{c}\Big)^2\,\Big(\frac{\partial}{\partial\tilde{\eta}^r_i}
\phi_{(1)j[(0)]}\Big)-\nonumber\\
&&\nonumber\\
&&-4\,\frac{v^r_i}{c}\,\Big(\frac{\vec{v}_i}{c}\cdot\frac{\partial}{\partial\vec{\tilde{\eta}}_i}
+\frac{\vec{v}_j}{c}\cdot\frac{\partial}{\partial\vec{\tilde{\eta}}_j}
\Big)\phi_{(1)j[(0)]})
\nonumber
\end{eqnarray}

and for $k>1$

\begin{eqnarray}
\tilde{F}^r_{ij[(2k)]}&=&-\frac{\partial\hat{n}_{(1)j[(2k)]}}{\partial\tilde{\eta}^r_i}+
2\frac{v^r_i}{c}\,\frac{\vec{v}_i}{c}\cdot\frac{\partial\hat{n}_{(1)j[(2k-2)]}}
{\partial\vec{\tilde{\eta}}_i}+
\frac{v^r_i}{c}\,\frac{\vec{v}_j}{c}\cdot\frac{\partial\hat{n}_{(1)j[(2k-2)]}}
{\partial\vec{\tilde{\eta}}_j}+\nonumber\\
&&\nonumber\\
&&+\sum_u\Big(\,\frac{v^u_i}{c}\,\frac{\partial\hat{\bar{n}}_{(1)(u)j[2k-1]}}
{\partial\tilde{\eta}^r_i}-
\frac{v^u_i}{c}\,\frac{\partial\hat{\bar{n}}_{(1)(r)j[(2k-1)]}}{\partial\tilde{\eta}^u_i}-
\sum_{j\neq i}\frac{v^u_j}{c}\,\frac{\partial\hat{\bar{n}}_{(1)(r)j[(2k-1)]}}
{\partial\tilde{\eta}^u_j}\Big)+\nonumber\\
&&\nonumber\\
&&+\frac{v^r_i}{c}\sum_{u,v}\frac{v^u_i}{c}\frac{v^v_i}{c}\,
\frac{\partial\hat{\bar{n}}_{(1)(u)j[(2k-3)]}}{\partial\tilde{\eta}^v_i}+\nonumber\\
&&\nonumber\\
&&+\sum_s\,\Big(\frac{v_i^s}{c}\Big)^2\,\Big(\frac{\partial}{\partial\tilde{\eta}^r_i}
(\hat{\Gamma}^{(1)}_{sj[(2k-2)]}+2\phi_{(1)j[(2k-2)]})-\nonumber\\
&&\nonumber\\
&&\qquad -2\frac{v^r_i}{c}\,\frac{\vec{v}_i}{c}\cdot\frac{\partial}
{\partial\vec{\tilde{\eta}}_i}(\hat{\Gamma}^{(1)}_{sj[(2k-4)]}
+2\phi_{(1)j[(2k-4)]})\Big)-\nonumber\\
&&\nonumber\\
&&-2\,\frac{v^r_i}{c}\,\Big(\frac{\vec{v}_i}{c}\cdot\frac{\partial}
{\partial\vec{\tilde{\eta}}_i}+\frac{\vec{v}_j}{c}\cdot
\frac{\partial}{\partial\vec{\tilde{\eta}}_j}
\Big)(\hat{\Gamma}^{(1)}_{sj[(2k-2)]}+2\phi_{(1)j[(2k-2)]})+\nonumber\\
&&\nonumber\\
&&+2\,\frac{v^r_i}{c}\,\frac{v_i^2}{c^2}\,\Big(\frac{\vec{v}_i}{c}\cdot
\frac{\partial}{\partial\vec{\tilde{\eta}}_i}+\frac{\vec{v}_j}{c}\cdot
\frac{\partial}{\partial\vec{\tilde{\eta}}_j}
\Big)(\hat{\Gamma}^{(1)}_{sj[(2k-4)]}+2\phi_{(1)j[(2k-4)]})\nonumber\\
&&\nonumber\\
&&-\frac{v^r_i}{c}\,\sum_s\,\Big(\frac{v_i^s}{c}\Big)^2\,\sum_{j\neq i}\,
\frac{\vec{v}_j}{c}\cdot\frac{\partial}{\partial\vec{\tilde{\eta}}_j}
(\hat{\Gamma}^{(1)}_{sj[(2k-4)]}+2\phi_{(1)j[(2k-4)]})\Big)\,\,
\Big]\nonumber
\end{eqnarray}

\begin{eqnarray}
\tilde{F}^r_{ij[(2k+1)]}&=&
\sum_u\Big(\,\frac{v^u_i}{c}\,\frac{\partial\hat{\bar{n}}_{(1)(u)j[(2k)]}}
{\partial\tilde{\eta}^r_i}-
\frac{v^u_i}{c}\,\frac{\partial\hat{\bar{n}}_{(1)(r)j[(2k)]}}{\partial\tilde{\eta}^u_i}-
\sum_{j\neq i}\frac{v^u_j}{c}\,\frac{\partial\hat{\bar{n}}_{(1)(r)j[(2k)]}}
{\partial\tilde{\eta}^u_j}\Big)+\nonumber\\
&&\nonumber\\
&&+\frac{v^r_i}{c}\sum_{u,v}\frac{v^u_i}{c}\frac{v^v_i}{c}\,
\frac{\partial\hat{\bar{n}}_{(1)(u)j[(2k-2)]}}{\partial\tilde{\eta}^v_i}
\label{5.13}
\end{eqnarray}

\medskip

Let us remark that the force ${\tilde F}^r_i$ contains at the 0PN
order the Newton force of Newtonian gravity. In Eqs.(\ref{5.13}) it
is possible to see that the terms depending on the inertial gauge
variable (the non-local York time) ${}^3{\tilde {\cal K}}_{(1)}$,
absent in the Euclidean 3-spaces of Newton gravity, are present at
the order 0.5PN. Moreover the force ${\tilde F}^r_i$ contains both
even and odd terms at all the orders. See Appendix A for the 1PN
expression of the ADM Poincare' generators with terms till order
$O(\zeta^2)$ included.

\medskip

In the standard approach in harmonic gauges the first odd terms
start at 2.5PN order: they are connected to the breaking of
time-reversal invariance due to the choice of the no-incoming
radiation condition and to the effect of back-reaction in presence
of gravitational self-force with the associated (either Hadamard or
dimensional) regularization (see the review in Ref.\cite{6}). In our
approach the Grassmann regularization eliminates the self-force but
back-reaction is present due to the constancy of the ADM energy and
produces the correct energy balance for the emission of GW's as
shown in paper II.\medskip

Since we are in a non-harmonic gauge, we use a Grassmann
regularization and, moreover, we are not introducing ad hoc
Lagrangians for the particles, it is not possible to make
comparisons with the standard results known till 3.5PN order
\cite{6} (where also the hereditary terms are present: we will need
higher orders in the HPM expansion to see these terms, if them are
present in our non-harmonic gauges).

\subsection{ The HPM Binaries at the 1PN Order}

Let us now consider the 1PN two-body problem, which is relevant for
the treatment of binary systems \footnote{For binaries one assumes
${v\over c} \approx \sqrt{{{R_m}\over {l_c}}} << 1$, where $l_c
\approx |{\vec {\tilde r}}|$ with ${\vec {\tilde r}}(t)$ being the
relative separation after the decoupling of the center of mass.
Often one considers the case $m_1 \approx m_2$. See chapter 4 of
Ref. \cite{6} for a review of the emission of GW's from circular and
elliptic Keplerian orbits and of the induced inspiral phase implying
a secular change in the semi-major axis, in the ellipticity and in
the period, during which the waveform of GW's increases in amplitude
and frequency producing a characteristic {\it chirp}. If we would
add terms of higher PN order from Eqs.(\ref{5.7}), we would get the
analogue in the HPM linearization of the standard 3.5PN calculations
for the inspiral phase before merging and ring-down (see section 5.6
of Ref.\cite{6} and Ref.\cite{37} for a review; see also
Ref.\cite{38}). Again the Grassmann regularization gives different
results for the back-reaction. Instead the PM equations of motion
Eqs.(\ref{5.3}) and (\ref{5.4}) should be used to take under control
the relativistic recoils during the inspiral phase.} as shown in
Chapter VI of Refs.\cite{6}  based on Ref.\cite{39}.

\medskip

Since there is no convincing evidence of dark matter in the Solar
System and near the galactic plane of the Milky Way \cite{40}, we
shall ignore the 0.5PN terms containing the non-local York time in
the study of binary systems of stars in some galaxy. Instead in the
next Section we will see that the non-local York time can be
relevant in the 0.5PN simulation of dark matter at the level of mass
and rotation curves of a whole galaxy.

\medskip

If we ignore the York time, the 1PN equations of motion for the
binary implied by Eqs.(\ref{5.13}) and (\ref{5.12}) are (we consider
only positive energy, i.e. $\eta_1, \eta_2 \mapsto <\eta_1 >, <
\eta_2 > = + 1$)

\bea
 &&m_1\,  a^r_1(t) \cir\nonumber\\
 &\cir& -G\, m_1m_2\, \Big[\,\,
 {{{\tilde \eta}^r_1(t) - {\tilde \eta}_2^r(t)}\over {|{\vec {\tilde \eta}}_1(t)
 - {\vec {\tilde \eta}}_2(t)|^3}}\, \Big(1 + {{{\vec v}_1^2(t)}\over {c^2}}+2 {{{\vec v}_2^2(t)}\over {c^2}}
 - 4 {{{\vec v}_1(t) \cdot {\vec v}_2(t)}\over{c^2}} -\frac{3}{2}\, {{\Big({{{\vec v}_2(t)}\over c} \cdot
 ({\vec {\tilde \eta}}_1(t) - {\vec {\tilde \eta}}_2(t))\Big)^2}\over
 {|{\vec {\tilde \eta}}_1(t) - {\vec {\tilde \eta}}_2(t)|^2}}
 \Big) +\nonumber \\
 &-& \frac{v^r_1(t) - {v}^r_2(t)}{2}\Big(
 \,4\,{{{{{\vec v}_1(t)}\over c} \cdot
 ({\vec {\tilde \eta}}_1(t) - {\vec {\tilde \eta}}_2(t))}\over
 {|{\vec {\tilde \eta}}_1(t) - {\vec {\tilde \eta}}_2(t)|^2}}-
 \,3\,{{{{{\vec v}_2(t)}\over c} \cdot
 ({\vec {\tilde \eta}}_1(t) - {\vec {\tilde \eta}}_2(t))}\over
 {|{\vec {\tilde \eta}}_1(t) - {\vec {\tilde \eta}}_2(t)|^2}}
 \Big)\,\,\Big] = {\tilde F}^r_{1(1PN)}(t),\nonumber \\
 &&{}\nonumber \\
 &&\nonumber\\
 &&m_2\,  a^r_2(t) \cir\nonumber\\
 &\cir& +G\, m_1m_2\, \Big[\,\,
 {{{\tilde \eta}^r_1(t) - {\tilde \eta}_2^r(t)}\over {|{\vec {\tilde \eta}}_1(t)
 - {\vec {\tilde \eta}}_2(t)|^3}}\, \Big(1 + {{{\vec v}_2^2(t)}\over {c^2}}+2 {{{\vec v}_1^2(t)}\over {c^2}}
 - 4 {{{\vec v}_1(t) \cdot {\vec v}_2(t)}\over{c^2}} -\frac{3}{2}\, {{\Big({{{\vec v}_1(t)}\over c} \cdot
 ({\vec {\tilde \eta}}_1(t) - {\vec {\tilde \eta}}_2(t))\Big)^2}\over
 {|{\vec {\tilde \eta}}_1(t) - {\vec {\tilde \eta}}_2(t)|^2}}
 \Big) +\nonumber \\
 &-& \frac{v^r_2(t) - {v}^r_1(t)}{2}\Big(
 \,4\,{{{{{\vec v}_2(t)}\over c} \cdot
 ({\vec {\tilde \eta}}_1(t) - {\vec {\tilde \eta}}_2(t))}\over
 {|{\vec {\tilde \eta}}_1(t) - {\vec {\tilde \eta}}_2(t)|^2}}-
 \,3\,{{{{{\vec v}_1(t)}\over c} \cdot
 ({\vec {\tilde \eta}}_1(t) - {\vec {\tilde \eta}}_2(t))}\over
 {|{\vec {\tilde \eta}}_1(t) - {\vec {\tilde \eta}}_2(t)|^2}}
 \Big)\,\,\Big] = {\tilde F}^r_{2(1PN)}(t).
 \label{5.14}
 \eea

The last two lines in each equation correspond to the
gravito-magnetic force generated by the shift functions. These
equations identify the 1PN forces ${\tilde F}^r_{i(1PN)}(t)$.

\bigskip

Let us now reformulate this two-body problem in the canonical basis
of center-of-mass and relative variables used in
Eqs.(\ref{5.5})-(\ref{5.10}) by restricting us to the lowest order
${\vec \eta}_i(\tau) = {\vec {\tilde \eta}}_i(t) = {\vec {\tilde
\eta}}_{(o)i}(t)$. Since we have ${\vec \kappa}_{(1)i}(\tau) =
{{m_i\, {\vec v}_i(t)}\over {\sqrt{1 - ({{{\vec v}_i(t)}\over
c})^2}}}$ with ${\vec \kappa}_{(1)1}(\tau) + {\vec
\kappa}_{(1)2}(\tau) = 0$, from Eq.(\ref{5.5}) we get ${\vec
\pi}_{(1)12}(\tau) = \mu\, \Big({{{\vec v}_1(t)}\over {\sqrt{1 -
({{{\vec v}_1(t)}\over c})^2}}} - {{{\vec v}_2(t)}\over {\sqrt{1 -
({{{\vec v}_2(t)}\over c})^2}}}\Big)$ with ${\dot {\vec
\eta}}_{(o)1}(\tau) = {{{\vec v}_1(t)}\over c} = ({{m_2}\over M} -
{\tilde A}_{(o)}(t))\, {{{\vec v}_{(rel)(o)12}(t)}\over c}$ and
${\dot {\vec \eta}}_{(o)2}(\tau) = {{{\vec v}_2(t)}\over c} = -
({{m_1}\over M} + {\tilde A}_{(o)}(t))\, {{{\vec
v}_{(rel)(o)12}(t)}\over c}$, where we introduced the definition
${\vec v}_{(rel)(o)12}(t) = {{d\, {\vec \rho}_{(o)12}(t)}\over
{dt}}$. In writing ${\dot {\vec \eta}}_{(o)i}(\tau)$ we ignored the
term $- {1\over c}\, {{d\, {\tilde A}_{(o)}(t)}\over {dt}}\, {\vec
\rho}_{(o)12}(t)$: it is of higher order because it depends on
${\dot {\vec \pi}}_{(1)12}(\tau)$.\medskip

Then at the 1PN order we get

\bea
 {\vec \pi}_{(1)12}(\tau) &=& \mu\, \Big({\vec v}_1(t)\, \Big[1 + {1\over 2}\,
 ({{{\vec v}_1(t)}\over c})^2\Big] - {\vec v}_2(t)\, \Big[1 + {1\over 2}\,
 ({{{\vec v}_2(t)}\over c})^2\Big] \Big) =\nonumber \\
 &=& \mu\, {\vec v}_{(rel)(o)12}(t)\, \Big(1 + {1\over 2}\,
 \Big[({{\mu}\over {m_1}} - {\tilde A}_{(o)}(t))^3 + ({{\mu}\over {m_2}}
 + {\tilde A}_{(o)}(t))^3\Big]\, \Big({{{\vec v}_{(rel)(o)12}(t)}\over c}\Big)^2
 \Big).\nonumber \\
 &&{}
 \label{5.15}
 \eea

Since the 1PN limit of the function ${\tilde A}_{(o)}(t)$ of
Eq.(\ref{5.6}) is ${{\mu\, (m_1 - m_2)}\over {2 M^2}}\, ({{{\vec
v}_{(rel)(o)12}(t)}\over c})^2$ (so that from Eq.(\ref{5.6}) we have
${\vec {\tilde \eta}}_{(o)12}(t) \approx 0 + O({{v^2}\over
{c^2}})$), we get ${\vec \pi}_{(1)12}(\tau) = \mu\, {\vec
v}_{(rel)(o)12}(t)\, \Big[1 + {{m_1^3 + m_2^3}\over {2\, M^3}}\,
({{{\vec v}_{(rel)(o)12}(t)}\over c})^2\Big]$ as the 1PN limit of
the relative momentum.\medskip

Then from Eqs.(\ref{5.8}) the 1PN limit of the ADM energy and
angular momentum in the rest-frame is (at this order there is no
dependence on the York time)

\bea
 {\hat E}_{ADM(1PN)} &=& \sum_i\, m_i\, c^2 + \mu\, \Big({1\over
 2}\, {\vec v}^2_{(rel)(o)12}(t) \,\Big[ 1 +
  {{m_1^3 + m_2^3}\over {M^3}}\, ({{{\vec v}_{(rel)(o)12}(t)}\over
 c})^2\Big] -\nonumber \\
 &-& {{G\, M}\over {|{\vec \rho}_{(o)12}(t)|}}\, \Big[ 1 + {1\over
 2}\, \Big((3 + {{\mu}\over {M}})\, {{{\vec v}^2_{(rel)(o)12}(t)}\over
 {c^2}} + {{\mu}\over M}\, ({{{\vec v}_{(rel)(o)12}(t)}\over c} \cdot
 {{ {\vec \rho}_{(o)12}(t) }\over {|{\vec \rho}_{(o)12}(t)|}})^2
 \Big)\Big]\Big),\nonumber \\
 &&{}\nonumber \\
 {\hat J}^{rs}_{ADM(1PN)} &=& \Big(\rho^r_{(o)12}(t)\, v^s_{(rel)(o)12}(t) -
 \rho^s_{(o)12}(t)\, v^r_{(rel)(o)12}(t)\Big)\nonumber \\
 &&\Big[1 + {{m_1^3 + m_2^3}\over
{2\, M^3}}\, ({{{\vec v}_{(rel)(o)12}(t)}\over c})^2\Big],
 \label{5.16}
 \eea

\noindent while from Eqs.(\ref{5.10}) the equation of motion for the
relative variable is

\bea
 {{d\, {\vec v}_{(rel)(o)12}(t)}\over {dt}} &\cir& {1\over {\mu}}\,
 \Big({{m_2}\over M}\, {\tilde F}^r_{1(1PN)}(t)
 - {{m_1}\over M}\, {\tilde F}^r_{2(1PN)}(t)\Big)(t, {\vec {\tilde
 \eta}}_{i(o)}(t), {\vec v}_i(t)) =\nonumber \\
 &=&  - G\, M\, {{{\vec \rho}_{(o)12}(t)}\over
 {|{\vec \rho}_{(o)12}(t)|^3}} \Big[1+ \Big(1+3\frac{\mu}{M}\Big)
 \frac{v^2_{(rel)(o)12}(t)}{c^2}
 -\frac{3\mu}{2M}\,\Big(\frac{{\vec v}_{(rel)(o)12}(t)\cdot
 {\vec \rho}_{(o)12}(t)\big)}{|{\vec \rho}_{(o)12}(t)|}\Big)^2\Big]+\nonumber\\
 &-& \frac{G\, M\,}{|{\vec \rho}_{(o)12}(t)|^3}\Big(
 4-\frac{2\mu}{M}
 \Big){\vec v}_{(rel)(o)12}(t)\,\frac{{\vec v}_{(rel)(o)12}(t)\cdot
 {\vec \rho}_{(o)12}(t)\big)}{|{\vec \rho}_{(o)12}(t)|}.
 \label{5.17}
 \eea

\noindent with the forces ${\tilde F}^r_{i(1PN)}(t)$ defined in
Eqs.(\ref{5.14}).\bigskip

This is the result (ignoring the 0.5PN contribution of the non-local
York time; for it see next Section) for the 1PN relative motion of
binaries in our HPM linearization in the 3-orthogonal gauges, where
the energy and angular momentum constants of motion are given by the
corresponding ADM generators (implying planar motion in the plane
orthogonal to the rest-frame ADM angular momentum).\medskip

Our 1PN equations (\ref{5.16}) and (\ref{5.17}) in the 3-orthogonal
gauges coincide with Eqs. (2.5), (2.13) and (2.14) of the first
paper Damour and Deruelle in Ref.\cite{39} (without $G^2$ terms
since they are $O(\zeta^2)$), which are obtained in the family of
harmonic gauges starting from an ad hoc 1PN Lagrangian for the
relative motion of two test particles (first derived by Infeld and
Plebanski \cite{41}) \footnote{This is also the starting point of
the effective one body description of the two-body problem of Refs.
\cite{42}.}. These equations are the starting point for studying the
post-Keplerian parameters of the binaries, which, together with the
Roemer, Einstein and Shapiro time delays (both near Earth and near
the binary) in light propagation, allow to fit the experimental data
from the binaries (see the second paper in Ref.\cite{39} and Chapter
VI of Ref.\cite{6}). Therefore these results are reproduced also in
our 3-orthogonal gauge with ${}^3K_{(1)}(\tau, \vec \sigma) = 0$.

\vfill\eject

\section{ Dark Matter as a Relativistic Inertial Effect due to York time}

In this Section we will see that the non-local York time can be
relevant in the simulation of dark matter at the level of the mass
of clusters of galaxies and of the rotation curves of
galaxies.\medskip

In the first Subsection we consider the 0.5PN equations of motion.
In the next two Subsections we will  consider  two of the main
signatures of the existence of dark matter in the observed masses of
galaxies and clusters of galaxies, namely the virial theorem
\cite{43,44} and weak gravitational lensing \cite{25,44,45}. Then in
the fourth Subsection we will reproduce the pattern of rotation
curves of spiral galaxies \cite{46}. In a final Subsection we
discuss which information we obtain on the York time from the dark
matter data.

\subsection{The 0.5 Post-Newtonian Limit of the Equations of Motion for the
Particles}

At the order $0.5PN$,  with the non-local York time  ${}^4{\tilde
{\cal K}}_{(1)}$ (with dimension $[{}^3{\cal K}_{(1)}] = [l]$ since
$[{}^3K_{(1)}] = [l^{-1}]$) taken into account,  Eqs.(\ref{5.13})
for the particles  become

\bea
  \eta_i\, m_i\, {{d^2\, {\tilde \eta}_i^r(t)}\over {dt^2}}
 &\cir& \eta_i\, m_i\, \Big[-G\, {{\partial}\over {\partial\,
 {\tilde \eta}_i^r}}\,  \sum_{j \not= i}\, \eta_j\, {{m_j}\over
 {|{\vec {\tilde \eta}}_i(t) - {\vec {\tilde \eta}}_j(t)|}} -
  {1\over c}\, {{d {\tilde \eta}^r_i(t)}\over {dt}} \Big(
  \partial^2_t{|}_{{\vec {\tilde \eta}}_i}\, {}^3{\tilde
  {\cal K}}_{(1)} +\nonumber \\
  &+& 2\, \sum_u\, v_i^u(t)\, {{\partial\, \partial_t{|}_{{\vec {\tilde \eta}}_i}\,
 {}^3{\tilde {\cal K}}_{(1)} }\over {\partial\, {\tilde \eta}_i^u}}
 + \sum_{uv}\, v_i^u(t)\, v^v_i(t)\, {{\partial^2\, {}^3{\tilde
 {\cal K}}_{(1)}}\over {\partial\, {\tilde \eta}_i^u\, \partial\, {\tilde \eta}_i^v}}
 \Big) (t, {\vec {\tilde \eta}}_i(t)) \Big].\nonumber \\
 &&{}
 \label{6.6}
 \eea

In these equations we can replace the Grassmann variables with their
mean value $< \eta_i > = 1$, i = 1,..,N, for positive energy
particles.

\bigskip

Therefore at the order 0.5PN the  double rate of change in time of
the trace of the extrinsic curvature, the arbitrary inertial gauge
function parametrizing the family of 3-orthogonal gauges, creates PN
damping terms with damping coefficients

\bea
 &&\gamma_i(t, {\vec {\tilde \eta}}_i(t)) =
 \Big(\partial_t^2{|}_{{\vec {\tilde \eta}}_i} +
 2\, \sum_u\, v_i^u(t)\, \partial_u\, \partial_t{|}_{{\vec
 {\tilde \eta}}_i} + \sum_{uv}\, v_i^u(t)\, v_i^v(t)\, \partial_u\,
 \partial_v \Big)\, {}^3{\tilde {\cal K}}_{(1)}(t,
 {\vec {\tilde \eta}}_i(t)).\nonumber \\
 &&{}
 \label{6.7}
 \eea

\noindent For instance the first term corresponds to a {\it damping}
when $\partial_t^2{|}_{{\vec {\tilde \eta}}_i}\, {}^3{\cal
K}_{(1)}(\tau, {\vec \eta}_i(\tau)) > 0$, but it is an {\it
anti-damping} when $\partial_t^2{|}_{{\vec {\tilde \eta}}_i}\,
{}^3{\cal K}_{(1)}(\tau, {\vec \eta}_i(\tau)) < 0$. Since we have
$[c^2\, \partial_{\tau}^2\, {}^3K_{(1)}(\tau, \vec \sigma)]{|}_{\vec
\sigma =   {\vec \eta}_i(\tau)} = [\triangle\, \partial_t^2\,
{}^3{\tilde {\cal K}}_{(1)}(t, \vec \sigma)]{|}_{\vec \sigma = {\vec
{\tilde \eta}}_i(t)}$, the anti-damping (damping) effect is governed
by the acceleration of the change in time of the convexity
(concavity) of the instantaneous 3-space $\Sigma_{\tau}$ near the
particle as an embedded 3-manifold of space-time. This is a inertial
effect, relevant at small accelerations of the particle, not
existing in Newton theory where the Euclidean 3-space is absolute
and absent in all the gauges with ${}^3K(\tau ,\vec \sigma) = 0$
(see for instance Ref.\cite{37} for the lowest order of PN harmonic
gauges). The other damping terms have similar interpretation but
with an extra dependence on the velocities.

\medskip

\noindent We can rewrite Eq.(\ref{6.7}) in the following form

 \beq
 \gamma_i(t, {\vec {\tilde \eta}}_i(t)) =\frac{d^2}{dt^2}\,{}^3{\tilde {\cal K}}_{(1)}(t,
 {\vec {\tilde \eta}}_i(t))+{\cal O}(\zeta^2),
\eeq

\noindent by using $ \frac{d}{dt}=\partial_t+v^u_i(t)\,\partial_u$
and by taking into account the fact that the accelerations
$dv^u_i(t)/dt$ are of order ${\cal O}(\zeta^2)$. As a consequence
Eq.(\ref{6.6}) can be written in the form

 \bea
 \frac{d}{dt}\Big[ \,m_i\Big(1+\frac{1}{c}\,\frac{d}{dt}\,{}^3{\tilde {\cal K}}_{(1)}(t,
 {\vec {\tilde \eta}}_i(t))\,\Big)\, {{d\, {\tilde \eta}_i^r(t)}\over {dt}}\Big]
 &\cir &\,-G\, {{\partial}\over {\partial\,
 {\tilde \eta}_i^r}}\,  \sum_{j \not= i}\, \eta_j\, {{m_i\,m_j}\over
 {|{\vec {\tilde \eta}}_i(t) - {\vec {\tilde \eta}}_j(t)|}}+\nonumber\\
 &&\nonumber\\
 &+&{\cal O}(\zeta^2)
 \label{6.6bis}
\eea

We see that the term in the non-local York time can be {\it
interpreted} as the introduction of an {\it effective (time-,
velocity- and position-dependent) inertial mass term} for the
kinetic energy of each particle: $m_i\, \mapsto\,
m_i\,\Big(1+\frac{1}{c}\,\frac{d}{dt}\,{}^3{\tilde {\cal
K}}_{(1)}(t, {\vec {\tilde \eta}}_i(t))\,\Big) $ in each
instantaneous 3-space. Instead in the Newton potential there are the
gravitational masses of the particles, equal to the inertial ones in
the 4-dimensional  due to the equivalence principle. Therefore the
effect is due to a modification of the effective inertial mass in
each non-Euclidean 3-space: it is the equality of the inertial and
gravitational masses of Newtonian gravity to be violated!

\medskip

\subsection{Dark Matter in Galaxy Masses from the Virial Theorem}

One of the experimental signatures of the existence of dark matter
comes from the use of the virial theorem for the determination of
the mass of clusters of galaxies \cite{43,44}. For a bound system of
N particles of mass $m$ (N equal mass galaxies) at equilibrium, the
virial theorem relates the average  kinetic energy $< E_{kin}
>$ in the system to the average  potential energy $< U_{pot} >$
in the system: $< E_{kin} > = - {1\over 2}\, < U_{pot} >$ assuming
Newton gravity. The equilibrium condition is supposed to be more
valid for clusters of galaxies rather than for galaxies (clusters of
stars). For the average kinetic energy of a galaxy in the cluster
one takes $< E_{kin} > \approx {1\over 2}\, m\, < v^2 >$, where $<
v^2 >$ is the average of the square of the radial velocity of single
galaxies with respect to the center of the cluster (measured with
Doppler shift methods; the velocity distribution is assumed
isotropic). The average potential energy of the galaxy is assumed of
the form $< U_{pot} > \approx - G\, {{m\, M}\over {\cal R}}$, where $M = N
m$ is the total mass of the cluster and ${\cal R}=\alpha\,R$ is a
''effective radius'' depending on the cluster size $R$ (the
angular diameter of the cluster and its distance from Earth are
needed to find $R$) and on the mass distribution on the cluster
(usually $\alpha\approx 1/2)$. Then the virial theorem implies $M \approx
{{\cal R}\over G}\, < v^2 >$. It turns out that this mass $M$ of the
cluster is usually at least an order of magnitude bigger that the
baryonic matter of the cluster (spectroscopically determined).

\medskip

If we consider the 0.5PN limit of the equations of motion for the
particles given in eq.(\ref{6.6}), we have to introduce a correction
to final form of the virial theorem due to the extra term depending
on the non local York time.

\medskip

Usually the derivation of virial theorem starts assuming that in a
self gravitating system at equilibrium we have

 \beq
 \frac{d^2}{dt^2}\sum_i\,m_i\mid\vec{\tilde{\eta}}_i(t)\mid^2=0.
 \eeq

This implies

 \beq
 \sum_i\,m_i\,v_i^2(t)+\sum_{i}\,m_i\,\vec{\tilde{\eta}}_i(t)
 \cdot\frac{d\vec{v}_i(t)}{dt}=0.
 \eeq

By using Eqs.(\ref{6.6}) and (\ref{6.7}) we get

 \beq
\sum_i\,m_i\,v_i^2(t)-G\sum_{i>j}\,\eta_j\, {{m_i\,m_j}\over
 {|{\vec {\tilde \eta}}_i(t) - {\vec {\tilde \eta}}_j(t)|}}-
 \frac{1}{c}\sum_i\,m_i\Big(\vec{\tilde{\eta}}_i(t)\cdot
 \vec{v}_i(t)\Big)\,\gamma_i(t, {\vec {\tilde \eta}}_i(t))=0.
 \label{virial1}
 \eeq

In the case $m_i=m$,  the mean square velocity is $\langle v^2
\rangle=\frac{1}{N}\sum_i\,v_i^2$, and the mean gravitational
potential energy for particle  ( with ${\cal R}=R/2$) has the form
$\langle U_{pot}\rangle=-\frac{1}{N}\sum_{i>j}\,\eta_j\,
G\,{{m_i\,m_j}\over {|{\vec {\tilde \eta}}_i(t) - {\vec {\tilde
\eta}}_j(t)|}}\approx G\,\frac{mM}{{\cal R}}$, with $M_{bar} = N m$
being the baryonic mass.\medskip

Then, if we define $\langle\Big(\vec{\tilde{\eta}}
\cdot\vec{v}\Big)\,\gamma(t, {\vec {\tilde \eta}})\rangle =
\frac{1}{N}\, \sum_i\, \Big(\vec{\tilde{\eta}}_i(t) \cdot
\vec{v}_i(t)\Big)\,\gamma_i(t, {\vec {\tilde \eta}}_i(t))$, we get
the following result from Eq.(\ref{virial1})

 \begin{equation}
 \frac{1}{2}m\langle v^2 \rangle = -\frac{1}{2}\,\langle
 U_{pot}\rangle+\frac{m}{2c}\,\langle\Big(\vec{\tilde{\eta}}\cdot\vec{v}\Big)\,\gamma(t,
 {\vec {\tilde \eta}})\rangle.
 \label{virial-finale}
 \end{equation}

Therefore for the measured mass $M$ (the effective inertial mass in
3-space) we have

\bea
 M &=& {{\cal R}\over G}\, < v^2 > = M_{bar} + {{\cal R}\over {G\,c}}\,
 \langle\Big(\vec{\tilde{\eta}}\cdot\vec{v}\Big)\,\gamma(t,
 {\vec {\tilde \eta}})\rangle =\nonumber \\
 &{\buildrel {def}\over =}& M_{bar} + M_{DM},
 \label{6.3}
 \eea

\noindent and we see that the non-local York time can give rise to a
dark matter contribution $M_{DM} = M - M_{bar}$.

\subsection{Dark Matter in Galaxy Masses from
Weak Gravitational Lensing}

Another experimental signature of dark matter is the determination
of the mass of a galaxy (or a cluster of galaxies) by means of weak
gravitational lensing \cite{25,44,45}. Let us consider a galaxy (or
a cluster of galaxies) of big mass $M$ (typically $M \approx
M^{12}_{sun}$) behind which a distant, bright object (often a
galaxy) is located. The light from the distant object is bent by the
massive one (the lens) and arrives on the Earth deflected from the
original propagation direction.\medskip

As shown in Ref.\cite{25} we have to evaluate Einstein deflection of
light, emitted by a source S at distance $d_S$ from the observer O
on the Earth, generated by the big mass  at a distance $d_D$ from
the observer O. The mass $M$, at distance $d_{DS}$ from the source
S, is considered as a point-like mass generating a 4-metric either
of the Schwarzschild type (Schwarzschild lens) or of the type of
Eq.(\ref{3.7}) (nearly point-like case). The ray of light is assumed
to propagate in Minkowski space-time till near $M$, to be deflected
by an angle $\alpha$ by the local gravitational field of M and then
to propagate in Minkowski space-time till the observer O. The
distances $d_S$, $d_D$, $d_{DS}$, are evaluated by the observer O at
some reference time in some nearly-inertial Minkowski frame with
nearly Euclidean 3-spaces (in the Euclidean case $d_{DS} = d_S -
d_D$). If $\xi = \theta\, d_D$ is the impact parameter of the ray of
light at $M$ and if $\xi >> R_s = {{2\, G\, M}\over {c^2}}$ (the
gravitational radius), Einstein's deflection angle is $\alpha =
{{2\, R_s}\over {\xi}}$ \cite{25} and the so-called Einstein radius
(or characteristic angle) is

\beq
 \alpha_o = \sqrt{2\, R_s\, {{d_{DS}}\over {d_D\, d_S}}} = \sqrt{{{4\, G\, M}\over {c^2}}\,
 {{d_{DS}}\over {d_D\, d_S}}}.
 \label{6.4}
 \eeq

A measurement of the deflection angle and of the three distances
allows to get a value for the mass $M$ of the lens, which usually
turns out to be much larger of its mass inferred from the luminosity
of the lens.\medskip

For the calculation of the deflection angle $\alpha = {{4\, G\,
M}\over {c^2}}$ one considers the propagation of ray of light in a
stationary 4-metric of the type of Eq.(\ref{3.8}) (with the
assumption that $\partial_{\tau}\, {}^3{\cal K}_{(1)}$ and
$\partial_r\, {}^3{\cal K}_{(1)}$ are slowly varying functions of
time) and uses a version of Fermat's principle (see Sections 3.3 and
4.5 of Ref.\cite{25}). In this description the spatial path $\vec
\sigma (l)$ ($d l = |\vec \sigma|$ is the Euclidean arc length in an
Euclidean 3-space) is the minimum of the variational principle(with
fixed end points) $\delta\, \int_1^2\, n\, dl = 0$, where the
effective (position- and direction-dependent) effective index of
refraction is $n = {}^4g_{\tau\tau} + \sum_r\, {}^4g_{\tau r}\, {{d
\sigma^r}\over {d l}}$. If one studies the Euler-Lagrange equations
for the variational principle, if one ignores gravito-magnetism and
if one chooses ${}^4g_{\tau\tau} = \sgn\, [1 - 2\, {w\over {c^2}} -
2\, \partial_{\tau}\, {}^3{\cal K}_{(1)}]$ as in Eqs.(\ref{3.8})
with the choice ${{2\, w}\over {c^2}} = - {{G\, M_{bar}}\over {c^2\,
|\vec \sigma|}}$ and with the definition $2\, \partial_{\tau}\,
{}^3{\cal K}_{(1)}\, {\buildrel {def}\over =}\, - {{G\, M_{DM}}\over
{c^2\, |\vec \sigma|}}$, then the resulting Einstein deflection
angle $\alpha$ turns out \cite{24} to be the modulus of the vector

\beq
 \vec \alpha = {{4\, G\, M}\over {c^2}}\, {{\vec \xi}\over {|\vec \xi|^2}}
 \qquad with\quad M = M_{bar} + M_{DM}.
 \label{6.5}
 \eeq

\noindent Here $\vec \xi$ is the impact parameter if the unperturbed
spatial trajectory of the ray is parametrized as $\vec \sigma (l) =
\vec \xi + l\, \vec e$ ($\vec e = {{d \vec \sigma(l)}\over {dl}}$ is
the unit tangent vector of the ray satisfying $\vec e \cdot \vec \xi
= 0$).\medskip

Therefore also in this case the measured mass $M$ is the sum of a
baryonic mass $M_{bar}$ and of a dark matter mass $M_{DM}$ induced
by the non-local York time at the location of the lens.

\subsection{The 0.5PN Two-Body Problem and the Rotation Curves of Galaxies}

To study Eqs.(\ref{6.6}) in the two-body case (i=1,2), we have to
define center-of-mass and relative variables in the gravitational
case with non-Euclidean 3-spaces, deviating from the Euclidean ones
by order $O(\zeta)$, at least at the 0.5PN order.\medskip

With the notation of Eqs.(\ref{5.5}) for the center-of-mass position
we put ${\tilde \eta}_{12}^r(t) = {\tilde \eta}^r_{(o)12}(t) +
{\tilde \eta}^r_{(1)12}(t)$, where ${\vec {\tilde \eta}}_{(o)12}(t)
= {{m_1\, {\vec {\tilde \eta}}_1(t) + m_2\, {\vec {\tilde
\eta}}_2(t)}\over M}$ ($M = m_1 + m_2$; $\mu = {{m_1\, m_2}\over M}$
is the reduced mass; $m_i = {M\over 2}\, (1 + (-)^{i+1}\, \sqrt{1 -
2\, {{\mu}\over M}})$) is the non-relativistic center of mass and
${\vec {\tilde \eta}}_{(1)12}(t) = O(\zeta)$ is a small
non-Euclidean correction.

\medskip

The relative position variable  is chosen as ${\tilde
\rho}_{12}^r(t) = {\vec
{\tilde \eta}}_{1}(t) - {\vec {\tilde \eta}}_{2}(t)$
\footnote{It should be defined as the tangent to the 3-geodesic of
$\Sigma_{\tau}$ joining the two points (see the next
Eq.(\ref{6.12})), which is parallel transported along it. See for
instance Ref.\cite{36}. At the orders $O(\zeta)$ and 0.5PN the above
definition is acceptable.} As a consequence we have ${\tilde
\eta}^r_1(t) = {\tilde \eta}^r_{(o)12}(t) + {\tilde
\eta}_{(1)12}^r(t) + {{m_2}\over M}\, {\tilde
\rho}_{12}^r(t)$ and ${\tilde \eta}^r_2(t) = {\tilde
\eta}^r_{(o)12}(t) + {\tilde \eta}_{(1)12}^r(t) - {{m_1}\over M}\,
{\tilde \rho}_{(12}^r(t))$.

\medskip

In Subsection VB also a decomposition ${\tilde\rho}_{12}^r(t) =
{\tilde\rho}_{(0)12}^r(t) + {\tilde \rho}_{(1)12}^r(t)$ was adopted.
This decomposition was motivated by the analogous decomposition of
the center-of-mass position, solution of the equation
$\hat{J}^{\tau\,r}_{ADM}\approx 0$. However the equation for the
relative motion (see below) do not permit to say if the
accelerations $d^2\tilde{\rho}_{(0)12}^r(t)/dt^2$ and
$d^2\tilde{\rho}_{(1)12}^r(t)/dt^2$ are of different order or if
they are of the same order. Therefore in this Subsection we do not
use this decomposition.

\medskip

As said after Eqs.(\ref{5.15}), the vanishing of the ADM Lorentz
boosts implies

\beq
 {\vec {\tilde \eta}}_{(o)12}(t) \approx 0 + O({{v^2}\over {c^2}}).
 \label{6.8}
 \eeq

\noindent It eliminates the non-relativistic 3-center of mass by
putting it in the origin of the 3-coordinates, ${\vec {\tilde
\eta}}_{NR}(t) \approx 0$. Therefore we have ${\tilde
\eta}_{12}^r(t) \approx {\tilde \eta}^r_{(1)12}(t) = O(\zeta)$.

\medskip

Then the sum and the difference of the two Eqs.(\ref{6.6}) gives the
following equations of motion for the center of mass position
${\tilde \eta}^r_{(1)12}(t)$ and for the relative variable ${\tilde
\rho}_{12}^r(t)$ (we use the notation $v^r(t)=d{\tilde
\rho}_{12}^r(t)/dt$)

\begin{eqnarray*}
  {{d^2\, {\tilde \eta}_{(1)12}^r(t)}\over {dt^2}} &\cir& -
  {{\mu}\over M}\, {1\over c}\,
 {{ d {\tilde \rho}_{12}^r(t)}\over {dt}}\, \gamma_{-}(t, {\vec {\tilde
 \rho}_{12}}(t),\vec{v}(t)) \nonumber \\
 &&{}\nonumber \\
  {{d^2\, {\tilde \rho}_{(1)12}^r(t)}\over {dt^2}}
  &\cir& - G\,  M\, {{{\tilde \rho}_{12}^r(t)}
 \over {|{\vec {\tilde \rho}_{12}}(t)|^3}} -
  {1\over  c}\, {{ d {\tilde \rho}_{12}^r(t)}\over {dt}}\,
 \gamma_{+}(t, {\vec {\tilde
 \rho}_{12}}(t),\vec{v}(t))  ,
 \end{eqnarray*}

\bea
  \gamma_{+}(t, {\vec {\tilde
 \rho}_{12}}(t),\vec{v}(t)) &=& {{m_1}\over M}\, \gamma_1(t, {{m_2}\over M}\, {\vec {\tilde
 \rho}_{12}}(t),\vec{v}(t)) + {{m_2}\over M}\,  \gamma_2(t, - {{m_1}\over M}\, {\vec {\tilde
 \rho}_{12}}(t),\vec{v}(t)),\nonumber \\
 &&\nonumber\\
 \gamma_{-}(t, {\vec {\tilde
 \rho}_{12}}(t),\vec{v}(t)) &=& \gamma_1(t, {{m_2}\over M}\, {\vec {\tilde
 \rho}_{12}}(t),\vec{v}(t)) - \gamma_2(t, - {{m_1}\over M}\, {\vec {\tilde
 \rho}_{12}}(t),\vec{v}(t)),\nonumber \\
 &&\nonumber\\
 \gamma_1(t, {{m_2}\over M}\, {\vec {\tilde \rho}_{12}}(t),\vec{v}(t)) &=&
  \Big(\partial_t^2{|}_{{\vec {\tilde \eta}}_1} +
 2\, {{m_2}\over M}\, \sum_u\, v^u(t)\,
 \partial_u\, \partial_t{|}_{{\vec {\tilde \eta}}_1} +\nonumber \\
 &+& ({{m_2}\over M})^2\, \sum_{uv}\, v^u(t)\,
 v^v(t)\, \partial_u\,
 \partial_v \Big)\, {}^3{\tilde {\cal K}}_{(1)}(t, {{m_2}\over M}\,
 {\vec {\tilde \rho}_{12}}(t)),\nonumber \\
 &&\nonumber\\
 \gamma_2(t, -{{m_1}\over M}\, {\vec {\tilde \rho}_{12}}(t),\vec{v}(t)) &=&
  \Big(\partial_t^2{|}_{{\vec {\tilde \eta}}_1} -
 2\, {{m_1}\over M}\, \sum_u\, v^u(t)\,
 \partial_u\, \partial_t{|}_{{\vec {\tilde \eta}}_1} +\nonumber \\
 &+& ({{m_1}\over M})^2\, \sum_{uv}\, v^u(t)\,
 v^v(t)\, \partial_u\,
 \partial_v \Big)\, {}^3{\tilde {\cal K}}_{(1)}(t,- {{m_1}\over M}\,
 {\vec {\tilde \rho}_{12}}(t)),.
 \label{6.9}
 \eea
\medskip

We want estimate the contribution of the York time to the rotation
curves. Motivated by the $1/c$ factor in front of this term, we can
treat the York time term $v/c\,\gamma_+$ in the relative motion
equation as a {\em perturbative term} added to usual Kepler problem
${{d^2\, {\tilde \rho}_{12}^r(t)}\over {dt^2}}\, \cir\, -  G\,  M\,
{{{\tilde \rho}_{12}^r(t)} \over {|{\vec {\tilde
\rho}_{12}}(t)|^3}}$. To realize this explicitly we take a solution
with circular trajectory such that $\mid {\vec {\tilde
\rho}_{12}}(t)\mid=R=constant$ and we make a decomposition of the
velocity $\vec{v}(t)=\vec{v}_o(t)+\vec{v}_1(t)$ such that
$\vec{v}_o(t)=v_o\,\hat{n}(t)$ is the keplerian velocity such that
$v_o = \sqrt{{{G\, M}\over {R}}} \rightarrow_{R \rightarrow
\infty}\, 0$ and where $v_1(t)$ is the first order perturbative
correction such that

 \beq
  \frac{dv^r_1(t)}{dt}=-{v^r_o\over  c}\, \hat{n}(t)\,
 \gamma_{+}(t, {\vec {\tilde \rho}_{12}}(t),\vec{v}^r_o(t)).
 \label{6.10}
\eeq

Then we get

\bea
 v_1^r(t)=-\frac{v_o^r}{c}\,\int^t dt_1\,\hat{n}(t_1)\,
 \gamma_{+}(t_1, {\vec {\tilde
 \rho}_{12}}(t_1),\vec{v}_o(t_1)).
 \label{6.11}
 \eea

\medskip

At the first order in the perturbation we get

\bea
 v^2(t)=v_o^2\,\left(\,
 1-\frac{2}{c}\,\hat{n}(t)\cdot\,\int_o^t dt_1\,\hat{n}(t_1)\,
 \gamma_{+}(t_1, {\vec {\tilde
 \rho}_{12}}(t_1),\vec{v}_o(t_1))
 \,\right).
 \label{6.12}
\eea

\medskip

Therefore, after having taken a mean value over a period $T$ (the
time dependence of the mass of a galaxy is not known) the effective
mass of the two-body system is

\bea
 M_{eff} &=& \frac{\langle v^2\rangle\,R}{G}= M\, \left(\,1 -\left\langle
 \frac{2}{c}\,\hat{n}(t)\cdot\,\int^t dt_1\,\hat{n}(t_1)\,
 \gamma_{+}(t_1, {\vec {\tilde
 \rho}_{12}}(t_1),\vec{v}_o(t_1))\right\rangle
 \right)) =\nonumber\\
 &&\nonumber\\
 &=& M_{bar} + M_{DM}.
 \label{6.13}
 \eea

Again the effective inertial mass in 3-space is the sum of the
baryonic matter $M_{bar} = M$ plus a dark matter term.

\bigskip

Let us consider the case $m_1 >> m_2$. Let $m_1$ be the visible mass
of a galaxy and let $m_2$ be the mass of either a star or a gas
cloud circulating around the galaxy outside outside its visible
radius. If the 3-space is Euclidean and the Keplerian orbit circular
we have that the velocity goes to zero when the distance from the
galaxy increases, since ${{d\, {\vec {\tilde r}}_{kepl,
circ}(t)}\over {dt}} = v_o\, \hat n(t) = \sqrt{{{G\, M}\over R}}
\hat n(t) \rightarrow_{R \rightarrow \infty} 0$. Instead from
observations one finds that the velocity tend to a constant (till
where it can be measured) and this so-called problem of the rotation
curves of galaxies supports the existence of {\it dark matter
haloes} around the galaxy (see for instance Ref.\cite{46} for a
review). Again dark matter is an inertial relativistic effect and
the experimental dark matter distributions can be used to get
informations on the non-local York time.

\medskip

This explanation of dark matter differs:\hfill\break
  1) from the non-relativistic MOND approach \cite{47} (where one
  modifies Newton equations);\hfill\break
   2) from modified gravity theories like the $f(R)$ ones
 (see for instance Refs.\cite{48}; here one gets a modification of
the Newton potential);\hfill\break
 3) from postulating the existence of WIMP particles \cite{49}.

 Let us also remark that the 0.5PN effect
has origin in the lapse function and not in the shift one, as in the
gravito-magnetic elimination of dark matter proposed in
Ref.\cite{50}.

\subsection{The Non-Local York Time from Dark Matter Data}

As a consequence of the non-Euclidean nature of 3-space in Einstein
space-times there is the possibility of describing part (or maybe
all) dark matter as a {\it relativistic inertial effect}. As we have
seen the three main experimental signatures of dark matter can be
explained in terms of the non-local York time ${}^3{\cal
K}_{(1)}(\tau, \vec \sigma)$, the inertial gauge variable describing
the general relativistic remnant of the gauge freedom in clock
synchronization.\medskip

The open problem is the determination of the non-local York time
from the data. From what is known from the Solar System and from
inside the Milky Way near the galactic plane, it seems that it is
negligible near the stars inside a galaxy. On the other hand, it is
non zero near galaxies and clusters of galaxies of big mass. However
only a mean value in time of time- and space-derivatives of the
non-local York time can be extracted from the data. At this stage it
seems that the non-local York time is relevant around the galaxies
and the clusters of galaxies where there are big concentrations of
mass and the dark matter haloes and that it becomes negligible
inside the galaxies where there is a lower concentration of mass.
Instead there is no indication on its value in the voids existing
among the clusters of galaxies.

\medskip

However to get an experimental determination of the York time
${}^3K_{(1)}(\tau, \vec \sigma) = \triangle\, {}^3{\cal
K}_{(1)}(\tau, \vec \sigma)$ we would need to know the non-local
York time on all the 3-universe: phenomenological parametrizations
of ${}^3{\cal K}_{(1)}(\tau, \vec \sigma)$ will have to be devised
to see the implications for ${}^3K_{(1)}(\tau, \vec \sigma)$. As
said in the Conclusions of paper II, a phenomenological
determination of the York time would help in trying to get a PM
extension of the Celestial reference frame (ICRS). This would be the
way out from the gauge problem in general relativity: the
observational conventions for matter would select a reference system
of 4-coordinates for PM space-times in the associated 3-orthogonal
gauge.

\vfill\eject

\section{Conclusions}

In this paper we ended the study of the PM linearization of ADM
tetrad gravity in the York canonical basis for asymptotically
Minkowskian space-times in the family of non-harmonic 3-orthogonal
gauges parametrized by the York time ${}^3K(\tau, \vec \sigma)$, the
trace of the extrinsic curvature of the 3-spaces. This inertial
gauge variable, not existing in Newton gravity, describes the
general relativistic remnant of the freedom in clock
synchronization: its fixation gives the final identification of the
instantaneous 3-spaces, after that their main structure has been
dynamically determined by the solution of the Hamilton equations
replacing Einstein equations. It turns out that at the PM level all
the quantities depend on the spatially non-local quantity ${}^3{\cal
K} = {1\over {\triangle}}\, {}^3K$ (the non-local York time) with
the only exception of the ADM Lorentz generators.\medskip

As matter we consider only N scalar point particles (without the
transverse electro-magnetic field present in papers I and II) with a
Grassmann regularization of the self-energies and with a ultraviolet
cutoff making possible the HPM linearization and the evaluation of
the PM solution for the gravitational field.\medskip

We studied in detail all the properties of these PM space-times
emphasizing their dependence on the gauge variable ${}^3{\cal K} =
{1\over {\triangle}}\, {}^3K$ (the non-local York time): Riemann and
Weyl tensors, 3-spaces, time-like and null geodesics, red-shift and
luminosity distance. Also the Ashtekar variables of PM space-times
were evaluated in the York canonical basis. We also studied the PM
equations of motion of the particles, their PN expansion and the PM
problem of the determination of center-of-mass and relative
variables for the particles.\medskip

All the main measurable quantities turn out to have a dependence on
the non-local York time. However it seems plausible that inside the
Solar system  and in every nearly isolated binary system this gauge
quantity is negligible. This is not  true at the astrophysical level
especially for galaxies and clusters of galaxies with a big mass.
\medskip

In Section VI we have shown that the main features of the
experimental signatures for dark matter (masses of clusters of
galaxies, rotation curves of spiral galaxies) can be explained in
terms of the non-local York time at the 0.5PN level in the PN
expansion.

\medskip

This opens the possibility {\it to explain dark matter inside
Einstein theory without modifications as a relativistic inertial
effect}: the determination of ${}^3{\cal K}_{(1)}$ from the mass and
the rotation curves of galaxies \cite{44,46} would give information
on how to find a PM extension of the existing PN Celestial reference
frame (ICRS) used as observational convention in the 4-dimensional
description of stars and galaxies.\medskip

Therefore what is called dark matter would be an indicator of the
non-Euclidean nature of 3-spaces as 3-sub-manifolds of space-time
(extrinsic curvature effect), whose internal 3-curvature can be very
small if it is induced by GW's. It is the Newtonian equality of
inertial and gravitational masses in Euclidean 3-space to be
violated, not their equality in the 4-dimensional space-time implied
by the equivalence principle.\medskip

This conclusion derives from the analysis of the {\it gauge problem
in general relativity} done in the Conclusions of paper II. The
gauge freedom of space-time 4-diffeomorphisms implies that a gauge
choice is equivalent to the choice of a set of 4-coordinates in the
atlas of the space-time 4-manifold and that the observables are
4-scalars. At the Hamiltonian level the gauge group is deformed and
the Hamiltonian observables are the Dirac observables (DO), which
generically are only 3-scalars of the 3-space. However, for the
tidal variables and the electro-magnetic field there is the
possibility (under investigation by using the Newman-Penrose
formalism \cite{51}) that  4-scalar DO's describing them could
exist.
\medskip

On the other side at the experimental level {\it the description of
baryon matter is intrinsically coordinate-dependent}, namely is
connected with the conventions used by physicists, engineers and
astronomers for the modeling  of space-time. As a consequence of the
dependence on coordinates of the description of matter, our proposal
for solving the gauge problem in our Hamiltonian framework with
non-Euclidean 3-spaces is to choose a gauge (i.e. a 4-coordinate
system) in non-modified Einstein gravity which is in agreement with
the observational conventions in astronomy. Since ICRS \cite{5} has
diagonal 3-metric, our 3-orthogonal gauges are a good choice. We are
left with the inertial gauge variable ${}^3{\cal K}_{(1)} = {1\over
{\triangle}}\, {}^3K_{(1)}$ not existing in Newtonian gravity. As
already said the suggestion is to try to fix ${}^3{\cal K}_{(1)}$ in
such a way to eliminate as much dark matter  as possible, by
reinterpreting it as a relativistic inertial effect induced by the
shift from Euclidean 3-spaces to non-Euclidean ones (independently
from cosmological assumptions). As a consequence, ICRS should be
reformulated not as a {\it quasi-inertial} reference frame in
Galilei space-time, but as a reference frame in a PM space-time with
${}^3K_{(1)}$ (i.e. the clock synchronization convention) deduced
from the data connected to dark matter. Then automatically BCRS
would be its quasi-Minkowskian approximation (quasi-inertial
reference frame in Minkowski space-time) for the Solar System. This
point of view could also be useful for the ESA GAIA mission
(cartography of the Milky Way) \cite{52} and for the possible
anomalies inside the Solar System \cite{14}.

\bigskip

Moreover our approach will require further developments in  the
following directions:\medskip

a) Find the second order of HPM to see whether in PM space-times
there is the emergence of hereditary terms (see Refs.\cite{6,53})
like the ones present in harmonic gauges. Like in standard
approaches (see the review in Appendix A of paper II) regularization
problems may arise at the higher orders.\medskip

b) Study  the PM equations of motion of the transverse
electro-magnetic field trying to find Lienard-Wiechert-type
solutions (see Subsection VB of paper II).\medskip

c) Dark energy in cosmology \cite{54}. Take a perfect fluid as
matter in the first order of HPM expansion \cite{55} adapting to
tetrad gravity the special relativistic results of Refs.\cite{56}.
Since in our formalism all the canonical variables in the York
canonical basis, except the angles $\theta^i$, are 3-scalars, we can
complete Buchert's formulation of back-reaction \cite{57} (see also
Ref.\cite{58}) by taking the spatial average of all the PM Hamilton
equations in our non-harmonic 3-orthogonal gauges. This will allow
to make the transition from the PM space-time 4-metric to an
inhomogeneous cosmological one (only conformally related to
Minkowski space-time at spatial infinity) and to reinterpret the
dark energy as a non-linear effect of inhomogeneities. The role of
the York time, now considered as an inertial gauge variable, in the
theory of back-reaction  and in the identification of what is called
dark energy \footnote{As we have seen the red-shift and the
luminosity distance depend upon the York time, and this could play a
role in the interpretation of the data from super-novae.} is
completely unexplored.\medskip

Let us remark that in the Friedmann-Robertson-Walker (FLW)
cosmological solution the Killing symmetries connected with
homogeneity and isotropy imply ($\tau$ is the cosmic time, $a(\tau)$
the scale factor) ${}^3K (\tau) = - {{\dot a(\tau)}\over {a(\tau)}}
= - H$, namely the York time is the Hubble constant. However at the
first order in cosmological perturbations we have ${}^3K = - H +
{}^3K_{(1)}$ with ${}^3K_{(1)}$ being an inertial gauge variable.
Instead in the spatial-averaging method of Ref.\cite{57} one gets
that the spatial average of the York time (a 3-scalar gauge
variable) gives the effective Hubble constant of that approach.
Therefore we will try to see if also dark energy can be considered
as an inertial effect of the York time \footnote{In paper I we
showed that in the York canonical basis the York time contributes
with a negative term to the kinetic energy in the ADM energy. It
would also play a role in study to be done on the reformulation of
the Landau-Lifschitz energy-momentum pseudo-tensor as the
energy-momentum tensor of a viscous pseudo-fluid. It could be
possible that for certain choices of the York time the resulting
effective equation of state has negative pressure, realizing in this
way a simulation of dark energy.} in the transition from
astrophysics to cosmology.

\vfill\eject

\appendix

\section{The PN expansion of the Weak ADM Poincare' Generators.}

The 1PN expansion of the ADM Poincare' generators of Eqs.
(\ref{2.10})-(\ref{2.13}) with $O(\zeta^2)$ order included is

\begin{eqnarray*}
 {\hat E}_{ADM} &=& \sum_i\, \eta_i\, m_i\, c^2 +
 \sum_i\, \eta_i\, {1\over 2}\, m_i\, \Big[1 + {1\over 4}\,
 {{{\vec v}_i^2(t)}\over {c^2}}\, \Big]\, {\vec v}_i^2(t) +
 Mc^2\,  O(\zeta)\, O({{v^6}\over {c^6}})  -\nonumber \\
  &-&\sum_i\, \eta_i\, m_i c^2\, \Big(1 - {1\over 2}\,
  {{{\vec v}_i^2(t)}\over {c^2}}\Big)\, {{{\vec v}_i(t)}\over c}
  \cdot \vec \partial\, {}^3{\tilde {\cal K}}_{(1)}(t,
  {\vec {\tilde \eta}}_i(t)) -\nonumber \\
 &-& G\, \sum_{i > j}\, \eta_i\, \eta_j\, {{m_i\, m_j}\over {|{\vec
 {\tilde \eta}}_i(t) - {\vec {\tilde \eta}}_j(t)|}}\, \Big[1 +
 {3\over 2}\, {{{\vec v}_j^2(t)}\over {c^2}}
 - {7\over 2}\, {{{\vec v}_i(t) \cdot {\vec v}_j(t)}\over {c^2}}
 -\nonumber \\
 &-& {1\over 2}\, {{  {{{\vec v}_i(t)}\over c} \cdot ({\vec {\tilde \eta}}_i(t) - {\vec {\tilde
 \eta}}_j(t))\, {{{\vec v}_j(t)}\over c} \cdot ({\vec {\tilde \eta}}_i(t) - {\vec {\tilde
 \eta}}_j(t))}\over {|{\vec {\tilde \eta}}_i(t) - {\vec {\tilde
\eta}}_j(t)|^2}}\Big] +\nonumber \\
 &+& Mc^2\, O(\zeta^2)\,O({{v^4}\over {c^4}}) + Mc^2\, O(\zeta^3),
 \end{eqnarray*}

\begin{eqnarray*}
 {\hat J}^{rs}_{ADM} &=& \sum_i\, \eta_i\, m_i\, \Big(1 + {1\over 2}\,
 {{{\vec v}_i^2(t)}\over {c^2}}\Big)\, \Big({\tilde \eta}_i^r(t)\, v^s_i(t) -
 {\tilde \eta}_i^s(t)\, v_i^r(t)\Big) + Mc\, L\, O(\zeta)\, O({{v^5}\over {c^5}}) +\nonumber \\
 &+& \sum_i\, \eta_i\, \Big({\tilde \eta}^r_i(t)\, {{\partial}\over
 {\partial\, {\tilde \eta}^s_i}} - {\tilde \eta}^s_i(t)\, {{\partial}\over
 {\partial\, {\tilde \eta}^r_i}} \Big)\, m_i\, \Big(c\, (1 +
 {{{\vec v}_i^2(t)}\over {2\, c^2}})\, {}^3{\tilde {\cal K}}_{(1)} -\nonumber \\
 &-&2\, \sum_u\, v_i^u(t)\, {{\partial_u}\over {\triangle}}\,
 \Big[{\tilde \Gamma}_u^{(1)} - {1\over 4}\, \sum_c\, {{\partial_c^2}\over {\triangle}}\,
 {\tilde \Gamma}_c^{(1)}\Big]\Big)(t, {\vec {\tilde \eta}}_i(t)) +\nonumber \\
 &+&2\,  \sum_i\, \eta_i\, m_i\, \Big[v_i^r(t)\, {{\partial_s}\over {\triangle}}\,
 \Big({\tilde \Gamma}_s^{(1)} - {1\over 4}\,
 \sum_c\, {{\partial_c^2}\over {\triangle}}\,
 {\tilde \Gamma}_c^{(1)}\Big) -\nonumber \\
 &-& v_i^s(t)\, {{\partial_r}\over {\triangle}}\,
 \Big({\tilde \Gamma}_r^{(1)} - {1\over 4}\,
 \sum_c\, {{\partial_c^2}\over {\triangle}}\,
 {\tilde \Gamma}_c^{(1)}\Big)\Big](t, {\vec {\tilde \eta}}_i(t))
 +\nonumber \\
 &+& {{C^3}\over {4\pi\, G}}\, \int d^3\sigma\, \Big({}^3{\tilde K}_{(1)}\,
 \partial_r\, \partial_s\, (\Gamma^{(1)}_r - \Gamma_s^{(1)})\Big)(t, \vec \sigma)
 + Mc\, L\, O(\zeta^2)\, O({{v^5}\over {c^5}}) + Mc\, L\, O( \zeta^3),
 \end{eqnarray*}

\begin{eqnarray*}
  {\hat P}^r_{ADM} &=&\sum_i\, \eta_i\, m_i\, \Big( \Big[\Big(1 +
 {{{\vec v}_i^2(t)}\over {2\, c^2}}\Big) \, v_i^r(t) + Mc\,  O(\zeta)\, O({{v^5}\over {c^5}})
  -\nonumber \\
 &-&\sum_i\, \eta_i\, m_i\, \sum_a\, v_i^a(t)\, {{\partial_r\,
 \partial_a}\over {\triangle}}\, \Big(2\, {\tilde \Gamma}_a^{(1)} - {1\over 2}\,
 \sum_c\, {{\partial_c^2}\over {\triangle}}\, {\tilde \Gamma}_c^{(1)}\Big)
 (\tau, {\vec {\tilde \eta}}_i(t))  -\nonumber \\
 &-& \sum_i\, \eta_i\, m_i\, c\, (1 + {{{\vec v}_i^2(t)}\over {2\, c^2}})\,
 \partial_r\, {}^3{\tilde {\cal K}}_{(1)}(\tau, {\vec {\tilde \eta}}_i(t))
 + + Mc\,  O(\zeta^2)\, O({{v^5}\over {c^5}}) + Mc\,  O( \zeta^3)  \approx 0,
 \end{eqnarray*}

\begin{eqnarray*}
 {\hat J}^{\tau r}_{ADM} &=& - \sum_i\, \eta_i\, {\tilde
 \eta}^r_i(t)\, m_ic\, \Big(1 + {{{\vec v}_i^2(t)}\over {2\, c^2}}
  + {{{\vec v}_i^2(t)}\over {c^2}}\, {G\over {c^2}}\, \sum_{j \not= i}\,
 \eta_j\, {{m_j}\over {|{\vec {\tilde \eta}}_i(t) - {\vec {\tilde \eta}}_j(t)|}}\Big)
 -\nonumber \\
 &-&\int d^3\sigma\,\sigma^r\,\Big[ \frac{1}{2}\sum_a\,
 \sum_i\,\eta_i\,m_ic\,\frac{\sigma^a - {\tilde \eta}_i^a(t)}
 {4\pi\mid\vec{\sigma} - \vec{\tilde \eta}_i(t)\mid^3}
 \,\partial_a\Big({\tilde \Gamma}_a^{(1)} - \frac{1}{2}\,\sum_c\,
 \frac{\partial_c^2}{\Delta}\, {\tilde \Gamma}_c^{(1)}\Big)+\\
 &+&\frac{32\pi G}{c} \sum_{i\neq j}\,\eta_i\eta_j\,
 m_im_j\,\Big(1+ {{v_i^2(t) + v_j^2(t)}\over {2c^2}}\Big)
 \frac{(\vec{\sigma} - \vec{\tilde \eta}_i(t))\cdot
 (\vec{\sigma} - \vec{\tilde \eta}_j(t))}{16\pi^2\mid\vec{\sigma}
 - \vec{\tilde \eta}_i(t)\mid^3\mid\vec{\sigma} - \vec{\tilde \eta}_j(t)\mid^3}
 -\\
 &-&{2\over c}\, \sum_{a,b}\,\Big(\widetilde{M}_{ab}\,\partial_t\, {\tilde \Gamma}^{(1)}_b\Big)
 \sum_i\,\eta_i\,\frac{m_i\,v_i^a(t)\, (\sigma^a - {\tilde \eta}_i^a(t))}
 {4\pi\mid\vec{\sigma} - \vec{\tilde \eta}_i(t)\mid^3}+\\
 &+&{2\over c}\, \sum_{a\neq b}\,
 \frac{\partial_a\partial_b\partial_t}{\Delta}\Big(
 {\tilde \Gamma}^{(1)}_a + {\tilde \Gamma}^{(1)}_b -
 \frac{1}{2}\,\sum_c\,\frac{\partial_c^2}{\Delta}\,
 {\tilde \Gamma}^{(1)}_c\Big)
 \sum_i\,\eta_i\,\frac{m_i\,v_i^a(t)\, (\sigma^a -
 {\tilde \eta}_i^a(t))}{4\pi\mid\vec{\sigma} - \vec{\tilde \eta}_i(t)\mid^3}-\\
 &-&\frac{c^2}{8\pi
 G}\sum_{a,b}\,\Big(\widetilde{M}_{ab}\,\partial_t\, {\tilde \Gamma}^{(1)}_b\Big)
 \frac{\partial_a^2}{\Delta}\Big( {}^3{\tilde K}_{(1)} - \frac{4\pi
 G}{c^3}\sum_i\,\eta_i\,\frac{m_i\,\vec{v}_i(t) \cdot (\vec{\sigma} -
 \vec{\tilde \eta}_i(t))}{4\pi\mid\vec{\sigma} - \vec{\tilde \eta}_i(t)\mid^3}\Big)
 +\\
 &+&\frac{c^2}{8\pi G}\sum_{a\neq b}\,
 \frac{\partial_a\partial_b\partial_t}{\Delta}\Big(
 {\tilde \Gamma}^{(1)}_a + {\tilde \Gamma}^{(1)}_b -
 \frac{1}{2}\,\sum_c\,\frac{\partial_c^2}{\Delta}\, {\tilde
 \Gamma}_c^{(1)} \Big)\nonumber \\
 && \frac{\partial_a\partial_b}{\Delta}\Big(
 {}^3{\tilde K}_{(1)} - \frac{4\pi G}{c^3}\sum_i\,\eta_i\,
 \frac{m_i\,\vec{v}_i(t) \cdot (\vec{\sigma}
 - \vec{\tilde \eta}_i(t))}{4\pi\mid\vec{\sigma} - \vec{\tilde \eta}_i(t)\mid^3}\Big)+\\
 &+&\frac{c^3}{16\pi
 G}\sum_{a,b}\,\Big(\frac{\partial_a\partial_b}{\Delta}\, {}^3{\tilde K}_{(1)}\Big)^2 +
 \nonumber \\
 &+& \frac{G}{16\pi c^3}\sum_{a,b}\,\sum_{i\neq j}\eta_i\eta_j\,m_im_j
 \frac{\partial_a\partial_b}{\Delta}\Big(\frac{\vec{v}_i(t) \cdot (\vec{\sigma} -
 \vec{\tilde \eta}_i(t))}{\mid\vec{\sigma} - \vec{\tilde \eta}_i(t)\mid^3}\Big)
 \frac{\partial_a\partial_b}{\Delta}\Big(\frac{\vec{v}_j(t) \cdot (\vec{\sigma} -
 \vec{\tilde \eta}_j(t))}{\mid\vec{\sigma} - \vec{\tilde \eta}_j(t)\mid^3}\Big) -\\
 &-&\frac{1}{8\pi}\sum_{a,b}\,\Big(\frac{\partial_a\partial_b}{\Delta}\, {}^3{\tilde K}_{(1)}\Big)
 \sum_i\eta_i\,m_i\Big(1 + \frac{{\vec v}_i^2(t)}{2c^2}\Big)\frac{\partial_a\partial_b}{\Delta}
 \Big(\frac{\vec{v}_i(t) \cdot (\vec{\sigma} - \vec{\tilde \eta}_i(t))}
 {\mid\vec{\sigma} - \vec{\tilde \eta}_i(t)\mid^3}\Big) +\\
 &+&\frac{1}{2\pi}\sum_{a,b}\,\Big(\frac{\partial_a\partial_b}{\Delta}\, {}^3{\tilde K}_{(1)}\Big)
 \sum_i\eta_i\,m_i\Big(1 + \frac{{\vec v}_i^2(t)}{2c^2}\Big)\frac{v^b_i(t)\,(\sigma^a -
 {\tilde \eta}_i^a(t))}{\mid\vec{\sigma} - \vec{\tilde \eta}_i(t)\mid^3} -\\
 &-&\frac{G}{2\pi c^3}\, \sum_{i\neq j}\eta_i\eta_j\,m_im_j
 \frac{v^b_i(t)\,(\sigma^a - {\tilde \eta}^a_i(t))}{\mid\vec{\sigma} - \vec{\tilde \eta}_i(t)\mid^3}\,
 \frac{\partial_a\partial_b}{\Delta}\Big(\frac{\vec{v}_j(t) \cdot (\vec{\sigma} -
 \vec{\tilde \eta}_j(t))}{\mid\vec{\sigma} - \vec{\tilde \eta}_j(t)\mid^3}\Big) -\\
 &-&\frac{c^3}{72\pi G}\Big(
 {}^3{\tilde K}_{(1)}\Big)^2 - \frac{1}{8\pi}\,{}^3{\tilde K}_{(1)}\,\sum_i\,\eta_i\,m_i\Big(1 +
 \frac{{\vec v}_i^2(t)}{2c^2}\Big)\,\frac{\vec{v}_i(t) \cdot (\vec{\sigma} -
 \vec{\tilde \eta}_i(t))}{\mid\vec{\sigma} - \vec{\tilde \eta}_i(t)\mid^3}+\\
 \end{eqnarray*}

\bea
 &+&\frac{3}{16}\frac{G}{\pi c^3}\, \sum_{i\neq
 j}\,\eta_i\eta_j\,m_im_j\,
 \frac{\vec{v}_i(t) \cdot(\vec{\sigma} - \vec{\tilde \eta}_i(t))}
 {\mid\vec{\sigma} - \vec{\tilde \eta}_i(t)\mid^3}\,
 \frac{\vec{v}_j(t) \cdot (\vec{\sigma} - \vec{\tilde \eta}_i(t))}
 {\mid\vec{\sigma} - \vec{\tilde \eta}_j(t)\mid^3} +\nonumber \\
 &+&\frac{8\pi G}{c^3}\, \sum_{i\ne j}\,
 \eta_i\eta_j\,m_im_j\nonumber \\
 &&\frac{
 \vec{v}_i(t) \cdot\vec{v}_j(t)\, (\vec{\sigma} - \vec{\tilde \eta}_j(t))
 \cdot (\vec{\sigma} - \vec{\tilde \eta}_i(t)) +
 \vec{v}_i(t) \cdot (\vec{\sigma} - \vec{\tilde \eta}_j(t))\,
 \vec{v}_j(t) \cdot (\vec{\sigma} - \vec{\tilde \eta}_i(t))}
 {16\pi^2\, \mid\vec{\sigma} - \vec{\tilde \eta}_i(t)\mid^3
 \mid\vec{\sigma} - \vec{\tilde \eta}_j(t)\mid^3}
 \,\,\Big](t, \vec \sigma) +\nonumber \\
 &+&\frac{3}{2}\,\int d^3\sigma\,
 \sum_i\,\eta_i\frac{m_ic}{4\pi\mid\vec{\sigma} - \vec{\tilde \eta}_i(t)\mid}
 \,\partial_r\, {\tilde \Gamma}^{(1)}_r +\nonumber \\
 &+&\int d^3\sigma\,\partial_r\,\Big[ \frac{2\pi G}{c^3}\, \sum_{i\neq
 j}\,\eta_i\eta_j\,\frac{m_im_jc^2}{16\pi^2\mid\vec{\sigma} -
 \vec{\tilde \eta}_i(t)\mid\mid\vec{\sigma} - \vec{\tilde \eta}_j(t)\mid}
 \Big(1 + {{v_i^2(t) + v_j^2(t)}\over {2c^2}}\Big)
 \,\,\Big] (t, \vec \sigma) +\nonumber \\
 &+& Mc\, L\, O(\zeta)\, O({{v^4}\over {c^4}}) +
  Mc\, L\, O(\zeta^2)\, O({{v^4}\over {c^4}}) +
 Mc\, L\,  O(\zeta^3)  \approx 0.
 \label{5.18}
 \eea

Since we have $R_{\bar a}, \Gamma_a^{(1)} \, = {G\over {c^2}}\,
O({{v^2}\over {c^2}})$ the GW kinetic term in the ADM energy is of
order $Mc^2\, O({{v^4}\over {c^4}})$. The expression of the energy
has been obtained by making an integration over 3-space and by using
the integral given after Eq.(4.17) of paper II.

\vfill\eject

\end{document}